\documentclass{article}

\usepackage{microtype}
\usepackage{graphicx}
\usepackage{subfigure}
\usepackage{booktabs} %
\usepackage{multirow}
\usepackage{lipsum}  
\usepackage{xspace}
\usepackage{tcolorbox}
\usepackage{hyperref}

\usepackage{cuted}
\newcommand{\fixmemode}{final}

\usepackage[multiuser,inline,nomargin,\fixmemode]{fixme}

\usepackage[accepted]{arxiv}

\usepackage{amsmath}
\usepackage{amssymb}
\usepackage{mathtools}
\usepackage{amsthm}

\usepackage[capitalize,noabbrev]{cleveref}

\theoremstyle{plain}

\theoremstyle{definition}

\theoremstyle{remark}

\usepackage[textsize=tiny]{todonotes}
\def\onedot{.\xspace}
\def\eg{\emph{e.g}\onedot} 
\def\ie{\emph{i.e}\onedot}

\newcommand{\ours}{NEXUS\xspace}
\newcommand{\ouragent}{AutoCompliance\xspace}
\newcommand{\expert}{Human expert\xspace}
\newcommand{\perplexity}{Perplexity Pro\xspace}
\newcommand{\chatgpt}{ChatGPT-4o\xspace}

\newcommand{\ourdata}{Data Compliance\xspace}

\newcommand{\entity}{entity\xspace}
\newcommand{\entities}{entities\xspace}

\newcommand{\Entities}{Entities\xspace}

\newcommand{\children}{dependencies\xspace}
\newcommand{\child}{dependency\xspace}

\newcommand{\Children}{Dependencies\xspace}
\newcommand{\Child}{Dependency\xspace}

\newcommand{\score}{risk score\xspace}

\newcommand{\globalclass}{aggregate class\xspace}
\newcommand{\localclass}{individual class\xspace}
\newcommand{\lclass}{\localclass}
\newcommand{\gclass}{\globalclass}

\newcommand{\tb}{\textbf}
\newcommand{\tr}{\textrm}

\DeclareUnicodeCharacter{03B1}{\ensuremath{\alpha}}
\DeclareUnicodeCharacter{03C7}{\ensuremath{\chi}}

\newcommand{\reporturl}{https://nexus.lgresearch.ai/legalframework}

\newcommand{\numfoundlicenses}{8,072\xspace}  %

\icmltitlerunning{Do Not Trust Licenses You See: Dataset Compliance Requires Massive-Scale AI-Powered Lifecycle Tracing}

\begin{document}

\twocolumn[
\icmltitle{Do Not Trust Licenses You See: Dataset Compliance Requires \\Massive-Scale AI-Powered Lifecycle Tracing}

\icmlsetsymbol{equal}{*}

\newlength{\auSpacing}
\setlength{\auSpacing}{2em}

\begin{icmlauthorlist}
\icmlauthor{Jaekyeom Kim}{equal}\hspace{\auSpacing}
\icmlauthor{Sungryull Sohn}{equal}\hspace{\auSpacing}
\icmlauthor{Gerrard Jeongwon Jo}{}\hspace{\auSpacing}
\icmlauthor{Jihoon Choi}{}
\\
\icmlauthor{Kyunghoon Bae}{}\hspace{\auSpacing}
\icmlauthor{Hwayoung Lee}{}\hspace{\auSpacing}
\icmlauthor{Yongmin Park}{}\hspace{\auSpacing}
\icmlauthor{Honglak Lee}{}
\\
\vspace{1em}
LG AI Research
\end{icmlauthorlist}

\vskip 0.3in
]

\printAffiliationsAndNotice{\icmlEqualContribution} %

\ifthenelse{\equal{\fixmemode}{draft}}
  {%
    \begin{strip}
        \listoffixmes
    \end{strip}
  }
  {%
  }

\begin{abstract}
This paper argues that a dataset’s legal risk cannot be accurately assessed by its license terms alone; instead, tracking dataset redistribution and its full lifecycle is essential.
However, this process is too complex for legal experts to handle manually at scale. 
Tracking dataset provenance, verifying redistribution rights, and assessing evolving legal risks across multiple stages require a level of precision and efficiency that exceeds human capabilities.
Addressing this challenge effectively demands AI agents that can systematically trace dataset redistribution, analyze compliance, and identify legal risks.
We develop an automated data compliance system called NEXUS and show that AI can perform these tasks with higher accuracy, efficiency, and cost-effectiveness than human experts.
Our massive legal analysis of 17,429 unique entities and \numfoundlicenses license terms using this approach reveals the discrepancies in legal rights between the original datasets before redistribution and their redistributed subsets, underscoring the necessity of the data lifecycle-aware compliance.
For instance, we find that out of 2,852 datasets with commercially viable individual license terms, only 605 (21\%) are legally permissible for commercialization.
This work sets a new standard for AI data governance, advocating for a framework that systematically examines the entire lifecycle of dataset redistribution to ensure transparent, legal, and responsible dataset management.
\end{abstract}

\section{Introduction}\label{sec:introduction}

Ensuring legal compliance in AI training datasets is becoming an overwhelming challenge for human experts alone, as the complexity and scale of modern datasets far exceed traditional manual review capabilities~\cite{Roberts_2024}.
Unlike simple data repositories, these datasets are structured with multi-level hierarchies and interdependent components, where individual data points originate from diverse sources and are recursively merged, transformed, and redistributed~\cite{khan2022subjects, bhardwaj2024state}.
Despite this inherent complexity, compliance assessments have long relied on surface-level license terms, treating datasets as static entities rather than evolving systems with intricate dependencies~\cite{Lawler_Childress_Hartwig_2023}.
This approach is fundamentally inadequate, as redistribution and integration introduce new legal relationships that cannot be captured without analyzing dataset lineage entirely. 

The necessity of robust legal analysis in AI training datasets has been increasingly recognized~\cite{sandra}.
The construction and utilization of large-scale datasets raise legal issues concerning data usage rights, including copyright, ownership, and privacy.
Legal disputes, such as \citet{NYTvOpenAI2023} and \citet{GettyImagesvStabilityAI2023}, underscore the growing tensions between AI development and regulatory frameworks. Recent research~\citep{buick2024copyright} highlights the risks associated with AI training datasets, emphasizing the necessity of legal standards for responsible AI data usage. While there have been efforts to construct legal frameworks and leverage AI-driven approaches to address compliance challenges~\citep{rajba, tan2024licensegptfinetunedfoundationmodel}, existing methodologies remain limited in tracking the life-cycle of massive datasets and assessing multifaceted risks.

Our \ourdata framework addresses this challenge by moving beyond simple license verification to conduct a holistic legal risk assessment.
By integrating key aspects of copyright law, personal data protection, and unfair competition law, \ourdata evaluates datasets across 18 weighted criteria, considering not just explicit license terms but also data provenance, transformation processes, and redistribution pathways.

\begin{figure*}[!ht]
    \centering
    \includegraphics[width=1.0\textwidth]{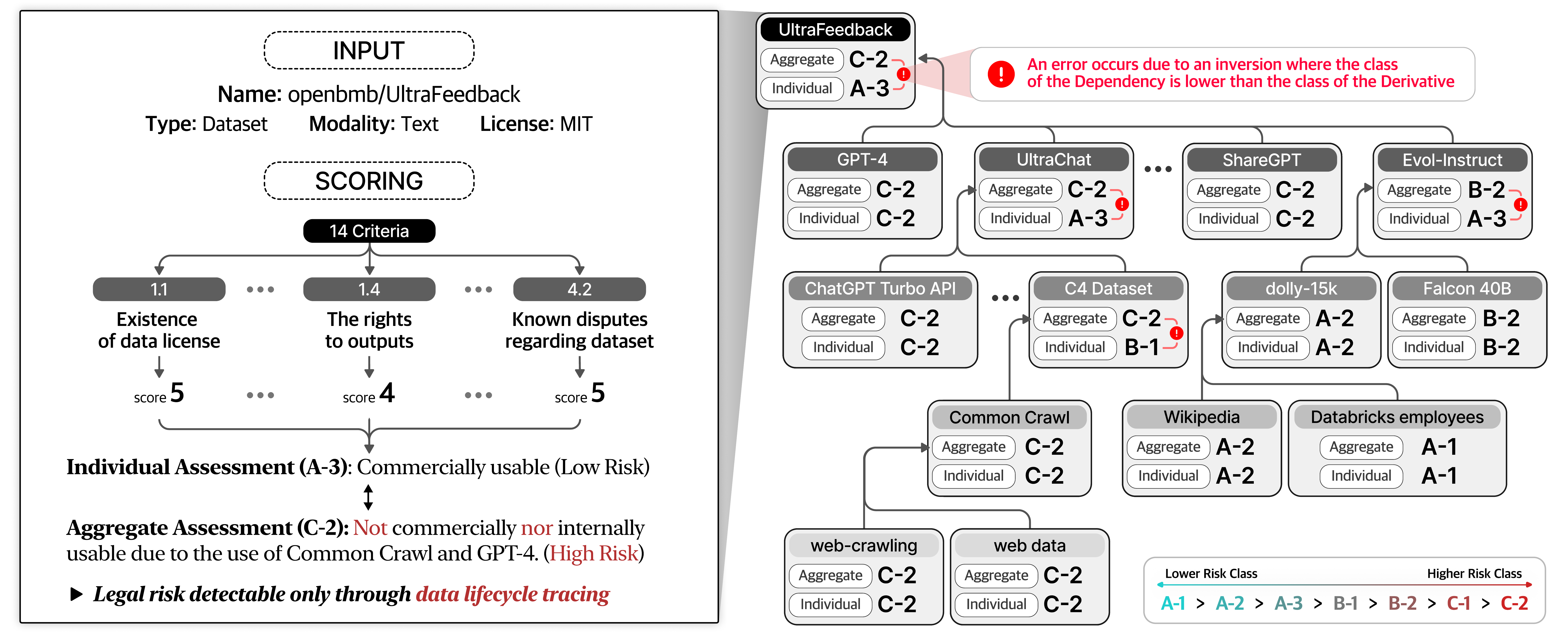}
    \vspace{-2em}    
    \caption{Data Compliance is a multi-layered legal risk assessment framework that evaluates entities through their full data lifecycle. The input includes dataset details such as name, URL, type, modality, and license, which are used to compute a score based on 14 criteria. The score is then used to determine the entity's individual class, and the aggregate class is computed by aggregating the individual classes of all dependencies.}
    \label{fig:class}
\end{figure*}

However, manual end-to-end compliance analysis exceeds human cognitive and physical capabilities~\cite{Samuel_2024}.
While some attempts have been made to trace and record dataset provenance manually~\cite{longpre2023dataprovenanceinitiativelarge}, such efforts fall short of achieving the necessary scalability. 
A scalable, AI-driven approach is essential to bridge this gap.

To this end, our automated data compliance agent, \ouragent streamlines dataset compliance assessment by systematically identifying dataset dependencies (\ie, the sources and tools used in dataset construction) and retrieving their corresponding licensing terms from public resources. Using \ourdata, \ouragent evaluates compliance at each hierarchical level, aggregating individual assessments into a comprehensive risk analysis that accounts for dataset inter-dependencies. This AI-driven methodology replaces the inefficiencies of manual review, ensuring better accuracy, scalability, and transparency in data compliance.

\ouragent assessed data compliance of \emph{17,429} entities and examined \emph{\numfoundlicenses} license terms, and the results reveal key insights about the limitations of current compliance practices. 
First, we show that surface-level license reviews are insufficient—compliance cannot be determined by checking individual license terms in isolation. 
For example, while direct license terms indicated 2,852 datasets were commercially viable, analysis of their dependencies revealed that only 605 (21.21\%) posed a legally permissible level of risk for commercialization.
Dataset redistribution and hierarchical dependencies create hidden risks that remain undetected without a full-lifecycle examination. %
Second, we highlight the limitations of human-driven legal assessments. 
Despite having access to detailed licensing information, human experts struggle to track multi-level dependencies, overlooking more than 35\% of critical dependencies. 
In contrast, \ouragent reduces this gap significantly, missing fewer than 19\%.
The sheer scale of modern datasets and the complexity of legal obligations make manual compliance assessment not just inefficient, but fundamentally unreliable. %

This paper advocates that the legal risk of AI training datasets cannot be determined solely by reviewing surface-level license terms; a thorough, end-to-end analysis of dataset redistribution is essential for ensuring compliance. Since such analysis is beyond human capabilities due to its complexity and scale, AI agents can bridge this gap by conducting it with greater speed and accuracy. Without automation, critical legal risks remain largely unexamined, jeopardizing ethical AI development and regulatory adherence. We urge the AI research community to recognize end-to-end legal analysis as a fundamental requirement and to adopt AI-driven approaches as the viable path to scalable dataset compliance.

\section{Legal Risk Framework of NEXUS}
\label{sec:data_compliance_framework}

\subsection{\ourdata}\label{sec:legal_framework}
The legal risks associated with using third-party data for AI training cannot be adequately assessed through a superficial review that merely examines the licenses attached to AI training data \cite{katzy2024exploratory}.
It is necessary to comprehensively consider relevant laws including copyright law, personal information protection law, and unfair competition prevention law based on international law to ensure safe data utilization, while also evaluating potential legal risks that may arise from the use of the developed AI model in a multidimensional manner \cite{hacker2021legal}.
Furthermore, to assess legal risks comprehensively, one must consider not only the licenses of individual \children comprising the dataset but also the dataset's structure and context.
An accurate legal risk assessment can only be achieved by exploring the network structure formed by lower-level \children and understanding the complex interconnections of the dataset by considering the origin, collection methods, processing procedures, and redistribution relationships of individual \children.
This comprehensive approach is necessary because datasets are not merely collections of data but rather complex networks intertwined through various relationships \cite{roh2019survey}.

In this work, to facilitate comprehensive recognition and management of legal risks through \ouragent, we first establish \ourdata as a legal framework.
The legality of using specific data for AI training is still actively debated across various countries, and judicial conclusions may vary by jurisdiction or governing law across different nations or institutions \cite{blaszczyk2024artificial}.
However, some countries have already issued judicial decisions regarding copyright law violations in AI outputs, and certain aspects of potential legal issues or disputes that may arise during dataset utilization have achieved international consensus, such as copyright law \cite{jaramillo2024comparative}.
We define our \ourdata framework based on various precedents and legal grounds at this point in time.

The significance of \ourdata proposed in this research lies in its ability to preemptively assess and prevent practical legal risks that various stakeholders may face at present, including (i) scientists developing AI models using data, (ii) users utilizing AI models, and (iii) companies providing services using AI models, while establishing a foundation for responsible AI development and utilization.
With \ourdata, we aim to evaluate legal risks comprehensively from a dynamic perspective encompassing the entire data lifecycle.
It enables analysis of legal issues that may arise at each stage of distribution, including collection methods, processing and modification, AI model training, service provision, potential inclusion of personal information, and disputed data.
We also present a report that provides further details of our legal assessment framework\footnote{\url{\reporturl}}.

\subsection{Legal Risk Assessment of Individual \Entities}\label{sec:legal_framework_individual}
\begin{table}[!ht]
\footnotesize
\centering
\begin{tabular}{p{0.95\linewidth}} 
\toprule    
\textbf{1. Risk Related to Data License} \\  
\midrule
\textit{1.1.} The \emph{existence of a license} to use the data  \\ 
\textit{1.2.} Authorization to \emph{modify data} and produce derivative works  \\ 
\textit{1.3.} The \emph{potential for dispute} arising from the outputs  \\ 
\textit{1.4.} The \emph{rights to outputs}  \\ 
\textit{1.5.} The existence of an \emph{obligation to notify data usage}  \\ 
\midrule 

\textbf{2. Risks Related to Data Use Period and Territory} \\  
\midrule
\textit{2.1.} \emph{Restrictions} on data use period \\ 
\textit{2.2.} Whether the data license is \emph{revocable}  \\ 
\textit{2.3.} \emph{Restrictions} on AI model service period  \\ 
\textit{2.4.} \emph{Restrictions} on data use territory  \\ 
\midrule     

\textbf{3. Risks Related to Personal Information and Data Security} \\  
\midrule
\textit{3.1.} Whether \emph{personal data} is included in AI training data  \\ 
\textit{3.2.} Whether \emph{data subjects have consented} to the use of their data  \\ 
\textit{3.3.} Whether \emph{pseudonymized data} is included in AI training data  \\ 
\textit{3.4.} Whether \emph{personal data may be entrusted} or provided to third parties  \\ 
\textit{3.5.} Whether the \emph{scope of data users} is limited  \\ 
\midrule 

\textbf{4. Additional Legal Risk} \\  
\midrule
\textit{4.1.} Risks in the \emph{data collection process}  \\ 
\textit{4.2.} Known \emph{disputes involving the use of the same dataset} in AI models  \\ 
\textit{4.3.} Other \emph{contract risks} associated with licenses  \\ 
\textit{4.4.} Type of \emph{license terms} \\ 
\bottomrule
\end{tabular}

\caption{The assessment criteria for our \ourdata framework. The detailed information on each criterion and the scoring methodology can be found in Appendix \ref{sec:legal_framework_appendix}.}
\label{tab:datacompliancepre}
\end{table}

For the comprehensive evaluation with our \ourdata framework, we start by defining an \emph{\entity} to be a single component that may pose legal terms or risks for its use.
There can be different types of \entities, such as datasets, data processing software, and AI models for generating and processing data, with more detailed discussions provided in \cref{sec:compre}.

Data Compliance consists of 18 assessment criteria, as outlined in \Cref{tab:datacompliancepre}, while this paper utilizes 14 of them. Among these, Criterion 1.1 does not contribute to score calculation but instead pre-determines the class itself.
We design each of our criteria to be answered with an integer on a scale of 1 to 5, with 5 indicating the lowest risk and 1 indicating the highest risk.
Given the integer answer $R_{c}(\text{\entity})$ to each criterion $c$ for the \entity, the \emph{\score} for the given \entity is calculated as
\begin{equation}
\begin{aligned}
R(\text{\entity}) = \sum_{c} w_{c} R_{c}(\text{\entity}) \label{eq:score}
\end{aligned}
\end{equation}
where $w_{c}$'s are the weights for the criteria described in \Cref{tab:datacompliance}, except for Criteria 1.1 and 4.4.
Then, based on the calculated \score, we determine the \emph{\localclass} for the \entity according to the categories described in \Cref{tab:datacompliancecat}, except for Criteria 3.2, 3.3, and 3.4, which are primarily used within enterprises and thus will not be introduced separately in this paper.

\subsection{Comprehensive Assessment of \Entities with Full Data Lifecycle Traces}\label{sec:compre}
Based on our assessment framework for individual \entities presented in \Cref{sec:legal_framework_individual}, we now extend our framework to take into account the full data lifecycles, making it a comprehensive framework for legal risk assessment, which we call \emph{\ourdata} framework.
Specifically, for a dataset of interest, we consider every \entity that is involved in the creation of the dataset.
We approach this challenge by defining the notion of \emph{\children} and \emph{license dependency graph}.
For each \entity, we define its \emph{\children} as the \entities that are directly used for building the \entity.
Again, the \children can have their own \children, which requires further investigations stemming from them.
This essentially leads us to construct the \emph{license dependency graph}, where the \entities are the nodes and the dependencies are the edges between them, with the \entity of interest being the \emph{root} of the graph.
The license dependency graph for the dataset intuitively captures the legal coherence of all subordinate \children and provides the full picture of the data lifecycles for the dataset.

For the legal risk assessment of each \entity in the license dependency graph, we define two types of legal risk classes.
The \tb{\localclass} refers to the class derived from the individual assessment solely about each \entity itself as described in \Cref{sec:legal_framework_individual}, without considering any of its dependencies.
The \textbf{\globalclass} is the class determined based on the \emph{aggregate score}, which is computed as
\begin{align}
    R_{\text{agg}}(\text{\entity}) = \sum_{c} w_{c} \min_{d \in D(\text{\entity})} R_{c}(d) \label{eq:aggregate_score}
\end{align}
where $w_{c}$'s are the same weights as in \Cref{eq:score} and $D(\text{\entity})$ is the set of nodes from the license dependency graph that stems from the \entity as the root, including the \entity itself.
Intuitively, the \globalclass takes into account the biggest risk from the license dependency graph for each assessment criterion.
The \globalclass should be used to accurately assess legal risks for datasets with multi-layered structures.
In case a dataset does not have any dependencies, its \localclass and \globalclass are identical.

As shown in Figure \ref{fig:class}, the \emph{inversion phenomenon} occurs when a dataset at a higher hierarchical level in the license dependency graph is assigned a lower \localclass than one of its direct and indirect dependencies. 
This suggests a misalignment in legal rights interpretation, where a derivative dataset appears legally safer than its source, potentially due to incomplete or inconsistent license tracking, leading to errors in understanding legal rights relationships or assessing risks based on available information about the higher-level \entity. 
For instance, if an \emph{inversion phenomenon} occurs where lower-level \entities in the graph prohibit commercial use of copyrighted works while the higher-level redistributed \entities state no restrictions on use, potential risks exist when making use of such datasets. 
To address and evaluate these risks, it is essential to clearly classify and organize \entities within a consistent framework.
To this end, we also determine and employ the type of each \entity, in the following manner.

\paragraph{Dataset.} When data is directly imported in part or whole from an existing dataset, or reconstructed for specific purposes, the used Dataset is defined as the \child. For example, this refers to pre-existing datasets like the MMLU dataset \cite{MMLU}.

\paragraph{Contents Service Provider.} When data is sourced from services that directly produce and provide specific content, the Content Service Provider is defined as the \child. For example, this refers to data created by content providers like The New York Times that provide news content.

\paragraph{Underspecified.} When data or content is not clearly specified and is expressed comprehensively or vaguely using pronouns or data characteristics, such expressions are defined as Underspecified. For instance, when something is simply described as ``a collection of books gathered from websites'', it is considered as collecting copyrighted works through crawling.

\paragraph{Platform Service Provider.} Platforms that don't directly create and supply content but provide functionality for collecting or modifying data are defined as \children. This includes cases where data is collected through platforms like Amazon Mechanical Turk.

\paragraph{AI Model.} When specific AI models are used to generate/augment data for dataset utilization, the used AI Model is defined as the \child. For example, this classification system evaluates restrictions such as prohibitions on creating competitive AI models when training data is generated using AI models like GPT-4.

\paragraph{Software/API.} When specific tools are used in the process of collecting, preprocessing, processing, and modifying data, the used Software/API is defined as the \child. This refers to tools used for actions like translating data through Google Translation API.

Platform Service Provider, AI Model, and Software/API are not classification systems that directly represent data or content itself.
However, they may impose restrictions on use purposes or outputs, and legal violations could occur if datasets are distributed beyond these restrictions.
Therefore, for \children of such types, only Criteria 1.1 and 1.3 are evaluated.

\begin{figure*}[!ht]
    \centering
    \includegraphics[width=1.0\textwidth]{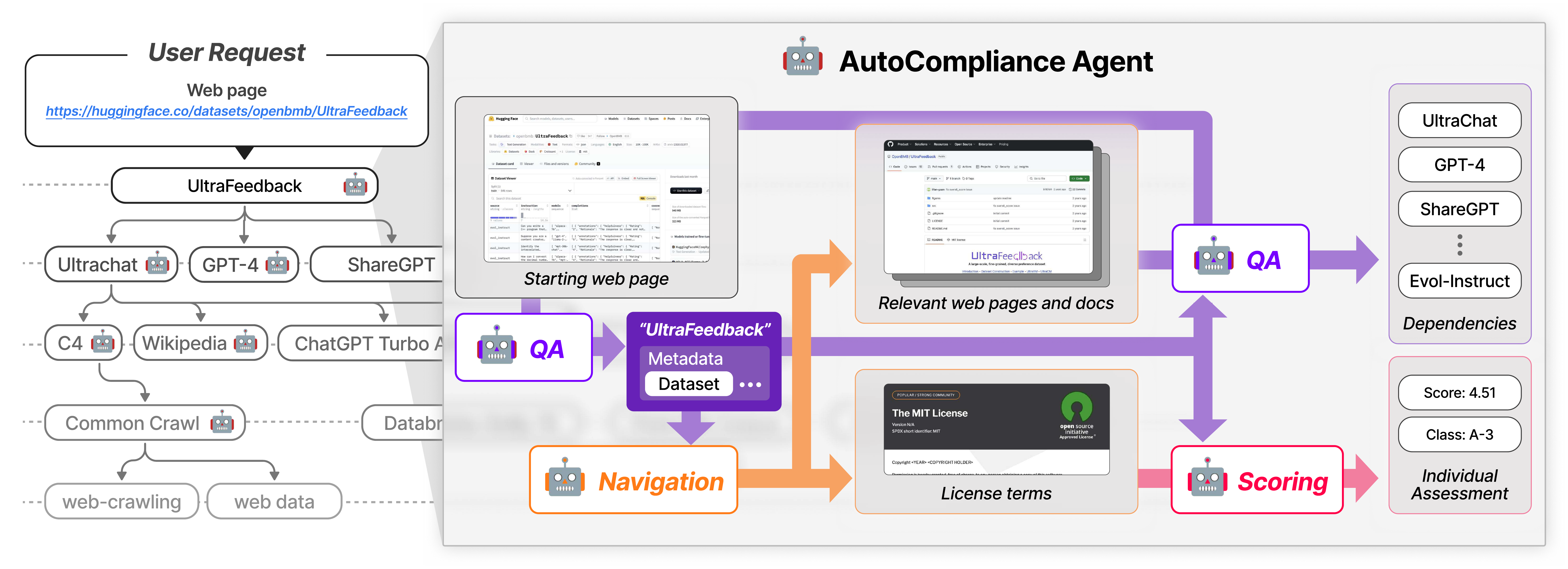}
    \vspace{-2em}    
    \caption{Overview of \ouragent. The user provides the starting web page for the target entity. From the web page, the QA module extracts the information of the target entity, such as name, type, and meta-data. Then, the agent finds the relevant resources on the web to identify the license terms and dependencies. Finally, it uses the target entity information and license terms to evaluate the legal score and individual class.}
    \label{fig:agent_overview}
\end{figure*}

\section{\ouragent: AI Agent for NEXUS}
\label{sec:method}

\subsection{Overview of \ouragent}

Given a user request with the URL for the dataset of interest, \ouragent considers the input dataset as the root \entity and starts performing the comprehensive legal assessment from the root. 
For each \entity, the agent searches for its license terms and its dependency \entities.
Based on a dependency \entity produced for each \entity, the agent recursively spawns a new task that starts from that dependency \entity.
Repeating this process constructs a license dependency graph with the dataset of interest given by the user being the root of the graph.
When there are no more graph expansions, the agent finalizes the license dependency graph and computes the data compliance scores for each \entity in the graph, which are then used to determine the individual and aggregate risk classes.

\Cref{fig:agent_overview} provides an overview of \ouragent and how it performs the comprehensive data compliance assessment based on our \tb{\ourdata} framework.
\ouragent consists of three main components: the navigation, question-answering (QA), and scoring modules.
The navigation module is responsible for navigating the web to find the web pages and documents for the license terms and dependency information, given each \entity.
The QA module performs reasoning tasks, including extracting the dependency \entities given the content of such web pages and documents as input.
The scoring module is a specialized module for inferring the data compliance scores and \localclass given the license terms and other metadata such as name, type, and modality for each \entity.

\subsection{Tracing Dataset Lifecycles with \ouragent}
\ouragent takes as input the URL of the starting web page for the target \entity to be assessed.
Based on the content of the starting web page, the agent first extracts the name and type of the \entity.
We categorize each \entity into one of the six types
(see \Cref{sec:compre}). 
The agent navigates the web to identify multiple relevant web pages for the target \entity considering its name and type.
Then, the agent finds the license terms and information about \children from the identified relevant web pages, including their names and URLs.
Based on the license term statements, our scoring module determines the data compliance scores and \localclass for the target \entity.
This recursive process is repeated until there exists no more dependency \entities to examine.
Note that our agent maintains its own database across different sessions for caching its examination results and avoids re-examining the same \entity multiple times.
Once the process is over, the \globalclass is computed by identifying the lowest scores for each of the 18 criteria across all the descendant nodes.

\subsection{Information Collected by \ouragent}\label{sec:infor}

For each \entity, which is a node in the license dependency graph, \ouragent gathers the following information.

\paragraph{\Child information.}
To identify the \child \entities of the \entity, \ouragent finds and detects the names and URLs of the \child \entities.
This information is used to expand the license dependency graph and spawn new corresponding tasks recursively.

\paragraph{License terms.}
The full license terms or terms of use texts for the \entity are collected.
These texts are used for determining the data compliance scores and \localclass.

\paragraph{Metadata.} 
For Dataset-type \entities, some metadata are used together for individual assessment.

\textbullet{  Modality:}
The modality of the dataset identified from eight categories (\eg, Text, Image).

\textbullet{  Tasks:}
The target AI tasks for which each dataset is designed.
\ouragent selects relevant tasks from a predefined set of 42 categories (\eg, Text Generation).

\textbullet{  Languages:}
The natural languages that are present in the dataset (\eg, English, Korean).

\textbullet{  Domains:}
The specific domains of application for the dataset, if there are any.
\ouragent classifies the dataset into one of five academic disciplines and provides detailed suggestions for relevant sub-disciplines or marks it as general-purpose.

\subsection{Evaluation and Analysis of \ouragent}
\paragraph{Tasks.}
We extracted the URLs of the top 1,000 most-downloaded\footnote{As of June 8th, 2024.} datasets on Hugging Face\footnote{https://huggingface.co/datasets?sort=downloads}, and then randomly sub-sampled 216 items to construct the test set.
\paragraph{Training details.}
We fine-tuned the EXAONE-3.5-32B-Instruct model~\cite{exaone-3.5} using our proprietary dataset.
Specifically, we trained three key components of \ouragent: the navigation module, the question-answering (QA) module, and the scoring module. 
The navigation and QA modules were fine-tuned with synthetic data samples, while the scoring module was fine-tuned with human-labeled data.

\paragraph{Evaluation metrics.}
We report the accuracy in finding the \children and license terms of a dataset with respect to the ground-truth labels.
The ground-truth labels are manually constructed as follows.
We asked five legal experts who are trained in similar tasks for at least 31 hours to find the \children and license terms for 216 unseen test tasks.
Then, they aggregated their answers and discussed together to produce the ground-truth labels.
For \children finding tasks, we measured the set accuracy, also known as Jaccard Index, as
\begin{align}
\text{Set Accuracy} &= \frac{|\tr{Pred} \cap \tr{GT}|}{|\tr{Pred} \cup \tr{GT}|}
\end{align}
where Pred is the set of \children predicted by the agent and GT is the set of ground-truth \children.
For license term finding tasks, we again compute the Set Accuracy between the ground-truth and predicted URLs of license terms.

\paragraph{Result: Accuracy of answers.}
\begin{table}[ht]
\centering
\begin{tabular}{lcc}
\toprule
 &\multicolumn{2}{c}{Set Accuracy ($\uparrow$)} \\
\cmidrule{2-3}
Name & \Children & License terms \\
\midrule
\tb{\ouragent} & \tb{81.04\%} & \tb{95.83\%} \\
\expert & 64.19\% & 87.73\% \\
ChatGPT-4o & 25.00\% & 39.81\% \\
Perplexity Pro & 28.24\% & 22.22\% \\
\bottomrule
\end{tabular}
\caption{Accuracy of finding the \children and license terms of the 216 datasets from the evaluation set.}
\label{tab:accuracy}
\end{table}

\Cref{tab:accuracy} compares the performance of three agents with \expert across two tasks: finding \children and license term. 
We consider three AI agents—\ouragent, \expert, \perplexity and \chatgpt\footnote{Both \chatgpt and \perplexity are 2024-09-30 versions.}—that have access to the web for a fair comparison.
Please refer to \Cref{app:prompts} for the prompts used to evaluate \chatgpt and \perplexity.
For measuring the human performance, we uniformly distributed the task among 5 professional lawyers who are trained in similar tasks for at least 20 hours.
The \ouragent significantly outperforms all other agents and \expert, achieving an accuracy of 81.04\% and 95.83\% in each task. 
In contrast, both \chatgpt and \perplexity show relatively low accuracy for Source and License tasks, respectively. 
These results highlight the superior performance of the \ouragent, demonstrating its efficacy in handling both tasks with remarkable accuracy, while also indicating a substantial performance gap between AI-based models and \expert in these domains.

\paragraph{Result: Time and cost efficiency.} The table illustrates a comparison between the \ouragent and \expert in terms of time and cost efficiency. The \ouragent significantly outperforms the \expert, completing tasks in just 53.1 seconds, compared to 2,418 seconds for the \expert. 
Additionally, the cost associated with the \ouragent is remarkably low at \$0.29, whereas the \expert incurs a cost of \$207.0. 
\ouragent ran on a single GCP a2-megagpu-16gpu node which costs \$14,225 per month, which translates to approximately \$0.29 per 53.1 seconds.
These results highlight the substantial advantages of using the \ouragent, both in terms of time savings and cost reduction, demonstrating its effectiveness in large-scale legal assessment.
\begin{table}[!t]
\centering
\begin{tabular}{lcc}
\toprule
Name        & Time (sec) & Cost (\$)  \\
\midrule
\tb{\ouragent}   & \tb{53.1}   & \tb{0.29} \\
\expert     & 2,418  & 207 \\
\bottomrule
\end{tabular}
\caption{Efficiency comparison between \ouragent and \expert.}
\label{tab:efficiency}
\end{table}

\section{Massive-Scale Analysis of Trending Datasets}
\label{sec:Empirical_Analysis}

The primary objective of this work is to proactively detect and clearly identify potential legal risks embedded in AI training datasets with \ouragent.
To uncover potential legal issues that are present in existing AI training datasets, we aim to perform an in-depth empirical analysis of trending datasets at scale.
The full assessment results of the target entities can be accessed on our NEXUS web repository\footnote{\url{https://nexus.lgresearch.ai}}.

\subsection{Selection of Datasets for Analysis}

For our empirical analysis, we selected a total of 3,612 datasets to be root entities (see \Cref{sec:datalist} for the full list).
This selection comprises the 3,000 most-downloaded datasets on Hugging Face\footnote{As of September 9th, 2024.} and the 612 datasets that are the \children of the data collection presented by \citet{longpre2023dataprovenanceinitiativelarge}. 

The inclusion of Hugging Face’s 3,000 most-downloaded datasets establishes a solid foundation to evaluate both the performance and legal implications of our work, as it provides a representative sample of contemporary dataset trends.
The additional 612 datasets, sourced from the ``Data Provenance Collection'' by \citet{longpre2023dataprovenanceinitiativelarge}, were selected to further improve the quality of the target dataset list.
With \ouragent, we analyzed these samples as well as all of their direct and indirect \children for legal risk assessment.

\subsection{Overview of Results}
Before presenting our full legal risk analysis, we first examine the statistical characteristics of the \entities identified by \ouragent.
Starting from the 3,612 target \entities, we identified a total of 17,429 unique entities, where 13,817 \entities appeared as the target \entities' direct or indirect \children.

For our empirical analysis, we consider an \entity and its license dependency graph to have a \emph{single-layered} structure if the \entity does not have any \children and a \emph{multi-layered} structure if it has one or more \children.
Out of the 3,612 target datasets, 2,086 (57.8\%) had multi-layered structures, whereas the other 1,526 (42.2\%) had single-layered structures with no dependencies.

\subsection{Statistical Risk Analysis of Dataset Distribution and Licensing}

\begin{figure}[!htp]
    \begin{center}
    \includegraphics[width=0.5\textwidth]{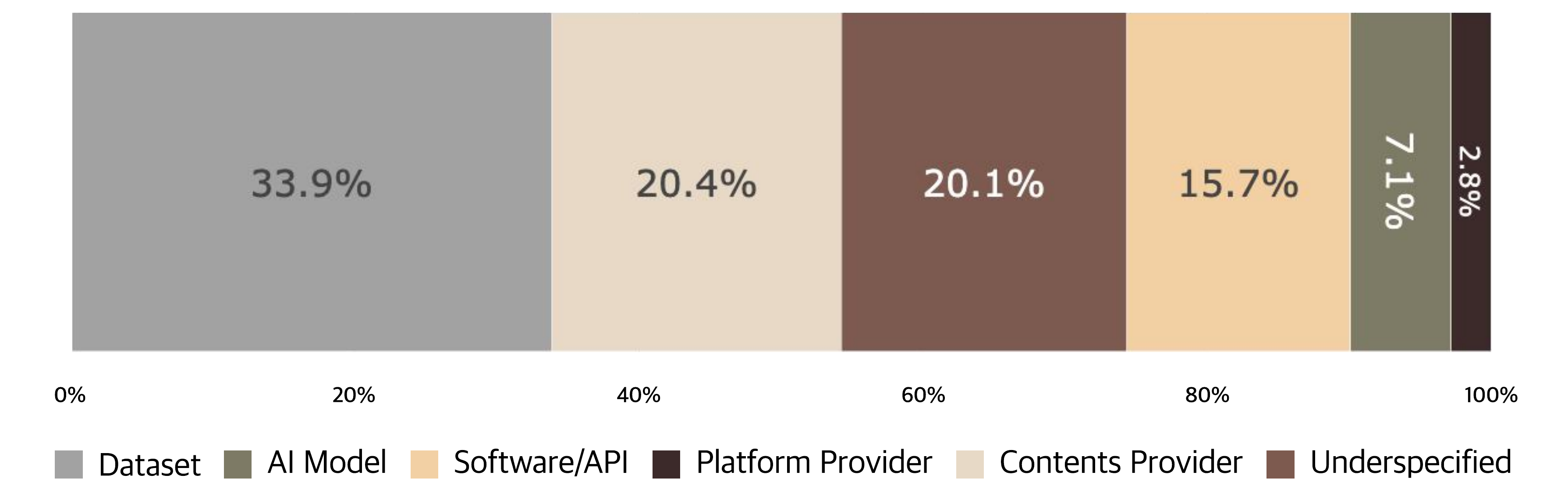}
    \end{center}
    \vspace{-1em}
    \caption{Distribution of the types of dependency \entities.}
    \label{fig:dist_entity_types}
\end{figure}

\paragraph{Types of \children \entities.}
Among the 13,817 \entities that appeared as \children, the Dataset type comprised the largest portion at 33.9\%.
Notably, a combined 25.6\% consisted of Software/API, AI models, and Platform Service Providers, all of which were used for generating, editing, or modifying data sources (see \Cref{fig:dist_entity_types}).
This indicates significant activity in modifying, editing, translating, and augmenting existing data, potentially leading to issues in quality control and source data tracking of human-created content.

\begin{table}[!ht]
\centering
\begin{tabular}{c c c c c c c c}
\toprule
Mean & Std & Min & 25\% & 50\% & 75\% & Max \\ \midrule
2.22 & 1.83 & 1 & 1 & 2 & 3 & 16 \\
\bottomrule
\end{tabular}
\caption{Depth statistics of the license dependency graphs for the target datasets.}
\label{tab:depth_statistics}
\end{table}

\paragraph{Complexity of dataset redistribution structure.}
For the analysis of the complexity of the dataset redistribution structure, we first define the \emph{depth} of a license dependency graph for each \entity to be the length of the longest path from the root in the graph, which can be an intuitive measure of the structure's scale and complexity.
To avoid potential biases that can come from the target datasets without any information on their construction processes, which are not rare, we focus only on the 2,086 target datasets with multi-layered structures.
\Cref{tab:depth_statistics} indicates that the target datasets with multi-layered structures have about three levels in their license dependency graphs on average, including the root \entities themselves.
The deepest dependency graph had a depth of 16, which highlights cases where excessive redistribution could obscure legal risk identification.

\paragraph{License terms.}
Based on our findings with \ouragent, among the 17,429 unique \entities, \numfoundlicenses \entities are provided with the corresponding license terms.
Manual review revealed that this gap was primarily due to the unavailability of license information on the web, rather than a limitation of \ouragent itself, which can be an issue that would likely challenge human experts as well.
This highlights a significant barrier to the legal and ethical use of data, platforms, and tools.
The absence of clear and accessible license information makes it difficult, if not impossible, to verify whether proper legal permissions have been granted.

\subsection{Legal Risk Assessment and Structural Implications}

\begin{figure}[!htp]
    \begin{center}
    \includegraphics[width=0.5\textwidth]{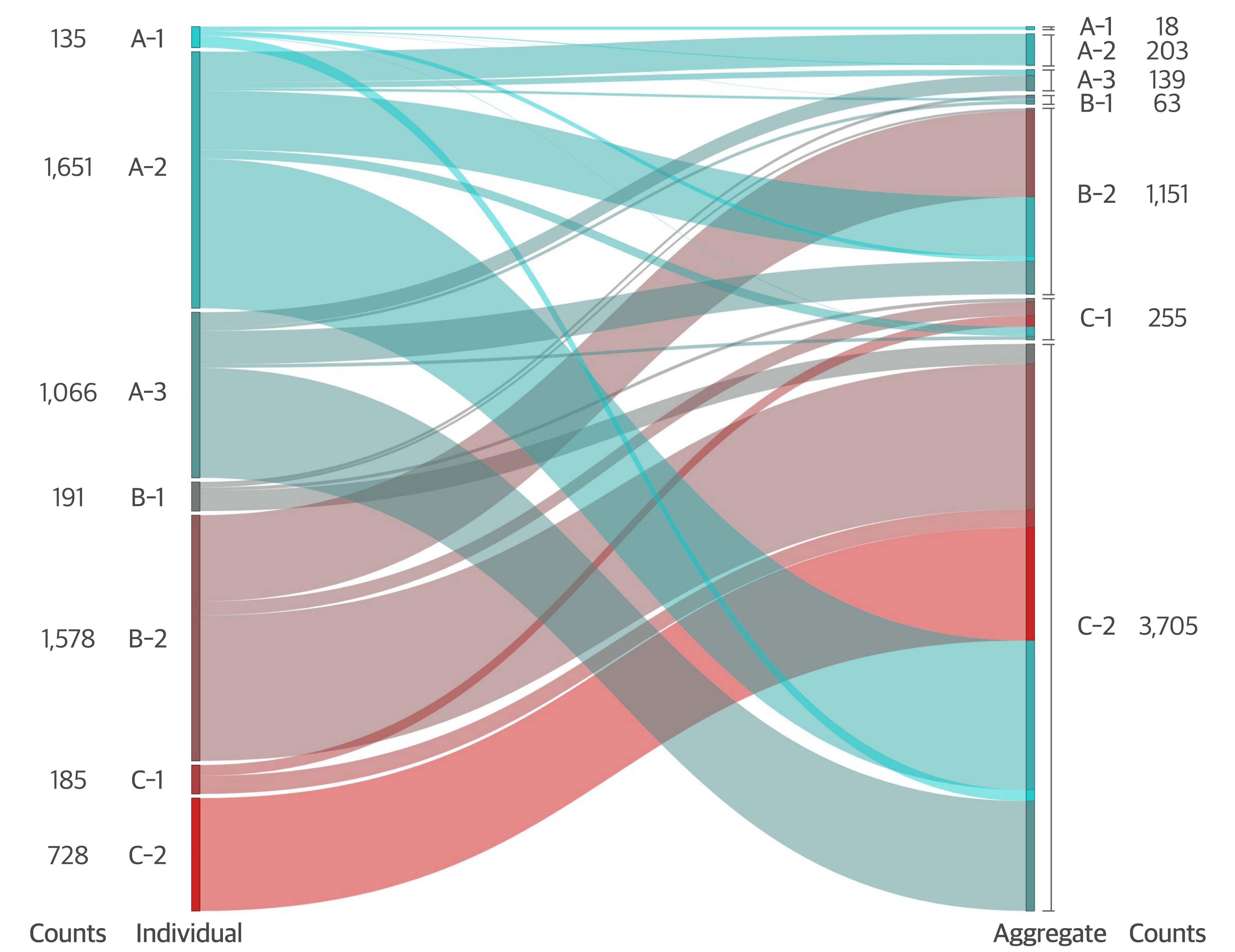}
    \end{center}
    \vspace{-1em}
    \caption{Discrepancies between the individual and aggregate classes of the analyzed \entities.}
    \label{fig:discrepencies}
\end{figure}

Leveraging our \ourdata framework, we assessed the legal risk levels of each entity, classifying them from A-1 (lowest risk) to C-2 (highest risk). 
Since entities without dependencies have identical individual and \gclass ratings, we focused on 5,534 entities with at least one dependency, among the 17,429 entities. 
\Cref{fig:discrepencies} shows the distribution of individual and \gclass scores for these datasets.

Notably, when a \child entity is assigned a high risk (\ie, a low score/class), any dataset derived from it (\eg, reconstructed or redistributed) inherits this risk. 
This implies that the class or scores of an entity should be the same as or lower than those of its \children, indicating higher risk. 
However, we observed a class/score inversion phenomenon, where this relationship is reversed. 
Such inversions can lead to errors in risk assessment, particularly when only the entity’s own license is reviewed, potentially underestimating the actual risk.

The presence of this inversion suggests a deeper issue in the risk assessment process. 
\Cref{fig:discrepencies} shows that many datasets classified in low-risk categories (\eg, A-1, A-2, A-3 in \lclass) are also categorized in higher-risk classes (\eg, C-1, C-2 in \gclass). 
This indicates that even if a dataset appears legally safe in isolation, there is a substantial possibility that the referenced \children could introduce higher legal risks.

To illustrate this, consider a dataset licensed under CC-BY-4.0 and classified as A-2 in \lclass due to the absence of identifiable personal information. 
Even so, if one of its referenced \children distributes data with personal information under a CC-BY-NC-4.0 license, the actual legal risk is likely to be higher. 
In such cases, the \gclass classification would likely place the dataset in a more hazardous category than A-2.

This issue becomes clearer when examining the analysis of derivative-direct dependency relationships. 
Out of 25,266 relationships analyzed, only 8,952 showed no class inversion. Thus, when an entity presents a legal risk, there is a 62.6\% chance that this risk is not explicitly reflected in the redistributed dataset.

\subsubsection{Inversion Phenomenon}
Inversion phenomena represent legal risks that are often obscured within the vertical hierarchical structures of data. Due to the complexity of dataset hierarchy, these risks can be easily overlooked, yet they are critical for ensuring thorough legal review. Each entity may experience inversion across 14 Data Compliance criteria (excluding criteria 3.2, 3.3, 3.4 and 4.4). To conservatively assess the legal integrity of an entity, it is essential that no inversions occur across any of these 14 criteria.

In this empirical study, out of 17,429 unique entities, 5,534 were identified as multi-layered entities (31.8\%), with 4,671 exhibiting inversion (84.4\%) and 863 classified as non-inversion (15.6\%).

\begin{figure}[!htp]
    \centering
    \vspace{-1.1em}
    \includegraphics[width=0.5\textwidth]{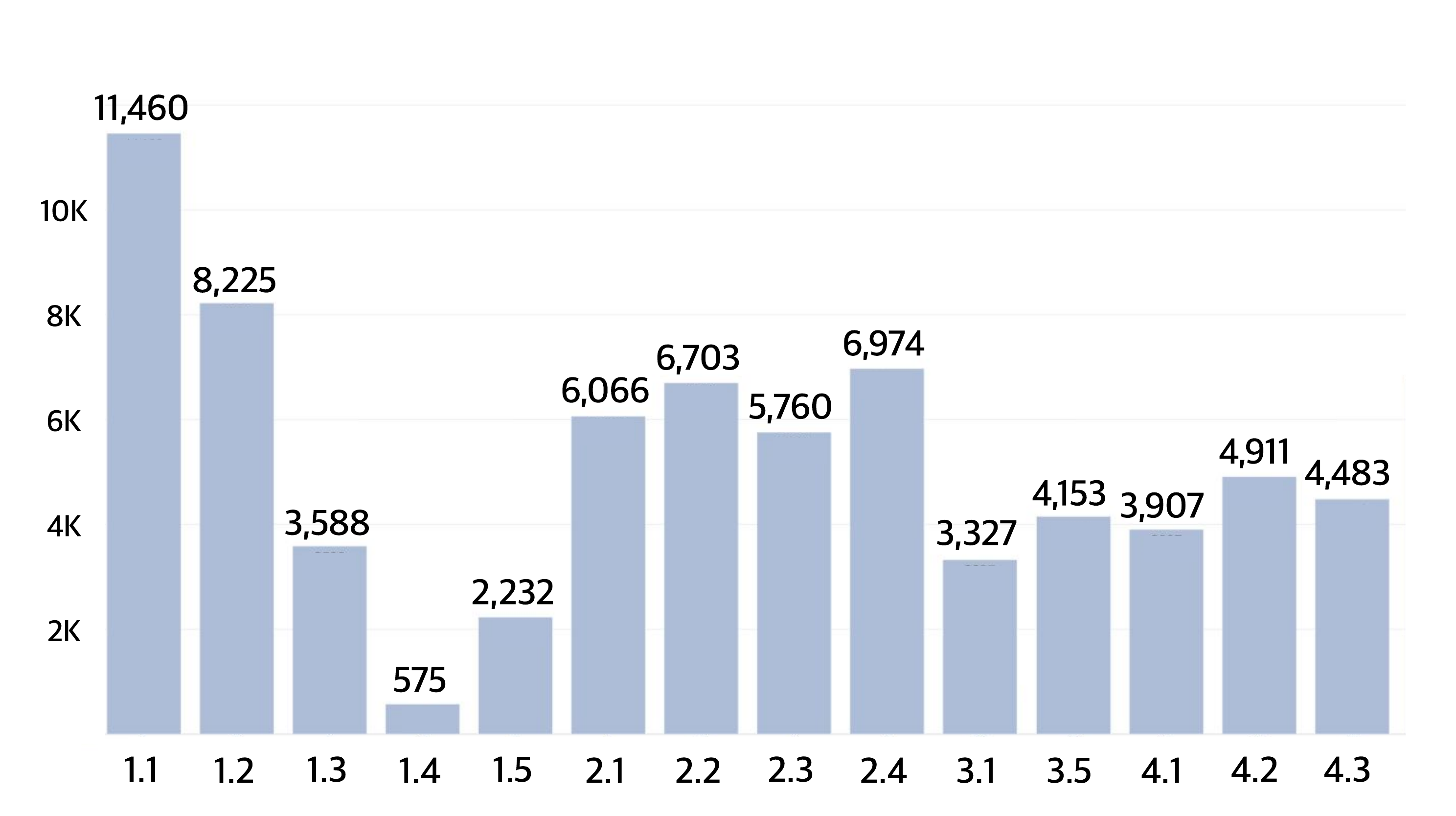}
    \vspace{-2em}
    \caption{Inversion occurrences across our 14 criteria.}
    \label{fig:distinversion_each}
\end{figure}

\paragraph{Across data compliance criteria.}
Figure \ref{fig:distinversion_each} illustrates the distribution of criteria that give rise to the Inversion Phenomenon.
The data reflects variations in inversion occurrences across multiple criteria, with a particularly notable concentration in Criterion 1.1 (data usage license) and 1.2 (data modification authorization.), which represent the highest and second-highest frequency of inversions, respectively.
This distribution warrants close legal scrutiny due to the fundamental nature of these two criteria in assessing the permissibility of data use and redistribution.

Criterion 1.1 pertains to the primary legal inquiry into whether a dataset is subject to a license that permits commercial utilization. While this criterion is often framed as a unidimensional assessment of whether a dataset can be exploited for commercial purposes, its legal significance extends far beyond a mere classification of usability. At its core, Criterion 1.1 addresses the most fundamental legal question—whether the requisite legal authority has been conferred upon the entity seeking to use the dataset. Without an unequivocal grant of such rights, any commercial or even non-commercial usage may constitute an unauthorized act, leading to potential copyright infringement, breach of contract, or violation of database rights, depending on the governing legal framework.

Criterion 1.2, on the other hand, bears the most significant weight in the determination of a dataset’s classification within the Data Compliance framework, as it pertains to the authority to modify the data. The right to modify a dataset is particularly critical in contexts such as AI training, data processing, and the creation of derivative works, where the ability to alter, refine, and repurpose the underlying data is not merely advantageous but legally indispensable. The absence of such a right introduces substantial compliance risks, as AI model training often necessitates data preprocessing, augmentation, and transformation, all of which may amount to acts of derivative authorship under copyright law. Any such modifications undertaken without express legal authorization may expose entities to intellectual property liability, licensing non-compliance, and potential contractual enforcement actions.

Given that these two criteria encapsulate the core legal determinants of data usability—the right to use and the right to modify—the high frequency of inversion occurrences within these criteria is of profound legal significance. The fact that criteria 1.1 and 1.2 exhibit the greatest number of inversions suggests that there exist systematic inconsistencies in the legal interpretation or application of licensing frameworks, thereby raising material concerns regarding the compliance of license requirements. These findings warrant further legal examination, as they underscore the potential for inadvertent non-compliance, regulatory exposure, and the necessity for enhanced due diligence in dataset governance and AI-related data processing.

\begin{figure}[!htp]
    \centering
    \hspace*{-0.25cm}
    \includegraphics[width=0.5\textwidth]{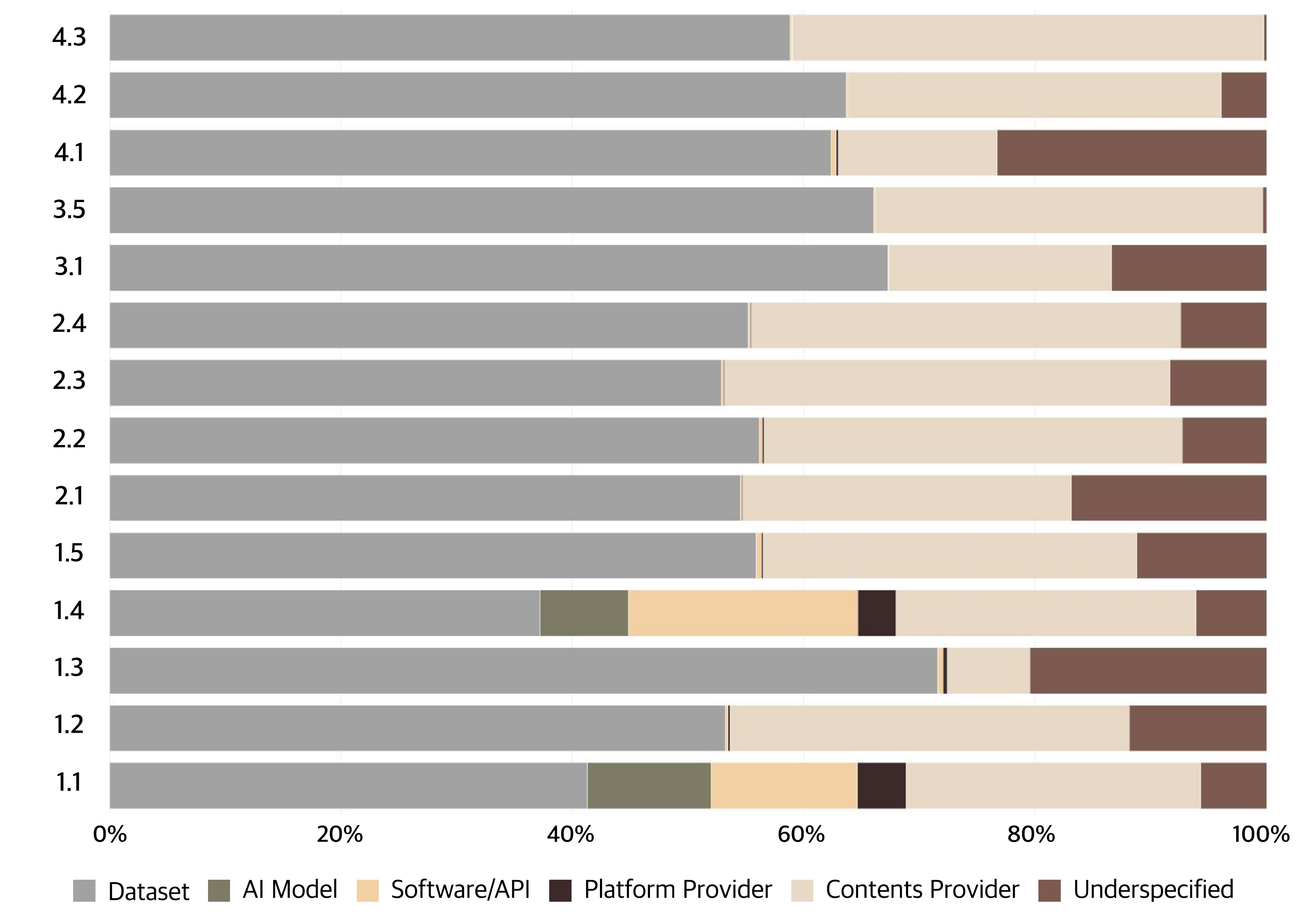}
    \vspace{-2em}
    \caption{Distribution of the entity types that caused the inversion of the scores.}
    \label{fig:distinversion}
\end{figure}

\paragraph{By entity types.}
\Cref{fig:distinversion} illustrates the entity types contributing to score inversion. 
A key observation is that not only datasets, but also AI models and Software/APIs used in processing datasets, can significantly influence the legal risks associated with the derived dataset. 
These entity types are often overlooked in manual data compliance assessments, highlighting the importance of considering both the AI models and Software/APIs involved in dataset construction. 
For instance, certain generative AI models, such as GPT and Llama, restrict the use of their output for developing competitive models in their Terms of Use~\citep{LLaMA3License, OpenAITerms}. Even when a dataset is redistributed under a permissive license, any license violations tied to the AI models or Software/APIs used in its creation could lead to legal disputes or injunctions. Consequently, users intending to utilize such datasets for training purposes must carefully assess these risks and ensure full compliance with the relevant terms of use.

\subsubsection{Compliance Considerations in Data Redistribution}
Throughout this analysis, we have examined multiple instances of data redistribution. However, beyond merely identifying redistribution patterns, it is crucial to assess whether the act of data distribution is legally compliant and whether upper-level datasets have been properly constructed based on their dependencies. Ensuring compliance in this context requires a thorough legal examination of whether a dataset is eligible for redistribution, whether such redistribution is subject to specific licensing conditions—including, but not limited to, Share-Alike obligations—or whether redistribution is expressly prohibited under the applicable legal or contractual framework. The primary objective of this evaluation is to quantify the extent of potential legal violations arising from data redistribution and to determine whether such violations constitute breaches of licensing terms, contractual obligations, or copyright laws.

From a legal standpoint, and as described in Criterion 4.4 of Data Compliance, datasets that qualify as copyright-protected works may only be redistributed if the distributing party possesses the legal authority to do so. This authority may derive from an explicit license granted by the copyright holder, statutory exceptions under applicable copyright laws, or contractual provisions permitting redistribution. In instances where redistribution is expressly prohibited or where redistribution occurs in a manner that contravenes licensing conditions—such as failing to adhere to share-alike provisions—such acts may give rise to legal liability. Depending on the jurisdiction, unauthorized redistribution may constitute copyright infringement, a breach of contract, or a violation of database rights, among other legal consequences. Given the complexities inherent in dataset licensing, it is essential to identify, categorize, and quantify instances of non-compliance with precision.

\begin{table}[H]
\centering
\begin{tabular}{p{0.44\linewidth}p{0.18\linewidth}p{0.18\linewidth}} 
\toprule
Type of License Term        & \Entities & Violations \\
\midrule
\textbf{Type 1:} Permitted for Distribution   & 8,781   & - \\
\textbf{Type 2:} Conditionally Permitted for Distribution   & 2,136  & \textbf{1,637} \\
\textbf{Type 3:} Not Permitted for Distribution & 6,512   & \textbf{8,268} \\
\bottomrule
\end{tabular}
\caption{Number of \entities and distribution violations under Criterion 4.4. of \ourdata.}
\label{tab:distribution}
\end{table}

In \cref{tab:distribution}, the entities excluding AI Model, Software/API, and Platform Provider were classified into Type 1, Type 2, and Type 3. This is to include only entities that compose the data itself, without directly constructing AI training data, and therefore excludes AI Model, Software/API, and Platform Provider to focus solely on assessing the legality of data distribution. Among the 17,429 entities, 8,781 (50.3\%) were identified as Type 1, where redistribution is permitted; 2,136 (12.2\%) were Type 2, where redistribution is conditionally permitted under share-alike condition; and 6,512 (37.4\%) were Type 3, where redistribution is not permitted. If redistribution occurs without complying with the conditions of Type 2 or Type 3, the act of redistribution itself may constitute a violation. To conduct a comprehensive legal assessment, we analyzed a total of 25,266 derivative-direct dependency relationships to determine the extent of redistribution violations.

Our findings indicate that 9,905 instances of non-compliant redistribution were identified. These instances were classified into two distinct categories based on the nature of the legal restrictions imposed:

The first category includes 8,268 (83.5\%) instances where redistribution is expressly prohibited and is legally restricted from being redistributed under its governing licensing terms. In these cases, the prohibition may stem from explicit contractual restrictions, exclusive rights held by the original data provider, or statutory limitations that preclude redistribution. Any unauthorized distribution of these datasets would likely constitute a clear violation of legal rights.

The second category consists of 1,637 (16.5\%) instances where redistribution is subject to specific licensing conditions, but its redistribution was rendered legally problematic due to inconsistencies in licensing terms between the dataset and its dependencies. These inconsistencies may arise when a dataset’s redistribution is contingent upon compliance with conditions such as Share-Alike or other requirements, which were not satisfied in the redistribution process. In such cases, redistribution is not explicitly prohibited but is effectively rendered non-compliant due to the failure to meet licensing prerequisites, creating potential legal exposure for the distributing entity.

Taken together, these two categories account for the total 9,905 instances of redistribution violations identified in our analysis. This assessment underscores the critical importance of ensuring compliance with licensing terms, contractual obligations, and applicable copyright laws when distributing datasets, particularly in cases where datasets rely on multiple dependencies with varying legal constraints. 

\section{Conclusion}\label{sec:conclusion}
The advancement of AI technology has intensified the focus on dataset legality and transparency. 
While AI performance critically depends on dataset quality—encompassing both technical and legal aspects—AI companies face significant challenges in ensuring data transparency. 
Data in the AI ecosystem, primarily sourced through internet crawling, open-source platforms, and commercial contracts, presents increasingly complex origins and rights structures that exceed human capacity for manual review.

The \ours addresses these challenges by providing real-time tracking and evaluation of datasets' legal risks throughout their lifecycle. The \ours's significance lies in its ability to systematically analyze complex data relationships and rights conflicts, particularly in managing the legal implications of dependencies within larger data structures. It reframes data as complex assets with embedded rights and responsibilities, rather than mere information collections.

While \ours effectively establishes new standards for dataset management and transparency, its risk assessment criteria serve as reference points that may vary across jurisdictions and AI applications.
Future developments should focus on expanding the system to adapt to evolving global regulatory environments while strengthening the integration of AI and legal review processes.
This approach promises to enhance the overall stability and trustworthiness of the AI ecosystem while providing practical tools for maintaining data transparency and legal compliance.

Since \ouragent agent recursively identifies the \children of a target entity and collects their license information, errors can propagate through the hierarchy, similar to how humans process and aggregate information. 
Our \ourdata currently treats all collected license information equally, without accounting for uncertainty. 
Future work could explore an error-aware \ourdata that quantifies and incorporates uncertainty in the collected data.

\newpage
\bibliography{references}
\bibliographystyle{icml2025}

\newpage
\appendix
\onecolumn

\section{Prompts Used for \chatgpt and \perplexity}
\label{app:prompts}

\begin{tcolorbox}[colframe=black, colback=white, coltitle=black, sharp corners=south, width=0.9\textwidth, boxrule=0.5mm, title=Prompts Used]
    \textbf{Intro:} \\
    In order to use the dataset for training AI models, it is necessary not only to check the dataset’s license but also to identify its sources and whether the original work complies with copyright laws, privacy regulations, etc. Therefore, it is crucial to analyze the data sources that make up the dataset and the methods used to compile it. Based on legal considerations, the sources (such as datasets, collective nouns, websites, etc.) that were used to create the data, as well as any AI models or services (such as translation tools or data cleansing tools, which might have licenses like Software Licenses from the original creators), should be identified. Starting from the dataset link, navigate the web and answer. Additionally, the linked papers on the website should be examined. \\

    \textbf{Output:} \\
    The prediction should be summarized as follows: \\
    \begin{itemize}
        \item \textbf{Data Source:} List the data sources of the dataset in comma-separated list (these could be services, AI models, collective nouns, website names, dataset names, etc.)
        \item \textbf{License Term Name:} Provide the name of the license term (such as CC-BY, MIT, etc) of the dataset.
        \item \textbf{License Term Link:} Provide the link to the license term (such as CC-BY, MIT, etc) of the dataset.
    \end{itemize}
    Only output the above. Do not output any explanation or details. \\

    \textbf{Your Task:} \\
    \begin{itemize}
        \item \textbf{Dataset Name:} NLLB
        \item \textbf{Dataset Link:} allenai/nllb
    \end{itemize}
\end{tcolorbox}

\section{Data Compliance}\label{sec:legal_framework_appendix}

\begin{table}[h]
\small
\centering
\begin{tabular}{p{0.22\linewidth}p{0.42\linewidth}p{0.09\linewidth}p{0.15\linewidth}} 
\toprule 
Clause & Criteria & Weight (\%) &
Class upper bounds\\ 
\midrule    
1. Risk Related to Data License & 1.1. The existence of a license to use the data & (Based on class scope) & Point 5: A-1, Point 3: B-2, Point 2: B-2, Point 1: C-2\\ 
 & 1.2. Authorization to modify data and produce derivative works & 10 \\ 
 & 1.3. The potential for dispute arising from the outputs & 15 & Point 1: B-1 \\ 
 & 1.4. The rights to outputs & 8 & Point 1: B-1\\ 
 & 1.5. The existence of an obligation to notify data usage & 3 \\ 
\midrule 
2. Risks Related to Data Use Period and Territory & 2.1. Restrictions on data use period & 7 & Point 1 : C-2 \\ 
 & 2.2. Whether the data license is revocable & 3 \\ 
 & 2.3. Restrictions on AI model service period & 5 \\ 
 & 2.4. Restrictions on data use territory & 4 \\ 
\midrule     
3. Risks Related to Personal Information and Data Security & 3.1. Whether personal data is included in AI training data & 9 & Point 2 \& 1: C-2\\ 
 & 3.2. Whether data subjects have consented to the use of their data & 3 & Point 1 : C-2\\ 
 & 3.3. Whether pseudonymized data is included in AI training data & 3 & Point 1 : C-1\\ 
 & 3.4. Whether personal data may be entrusted or provided to third parties & 5 & (If there is a plan for data processing through a third party) Point 1: B-1\\ 
 & 3.5. Whether the scope of data users is limited & 2 \\ 
\midrule 
4. Additional Legal Risk & 4.1. Risks in the data collection process & 8 \\ 
 & 4.2. Known disputes involving the use of the same dataset in AI models & 10 \\ 
 & 4.3. Other contract risks associated with licenses & 5 \\ 
 & 4.4. Type of license terms & - \\ 
\bottomrule
\label{tab:legal_detail}
\end{tabular}

\caption{Assessment criteria of \ourdata.}
\label{tab:datacompliance}
\end{table}
The final class is determined based on the score. Class A is generally considered commercially viable, Class B represents a level of risk suitable for internal research, and Class C is deemed unusable for any purpose, including AI training. As described in Table \ref{tab:legal_detail}, If a specific score is received in each criterion, it limits the range of the maximum class. For example, if a score of 1 is given in criteria 1.3, it becomes difficult to use a class higher than B-1.

\subsection{The existence of a license to use the data (Criteria 1.1)}

The highest class, A-1, is assigned to data that can be used without restriction for commercial purposes or falls within the public domain. Such datasets are evaluated as the safest for utilization due to their very low potential for legal disputes and absence of usage constraints. This applies to cases where data owners have explicitly authorized commercial use or where copyright protection has expired.

The second category, classs A-2 to B-2, applies to data explicitly authorized for internal research purposes. While these datasets can be safely utilized within their specified limitations, legal risks emerge when usage extends beyond these boundaries. In particular, additional permissions may be necessary for commercial applications or public disclosure.

The third category, classs B-1 to C-1, encompasses cases where license authorization is unclear. These datasets may be usable under fair use principles or Text and Data Mining (TDM) exemption provisions. The U.S. `Fair use' doctrine and TDM exemptions in Japan and the EU potentially permit online data collection and usage for purposes such as scientific research, making these exception clauses potentially applicable.

The lowest class, C-2, is assigned to data with explicit restrictions on specific uses such as AI training purposes. These datasets present high potential for legal disputes and may incur serious legal liability if utilized, thus their use should generally be avoided.
This grading system was designed with comprehensive consideration of international copyright conventions including the Berne Convention, national copyright laws, database protection regulations, and fair use principles. It particularly considers the principle of copyright protection without formalities, where copyright automatically exists upon creation without requiring any formal procedures. This characteristic makes it challenging to pre-determine whether data constitutes copyrightable work or qualifies for copyright exceptions such as fair use.

The framework also accounts for variations in data protection across different jurisdictions. For instance, the EU's Database Directive (96/9/EC) provides database rights, meaning databases can receive legal protection even without creative elements in individual data points. Such considerations highlight that data may receive various forms of legal protection even when not qualifying as copyrightable work, necessitating thorough risk assessment.

\subsection{Authorization to modify data and produce derivative works (Criteria 1.2)}

A derivative work refers to a creative work produced by translating, arranging, transforming, adapting, or creating visual works from the original work, and it is protected as an independent work. The exclusive right to create and use derivative works belongs to the original author, which is a fundamental right of authors recognized internationally.

The highest score of 5 points is assigned when all modification and transformation rights, including the creation of derivative works, are explicitly granted. This applies to cases where comprehensive modification rights licenses have been secured from the copyright holder, minimizing the possibility of legal disputes. Such extensive rights provide the safest legal status for utilization as AI training data.

3 points are awarded when the creation of derivative works is not permitted, but modifications or transformations that do not reach that level are possible. This includes cases where specific conditions for modification are specified, or where normative judgment is needed regarding the boundary between derivative works and general modifications. Legal interpretation may be required regarding the author's right of integrity, and protection scope varies by country – for instance, the United States recognizes the right of integrity only for visual art works.

2 points correspond to cases where the grant of data modification rights and derivative work creation rights is unclear. In such cases, the data may not qualify as copyrightable work, or fair use or Text and Data Mining (TDM) exemption provisions may apply. Particularly, license scope can vary according to contract terms, and the right to create derivative works is generally interpreted as not included unless explicitly permitted.

The lowest score of 1 point is assigned when all modifications and transformations, including the creation of derivative works, are explicitly prohibited. This represents cases with very high potential for legal disputes, particularly when modifications to works are explicitly restricted, such as with open source or Creative Commons License (CCL) ND (No Derivative Works) licenses. Such restrictions can also be implemented through contracts or terms of service, and legal risks may exist even when data is not protected as copyrightable work.

This scoring system was designed considering that data preprocessing in AI model training and model outputs could potentially constitute derivative works. Typically, preprocessing is essential for efficient data learning in the AI model training phase, and this preprocessing process might constitute the creation of derivative works or modifications/transformations of the data. Furthermore, there has not yet been clear legal determination on whether outputs generated from trained AI models can be interpreted as derivative works or modified/transformed results of the original data.

\subsection{The potential for dispute arising from the outputs (Criteria 1.3)}

This clause assesses potential legal risks that may occur even when proper data usage procedures have been followed.

The highest score of 5 points is awarded in cases where explicit consent has been obtained from the original author or when no outputs are generated. For example, this applies to AI models that only make judgments based on input data, such as image classification or speech recognition. These models do not generate results similar to the original data, thus minimizing the possibility of copyright infringement claims from original authors or third parties.

4 points are assigned when outputs are created in a form different from the original work. This includes cases where generative AI cannot easily produce directly similar outputs, such as generating text from voice data or creating graphs based on numerical data. Cases where new chemical structures are predicted from molecular structure data can also fall into this category.

3 points are awarded when outputs similar to the original author's work can be generated, but the likelihood is low. For example, this applies to cases where AI trained on various image styles generates completely new images that share similarities with certain artistic styles. While potential legal risk exists, the likelihood of actual disputes is relatively low.

2 points correspond to cases where portions of the original work may be included in the output. For example, this applies when large language models directly output phrases or expressions from training data. As seen in the \textit{Complex Systems, Inc. v. ABN AMRO Bank N.V.} case, legal dispute risks can increase when license relationships are complex.

The lowest score of 1 point is given when there is a high possibility of generating outputs similar to the original author's work. For example, this applies to cases where AI trained exclusively on a specific author's works generates very similar works to that author's style. As demonstrated in the \textit{Warner Chappell Music, Inc. v. Nealy} case, copyright holders can assert their rights even after a considerable time has passed.

The characteristics of different AI model types are also important considerations. Models that identify targets, such as Classification AI or Recognition AI, have low potential for legal disputes as they merely determine whether an image contains a ``dog'' when trained on dog images. However, Generative AI can create images similar to the trained ``dog'' images, potentially becoming subject to copyright infringement claims.

The form of output is also a crucial criterion. Outputs in forms where visual similarities can be easily confirmed, such as text or images, carry higher legal dispute risks. Conversely, outputs that are difficult to visually recognize or are transformed into different forms from the original, such as numerical data or chemical formulas, have relatively lower dispute potential. Scores are differentially assigned considering these various factors comprehensively.

\subsection{The rights to outputs (Criteria 1.4)}

This clause reflects the current reality where the legal status of AI outputs varies by country and clear standards have not yet been established.

The highest score of 5 points is awarded when ownership or intellectual property rights of the outputs clearly belong to the company. This includes cases where the data is not subject to legal protection or where rights to the outputs have been explicitly secured through contracts. For example, this applies when AI model usage contracts explicitly stipulate that ``all intellectual property rights to the outputs belong to the company,'' or when outputs are generated using public domain data. In such cases, legal risks in utilizing the outputs are minimized.

4 points correspond to cases where the company holds usage rights to the outputs. While not extending to ownership, explicit permission for use has been granted, allowing safe utilization within certain boundaries. For example, this applies to cases where limited licenses are granted, such as ``outputs may be used for commercial purposes but cannot be sublicensed to third parties.'' However, there may be restrictions on usage beyond the specified scope.

3 points are assigned when the existence of rights to the outputs is unclear. This applies to cases without clear contractual provisions, where legal risks are relatively low as long as outputs are not used secondarily. For example, this includes cases where AI service terms of use do not mention rights attribution for outputs. This requires careful attention as judgments may vary by country. Particularly in China, there exist lower court decisions recognizing image-generating AI outputs as copyrightable works, demonstrating how countries have different positions on the legal protection of AI outputs.

The lowest score of 1 point is given when the company explicitly has no rights to the outputs. For example, this applies to cases with explicit restrictions such as ``all rights to the outputs belong to the data provider'' or ``outputs may only be used for internal testing purposes.'' In these cases, secondary use of outputs is impossible, and particularly, using such outputs for training other AI models carries very high risk of legal disputes.

This scoring system considers the core legal issues of whether AI outputs can be copyrighted and how to determine authorship. Currently, most countries have not clearly established the legal status of AI outputs, particularly regarding rights attribution relationships among training data rights holders, AI model developers, and users. While some countries do not recognize AI outputs as copyrightable works, others are attempting to protect them as such, necessitating more cautious approaches for companies conducting international business.

\subsection{The existence of an obligation to notify data usage (Criteria 1.5)}

This clause reflects the current reality where the legal status of AI-generated content varies significantly by jurisdiction and lacks unified standards.

The maximum score of 5 points is assigned when ownership or intellectual property rights of the output are clearly attributed to the company. This includes cases where the data is not subject to legal protection or where rights to the output have been explicitly secured through contractual agreements. For example, this applies when an AI model usage contract explicitly states ``all intellectual property rights to the outputs belong to the company'' or when outputs are generated using public domain data. In such cases, legal risks associated with utilizing the outputs are minimized.

4 points are awarded when the company holds usage rights to the output. While this doesn't extend to full ownership, it represents cases where explicit permission for use has been granted, allowing for safe utilization within certain parameters. An example would be when a limited license is granted stating ``outputs may be used for commercial purposes but cannot be sublicensed to third parties.'' However, usage beyond the specified scope may be restricted.

3 points are assigned in cases where the existence of rights to the output is ambiguous. This applies to situations lacking clear contractual provisions, where legal risks remain relatively low as long as the output is not used derivatively. For instance, this score would apply when AI service terms of use do not specifically address rights attribution for outputs. This requires careful consideration as interpretations may vary by jurisdiction. Notably, China has seen lower court decisions recognizing AI-generated images as copyrightable works, highlighting the divergent approaches different countries take toward legal protection of AI outputs.

The minimum score of 1 point is given when the company explicitly lacks rights to the output. This includes cases with explicit restrictions such as ``all rights to the outputs belong to the data provider'' or ``outputs may only be used for internal testing purposes.'' In these situations, derivative use of the outputs is prohibited, and utilizing such outputs for training other AI models carries a particularly high risk of legal disputes.

This scoring system addresses core legal issues surrounding the copyright of AI outputs and the determination of authorship. Currently, most jurisdictions lack clear establishment of the legal status of AI-generated content, particularly regarding rights attribution among training data rights holders, AI model developers, and users. While some jurisdictions refuse to recognize AI outputs as copyrightable works, others are moving toward providing copyright protection. This divergence necessitates especially careful consideration for companies operating internationally.

\subsection{Restrictions on data use period (Criteria 2.1)}

This clause reflects the characteristics of data ownership and license agreements.

The highest score of 5 points is assigned when data can be used perpetually. This applies to cases where a company holds direct ownership of the data or has established a permanent license agreement. Examples include self-collected data or cases where an ``perpetual and irrevocable license'' has been explicitly granted.

4 points are awarded when either a sufficient period is guaranteed for AI model operation or there is no explicit time limitation. This includes cases where long-term licenses are granted, such as ``usable for 10 years,'' or when no time limitation is specified in the contract. However, it should be noted that in the United States, contracts without explicit duration provisions are generally interpreted as terminable at the will of either party.

3 points are assigned when there are data usage time limitations but restrictions on AI model usage remain unclear. This refers to situations where the contract explicitly specifies the usage period for the data itself, but it remains ambiguous whether the use of AI models trained on such data constitutes ``data usage.'' The risk of model usage being interpreted as data usage increases particularly when AI models can directly reproduce training data.

The lowest score of 1 point is given when the data usage period has already expired. This situation is equivalent to having no license and carries a very high risk of legal dispute. For example, this applies to cases where data ``usable until 2023'' continues to be used in 2024.

This clause takes into account the unique characteristics of AI models. Unlike conventional data usage, with AI training data, it remains legally ambiguous whether the use of the data itself can be distinguished from the use of trained models. Particularly, as contract interpretation should center on the parties' intentions under the principle of private autonomy, a crucial consideration is whether both parties intended to include AI model usage within the scope of data ``usage'' at the time of contracting.

Furthermore, differences in the sources of data usage rights must be considered. Data directly owned by companies, data with acquired usage rights through licensing, and open-source data each possess distinct legal characteristics. Specifically, while perpetual usage rights are possible in license agreements, this requires explicit contractual basis.

\subsection{Whether the data license is revocable (Criteria 2.2)}

Continuous contracts, which persist over a specified period, can be terminated when circumstances render the continuation of the contract untenable, such as when the foundational trust relationship is destroyed through material breach. License agreements, as a type of continuous contract, may include provisions for contract termination or license revocation.

The highest score of 5 points is assigned when the contract explicitly stipulates that the license grant is irrevocable. This scenario minimizes legal risks associated with continued data usage, as the data rights holder cannot subsequently revoke the license under the principle of private autonomy.

4 points are awarded when the license agreement either lacks explicit revocation provisions or permits termination only under specific conditions, such as bankruptcy or contract non-compliance. While these conditions can generally be mitigated through diligent fulfillment of contractual obligations, the possibility of revocation cannot be entirely eliminated.

3 points are assigned when the data rights holder maintains the discretionary authority to revoke the license. This situation presents significant legal risks, particularly regarding the ambiguous legal validity of previously trained AI models following license revocation, and may prevent continuous AI model updates.

This evaluation framework prioritizes the stable operation of AI models, with irrevocable licenses representing the optimal arrangement. When revocation clauses are necessary, it is crucial to define revocation grounds objectively and clearly, limiting them to circumstances within the company's control. Furthermore, to prevent subsequent legal disputes, contracts should explicitly specify whether previously trained AI models can continue to be used following license revocation.

\subsection{Restrictions on AI model service period (Criteria 2.3)}

AI model services may not be provided perpetually due to license agreement conditions for data and AI model usage, data transformation periods, third-party contract terms, and changes in relevant laws. These limitations pose economic risks, particularly given that AI model services are intended for company revenue generation. Therefore, specific criteria are needed to evaluate the stability of AI model service periods.

The highest score of 5 points is awarded when data is perpetually available and irrevocable, with no temporal restrictions on AI model service provision. This applies when the contract explicitly guarantees perpetual data usage, imposes no constraints on license duration, and confirms that AI model services can be provided continuously. In such situations, legal risks associated with data and AI model usage are minimized, enabling companies to maintain stable revenue structures.

4 points are assigned when conditions regarding data perpetuity or AI model service provision periods are unclear, or when specific termination conditions are included due to third-party contract terms. For instance, this applies to cases where AI model services might be discontinued due to conditions in agreements with co-developers or third parties. While these conditions cannot completely eliminate risks of contract termination or service interruption, risks can be managed through diligent fulfillment of contractual obligations and strict compliance with data usage conditions.

3 points are given when AI model service periods are uncertain in data license agreements or third-party contract terms, or when contract termination and revocation conditions lack clarity. In such cases, AI model services may face potential discontinuation at specific points, significantly impacting company service continuity and revenue structure. Particularly, when such contract conditions are ambiguous, constraints on AI model service provision periods can become obstacles to long-term operational planning.

The lowest score of 1 point is assigned when explicit limitations exist on data usage rights or AI model service provision periods. This includes cases where contract conditions limit data or AI model usage to specific periods, or where third-party agreements prevent perpetual AI model service provision. In such situations, AI model service sustainability is reduced, and companies likely face additional costs and legal risks related to securing alternative data or contract renewal.

This evaluation framework aims to effectively manage risks arising from AI model service period limitations. Contract conditions regarding data and AI model usage rights should be clearly and specifically defined, with perpetual usage being particularly desirable. When third-party contract conditions are involved, it is crucial to clearly define termination conditions and specify the possibility of AI model usage after service termination in the contract to prevent potential legal disputes.

\subsection{Restrictions on data use territory (Criteria 2.4)}

This framework evaluates constraints on service provision periods due to contractual restrictions and legal regulations. Considering that the ultimate purpose of AI models is profit generation through service provision, limitations on service periods constitute a significant economic risk factor.

The highest score of 5 points is awarded when AI model services can be provided perpetually. This applies to cases where data is perpetually available and irrevocable, with no legal or contractual restrictions on service provision. This includes scenarios where perpetual and irrevocable data usage rights are held, there are no restrictions on data transformation and processing, no service limitation clauses exist in third-party contracts, and the service is free from current and foreseeable future legal regulations.

3 points are assigned when AI model service period limitations cannot be clearly determined. This includes situations where the perpetual availability or revocability of data is unclear, or where termination conditions exist in contracts with essential third parties such as co-developers or technology partners. This also applies to cases where data usage periods are specified but their relationship with AI model services remains ambiguous, or where potential regulatory changes exist but their impact cannot be accurately predicted.

The lowest score of 1 point is given when clear restrictions exist preventing perpetual AI model service provision. This includes cases with explicit limitations on data usage or transformation periods, or where service periods are clearly restricted in third-party contracts. Additionally, this applies to situations where service restrictions are clearly anticipated due to forthcoming regulations, or where perpetual service provision is impossible due to technical or commercial reasons.

This clause aims to ensure AI service sustainability by comprehensively considering the perpetuity and stability of data usage rights, sustainability of contractual relationships, potential changes in regulatory environments, and technical and commercial sustainability. Particularly, given that AI services operate within complex contractual relationships involving various stakeholders, systematic evaluation and management of these multilayered factors are necessary to ensure service perpetuity.

\subsection{Whether personal data is included in AI training data (Criteria 3.1)}
Personal information encompasses all data that can identify a living individual, ranging from general identification information to sensitive private information such as portraits, voice recordings, thoughts, beliefs, and health data. This personal information is protected globally as a fundamental right, with explicit guarantees of informational self-determination in frameworks such as the EU Charter of Fundamental Rights and various national constitutions.

The highest score of 5 points is assigned when no personal information is included or when there are plans for anonymization and de-identification. Anonymized information is no longer treated as personal information and is thus exempt from relevant legislation such as personal information protection laws. This represents the safest form for AI model training, with minimal legal risks related to personal information processing.

4 points are awarded when personal information is included but there are plans for pseudonymization, or when the presence of personal information cannot be determined. Pseudonymization significantly reduces the risk of data breaches by making individual identification impossible without additional information. However, as pseudonymized data fundamentally remains personal information, the legality of its collection and processing is still required. Notably, there exists a risk of personal information exposure in AI outputs through methods such as prompt injection attacks, which could be interpreted as unauthorized personal information disclosure.

2 points are assigned when there is a high probability that the data contains personal information. Such cases require comprehensive review of the legality of data collection and provision processes, the company's authority to process personal information, and the necessity of pseudonymization or de-identification. Particularly important considerations include personal information protection measures during AI model training and prevention of personal information exposure in training outputs.

The lowest score of 1 point is given when personal information is clearly included. This necessitates strict examination of the legality of initial information collection procedures and whether consent for AI training purposes was included. Notably, if AI model training was not included in the original collection purpose, it may be necessary to filter out such personal information and proceed with training using only the remaining data. Furthermore, if personal information is included in AI model outputs or exposed to third parties, this could be interpreted as unauthorized information disclosure, potentially seriously infringing upon data subjects' rights.

This evaluation framework reflects the characteristics of modern AI systems. AI models can generate new outputs or recombine existing information based on trained data, potentially exposing personal information in unexpected ways. Therefore, the presence, acquisition circumstances, and processing methods of personal information in training data become crucial risk factors in AI system operation, necessitating systematic evaluation and management.

\subsection{Whether the scope of data users is limited (Criteria 3.5)}

This clause evaluates cases where data access permissions and controls are required by contract or statute.

The highest score of 5 points is assigned when data usage rights are not restricted to specific users. In such cases, necessary operators can freely access, process, and utilize data within the scope of intended purposes without the need for permission control. This represents the most flexible and efficient scenario for data utilization in AI model training processes. For example, this may apply to public datasets or company-generated data, where data scientists, engineers, and researchers can freely utilize the data as needed.

3 points are awarded when data usage is restricted to specific users. Such restrictions may arise from explicit contractual provisions or statutory requirements. Particularly for data containing personal information, these access restrictions and controls are more stringently required. Article 5 of the GDPR stipulates that personal data processing should be limited to what is necessary for the purpose (Article 5(1)(c)) and must ensure appropriate security (Article 5(1)(f)), clearly establishing the necessity to restrict the scope and participants in personal data processing.

From a legal stability perspective, serious legal issues may arise when unauthorized individuals access and process data restricted to specific users. For example, if a license agreement specifies that ``only designated researchers may access the data,'' use by other employees constitutes a contract breach. Additionally, when permission is purpose-based, AI model training as a specific purpose may exceed the permitted scope.

For data containing personal information, the importance of access restrictions is further emphasized. Personal information protection laws across jurisdictions require differential assignment and control of access rights as essential measures for ensuring the security of personal information processing. For instance, medical data may require access rights to be granted only to specific research teams, maintaining access logs, and implementing technical and administrative measures to prevent data breaches.

\subsection{Risks in the data collection process (Criteria 4.1)}

Given the requirement for large datasets in AI model training, collecting publicly available online data through mechanical methods such as web crawling has become standard practice. However, such collection processes may involve legal risks.

The highest score of 5 points is assigned when data acquisition methods present no particular issues. This applies to cases where companies generate raw data directly or receive explicit permission from rights holders. It also includes collection of data without rights holders, such as meteorological data or national statistics.

3 points are awarded when data is acquired through web crawling or similar methods. Web crawling, which uses automated programs to extensively collect website data, has limitations in reviewing the legality of individual data points. Legal disputes may arise particularly when collected data is protected as copyrighted work or databases. Cases such as \textit{hiQ v. LinkedIn} in the US and \textit{Ryanair Ltd. v. PR Aviation BV} in the EU well illustrate the legal issues surrounding web crawling.

2 points are given when the data acquisition method is unknown. This presents a high risk of legal disputes due to the possibility of illegal data collection by the original collector. Particularly with data containing personal information, it becomes impossible to verify compliance with strict collection standards under various national personal information protection laws.

The lowest score of 1 point is assigned when data is acquired through circumventing Robots.txt or other problematic methods. Robots.txt represents a website administrator's explicit refusal of crawling, and circumventing it may lead to both moral criticism and legal disputes. This also applies to cases where data is acquired through illegal means, such as Shadow Libraries.

This evaluation framework considers the legal stability of data collection. Particularly for personal information, GDPR and national personal information protection laws require clear collection purposes, minimal collection scope, and data subject consent. Furthermore, as case law and legislation regarding web crawling continue to evolve across jurisdictions, the legality of data collection methods becomes a crucial risk factor in AI model development.

\subsection{Known disputes involving the use of same dataset in AI models (Criteria 4.2)}

This system assesses legal dispute cases involving other AI models trained using identical data, databases, or data acquired from the same licensor.

The highest score of 5 points is assigned when there are no known data disputes. This indicates a minimized state of legal risk that needs to be considered prior to data usage. Such cases can be viewed as having clear legality and rights relationships, with low potential for disputes.

4 points are awarded when known data disputes exist but are limited to small-scale disputes targeting individuals. In such cases, considering the company's data legality review results, even if similar disputes arise, the risk from losing such cases is deemed minimal.

3 points are given when disputes exist with claims exceeding 1 billion won. Disputes of this scale typically involve substantial datasets, highly important data, or parties taking an aggressive stance toward legal resolution. Therefore, if similar disputes arise, they are likely to lead to significant litigation.

The lowest score of 1 point is assigned when large-scale disputes exist with claims exceeding 10 billion won. The existence of such large-scale disputes suggests that the use of such data may pose significant legal risks to the company. Particularly, as commercial utilization of AI models intensifies, the possibility of such large-scale disputes may increase further.

This evaluation framework indirectly assesses the legal stability of data through existing dispute cases. The existence of previous disputes suggests potential errors in determining data legality or incorrect identification of legitimate rights holders. Additionally, the data may fall into a legal gray zone where interpretation is ambiguous, given that AI model-related legislation and case law are not yet fully developed. Therefore, the scale and nature of existing dispute cases become important indicators for predicting potential future legal risks.

\subsection{Other contract risks associated with licenses (Criteria 4.3)}

License agreements can include various obligations beyond simple fee payments, and legal risks may vary depending on the nature and scope of these obligations.

The highest score of 5 points is assigned when there are no known additional contractual risks. This represents a state where contractual performance uncertainty is minimized due to the absence of additional obligations or restrictions beyond basic license conditions. In such cases, data can be freely utilized, and the possibility of legal disputes due to contract breaches is low.

4 points are awarded when stringent management systems are required, such as data security or confidentiality obligations. These obligations inherently carry risks of unexpected non-compliance due to human or material management deficiencies. Additionally, complex obligations may prevent data from being used in desired forms at desired times.

3 points are given when significant risks exist, such as unlimited liability or provisions allowing licensors to conduct unrestricted audits. Particularly when licensees bear full responsibility for data usage, they may be liable even for portions attributable to the licensor's fault in disputes with third parties. Furthermore, licensors' unrestricted audit rights can easily expose minor deficiencies in contract performance, increasing the risk of legal disputes.

This evaluation framework considers the complexity and risks of license agreements. The more numerous and stringent the obligations, the higher the possibility of non-compliance, which can lead to legal disputes. Particularly when licensors have extensive audit rights over the entire contract content, this can become a potential source of disputes.

    \subsection{Type of license terms (Criteria 4.4)}

This framework classifies data license conditions for AI models as follows, reflecting the multilayered structure where datasets are formed through combinations of multiple sub-datasets. Datasets used in large-scale AI training typically form a final 'Derivative set' through horizontal and vertical combinations of multiple datasets.

Type 1 represents cases where data can be freely distributed, used, modified, combined, and derived without any conditions. This is the most liberal form of license, minimizing legal constraints on data utilization. This type represents the ideal form of license for AI model development.

Type 2 includes cases where data use and distribution are permitted but must meet certain conditions. This is subdivided into five specific subtypes:

First, cases requiring notice of author, source, copyright, and license information. This is a fundamental condition included in most open source licenses, typically represented by MIT, BSD, and CC-BY. This is the most common and easily satisfiable condition, requiring only display of relevant information during data distribution.

    Second, cases requiring notification of modifications. Licenses like Apache-2.0 and GPL series fall into this category, requiring specification of modification facts and content. While this enables tracking of data modification history, it risks difficulty in verifying previous modification histories for externally collected data.

Third, cases requiring author permission for creating Larger Works. This applies when combining datasets with different licenses requires original author consent, as in GPL v2.0 and LGPL v2.1. This represents a significant constraint that can practically limit dataset combination.

Fourth, cases requiring application of the same license to entire derivative works. GPL series and CC-BY-SA are representative examples, presenting significant factors that can cause license conflicts. Particularly when combining datasets with different licenses, the requirement to apply the same license to the whole can make practical data combination impossible.

Fifth, cases allowing only data sub-licensing. This format, commonly found in commercial licenses, permits only primary provision and prohibits redistribution. This can affect the distributability of the entire Derivative set.

Type 3 represents cases where data use is possible but distribution, modification, combination, and derivation are prohibited. This severely restricts practical data utilization and includes cases where license conditions are practically difficult to satisfy. Such data has very limited utility in AI model development.

The practical importance of this classification system is particularly evident in Derivative set composition. When combining datasets with different license conditions, serious legal risks can arise, especially when Type 2 conditions conflict. For example, combining datasets with GPL series licenses and Apache licenses may make legitimate dataset composition impossible due to license condition conflicts.

Based on this, Type 1 was considered to present no significant distribution risks regardless of Derivative and Children node states. Within Type 2, considering our research's process of tracing Children nodes, we judged that direct/indirect attribution was made to original sources. Regarding licenses with Share-Alike mentioned fourth, requiring identical Derivative and Children nodes, our research conducted analysis to the level where Agents detect Share-Alike. For CCL-type cases, we considered no risks exist in Type 2 when using either later versions or licenses with Share-Alike obligations. Type 3 cases were considered risky for Derivative nodes distributed from Children nodes, as they either prohibit dataset redistribution or may be interpreted as lacking sufficient rights.

\subsection{Class Category of Data Compliance}
\label{app:legal_risk_category}
In this framework, the classification of the final score is determined according to the table \ref{tab:datacompliancecat} below. However, the classification system and the Legal Risk associated with each category serve merely as a reference, established by legal professionals and IP experts in the AI industry, considering potential risks that may arise in AI models that are developed and used globally. This classification was formulated after evaluating over 1,000 datasets to categorize the overall risk.
\begin{table}[!ht]
\small
\centering
\begin{tabular}{p{0.07\linewidth}p{0.07\linewidth}p{0.12\linewidth}p{0.61\linewidth}} 
\toprule 
Class & Score & Category & Legal Risk \\ 
\midrule    
A-1 & 4.90 & License/Privacy : Risk Free & There is virtually no risk of legal disputes being filed in relation to data by original authors, licensors, data subjects, or related organizations even if the data is disclosed through in-house services or AI model services in public cloud. \\ 
\midrule    
A-2 & 4.57 & License/Privacy : Low Risk & While license or privacy issues exist with low likelihood of violation, there are no known cases where these issues have escalated to litigation, arbitration, or regulatory interventions. \\ 
\midrule    
A-3 & 4.22 & License/Privacy : Low Risk & While license or privacy issues exist with some likelihood of
violation, there are no known cases where these issues have escalated to litigation, arbitration, or regulatory interventions. \\ 
\midrule    
B-1 & 3.73 & License/Privacy : Moderate Risk & License or privacy issues exist with high likelihood of violation, and there are a few cases that have escalated to litigation, arbitration, dispute or regulatory interventions. The company faces a slight risk of becoming involved in the dispute. \\ 
\midrule    
B-2 & 3.51 & License/Privacy : Moderate Risk & License or privacy issues exist with high likelihood of violation, and there are some number of cases that have escalated to litigation, arbitration, dispute or regulatory interventions. The company faces a slight risk of becoming involved in the dispute. \\ 
\midrule    
C-1 & 3.18 & License/Privacy : High Risk & License or privacy issues exist with substantially high likelihood of violation, and there are a sizeable number of cases that have escalated to litigation, arbitration, dispute or regulatory interventions. The company faces a risk of becoming involved in the dispute. \\ 
\midrule    
C-2 & - & License/Privacy : High Risk & License or privacy issues exist with substantially high likelihood
of violation, and there are cases that have escalated to litigation, arbitration, dispute or regulatory interventions involving substantial financial stakes. The company faces a notable risk of becoming involved in the dispute. \\ 
\bottomrule
\end{tabular}
\caption{Class categories for \ourdata.}
\label{tab:datacompliancecat}
\end{table}

\newpage
\section{Detailed Results of Massive-Scale Analysis}
\label{sec:detailsofinformation}

\subsection{Licenses for \Entities}

\begin{figure}[!htp]
    \begin{center}
    \includegraphics[width=0.5\textwidth]{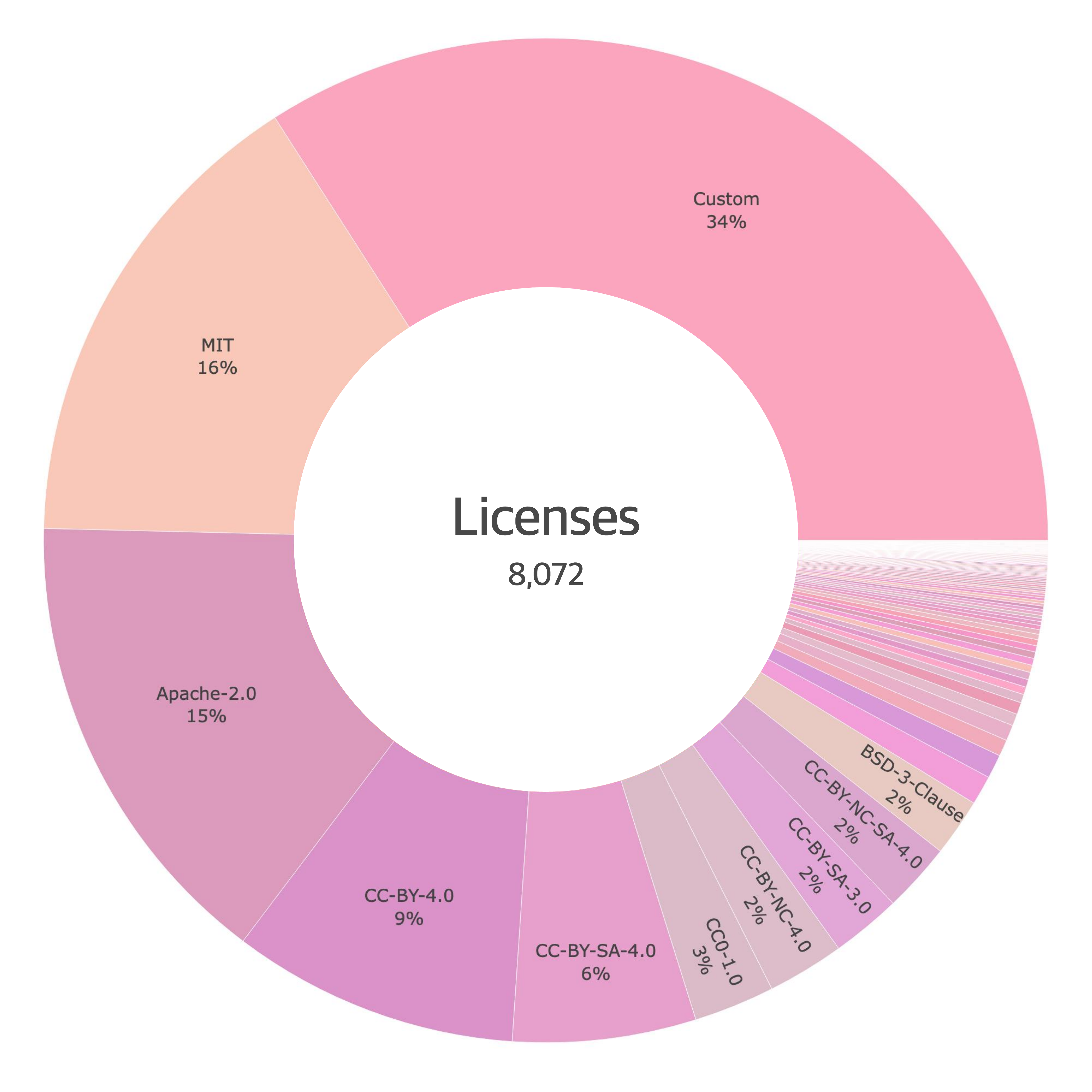}
    \end{center}
    \vspace{-1.9em}
    \caption{Distribution of licenses with their names.}
    \label{fig:dist_licenses}
\end{figure}

Out of 17,429 unique \entities, \numfoundlicenses \entities came with the corresponding license terms according to \ouragent.
As shown in \Cref{fig:dist_licenses}, the use of permissive licenses like MIT, Apache-2.0, CC-BY-4.0, and CC-BY-SA-4.0 is common.
About 34\% of the \entities had their own license terms without identified license names.
Note that \ouragent performs legal scoring and assessment based on the full texts of the terms instead of the license names, and this identification of the license names is just for an informational purpose.

\paragraph{Identified license names.}

Adobe-2006, AFL-3.0, AGPL-3.0-or-later, ANTLR-PD, Apache-2.0, Artistic-1.0, Artistic-1.0-Perl, BSD, BSD 2-Clause, BSD-2-Clause, BSD-3-Clause, BSD-Protection, C-UDA-1.0, CC, CC BY-SA 4.0, CC-BY-2.0, CC-BY-2.0-FR, CC-BY-2.0-UK, CC-BY-2.5, CC-BY-2.5-AU, CC-BY-3.0, CC-BY-3.0-IGO, CC-BY-3.0-US, CC-BY-4.0, CC-BY-NC-2.0, CC-BY-NC-3.0, CC-BY-NC-4.0, CC-BY-NC-ND-3.0, CC-BY-NC-ND-4.0, CC-BY-NC-SA-1.0, CC-BY-NC-SA-2.0, CC-BY-NC-SA-3.0, CC-BY-NC-SA-4.0, CC-BY-ND-2.1-JP, CC-BY-ND-4.0, CC-BY-SA-2.0, CC-BY-SA-2.1-JP, CC-BY-SA-3.0, CC-BY-SA-4.0, CC-PDDC, CC0-1.0, CDLA-Permissive-1.0, CDLA-Permissive-2.0, CDLA-Sharing-1.0, CeCILL-2.0, CeCILL-2.1, CreativeML-OpenRail-M, Custom, Elastic-2.0, EUPL-1.1, GFDL-1.1-only, GFDL-1.3-or-later, GPL-2.0-only, GPL-2.0-or-later, GPL-3.0, GPL-3.0-only, GPL-3.0-or-later, IDL-Train, LDC User Agreement for Non-Members, LGPL-2.1-only, LGPL-2.1-or-later, LGPL-3.0-only, LGPL-3.0-or-later, LGPLLR, License.md, Llama2, LPPL-1.3c, MIT, MIT-0, MIT-CMU, MIT-Modern-Variant, Mixed-Open-Data-License, MPL-2.0, MS-PL, MTLL, NCGL-UK-2.0, NIST-PD, O-UDA-1.0, ODBL, ODBL-1.0, ODC-By, ODC-By-1.0, Open Government Licence, OpenRail, PDDL-1.0, PolyForm-Noncommercial-1.0.0, PostgreSQL, PSF-2.0, Public Domain, Python-2.0, Unlicense, W3C-20150513, WTFPL, X11

\subsection{Metadata for \Entities}

\begin{table}[!ht]
\centering
\begin{tabular}{l c c}
\toprule
Metadata Type & Categories & Count \\ \midrule
Modality & 8 & 8,224 \\
Task & 47 & 6,932 \\
Language & 655 & 5,605 \\
Applications (General-purpose) & 1 & 5,191 \\
Applications (Specific-purpose) & 15 & 3,094 \\
\bottomrule
\end{tabular}
\caption{Overview of collected metadata.}
\label{tab:num_metadata_extracted}
\end{table}

We collected the metadata for the 8,285 Dataset-typed \entities and provide its overview in \Cref{tab:num_metadata_extracted}.

\paragraph{Modalities.}
Audio, Code, Geospatial, Image, Numeric, Text, Time-series, Video.

\begin{figure*}[h]
    \centering
    \vspace{-3.5mm} %
    \includegraphics[width=0.6\textwidth]{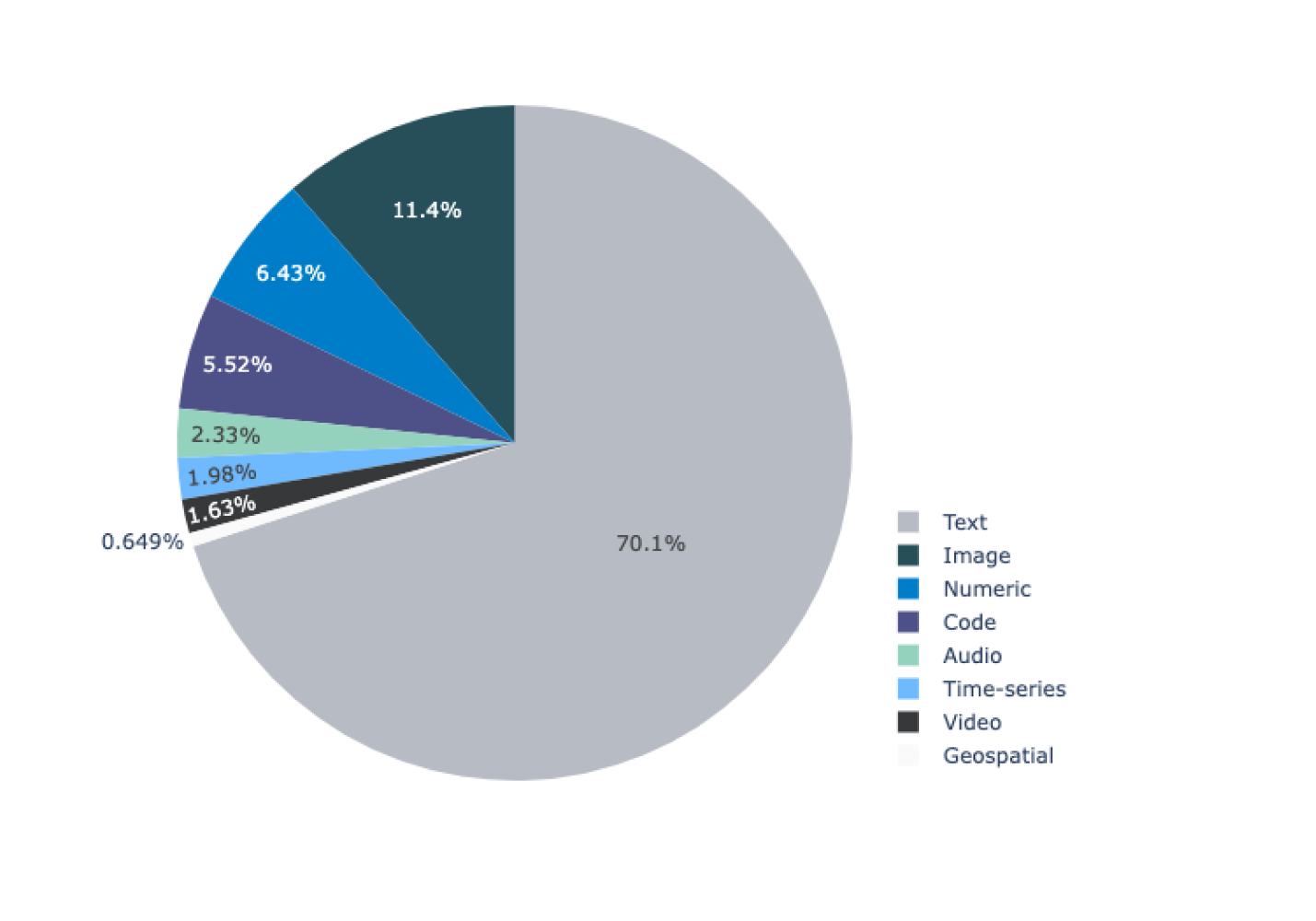}  
    \vspace{-10mm} %
    \caption{Distribution of modality subtypes.}
    \label{fig:dist_modality_subtypes}
\end{figure*}

Since multiple modalities can be selected for a single dataset, the total count exceeds the number of datasets. Most recent datasets are focused on LLM training, which contributes to the higher number of Text modalities compared to others. Additionally, the inherent inclusion of text in most datasets creates the impression that Text is significantly more prevalent than other subtypes.

\paragraph{Tasks.}
Audio Classification, Audio-To-Audio, Automatic Speech Recognition, Code, Depth Estimation, Explanation Generation, Feature Extraction, Fill-Mask, Graph Machine Learning, Image Classification, Image Feature Extraction, Image Retrieval, Image Segmentation, Image-To-Image, Image-To-Text, Mask Generation, Multiple Choice, Object Detection, Question Answering, Reinforcement Learning, Robotics, Sentence Similarity, Sentiment Analysis, Summarization, Table Question Answering, Table-To-Text, Tabular Classification, Tabular Regression, Text Classification, Text Generation, Text Retrieval, Text-To-Audio, Text-To-Image, Text-To-Speech, Text-To-Video, Text2Text Generation, Time Series Forecasting, Token Classification, Token Generation, Translation, Video Classification, Video-Text-To-Text, Visual Question Answering, Voice Activity Detection, Zero-Shot Classification, Zero-Shot Object Detection.

\begin{figure*}[h]
    \centering
    \includegraphics[width=0.6\textwidth]{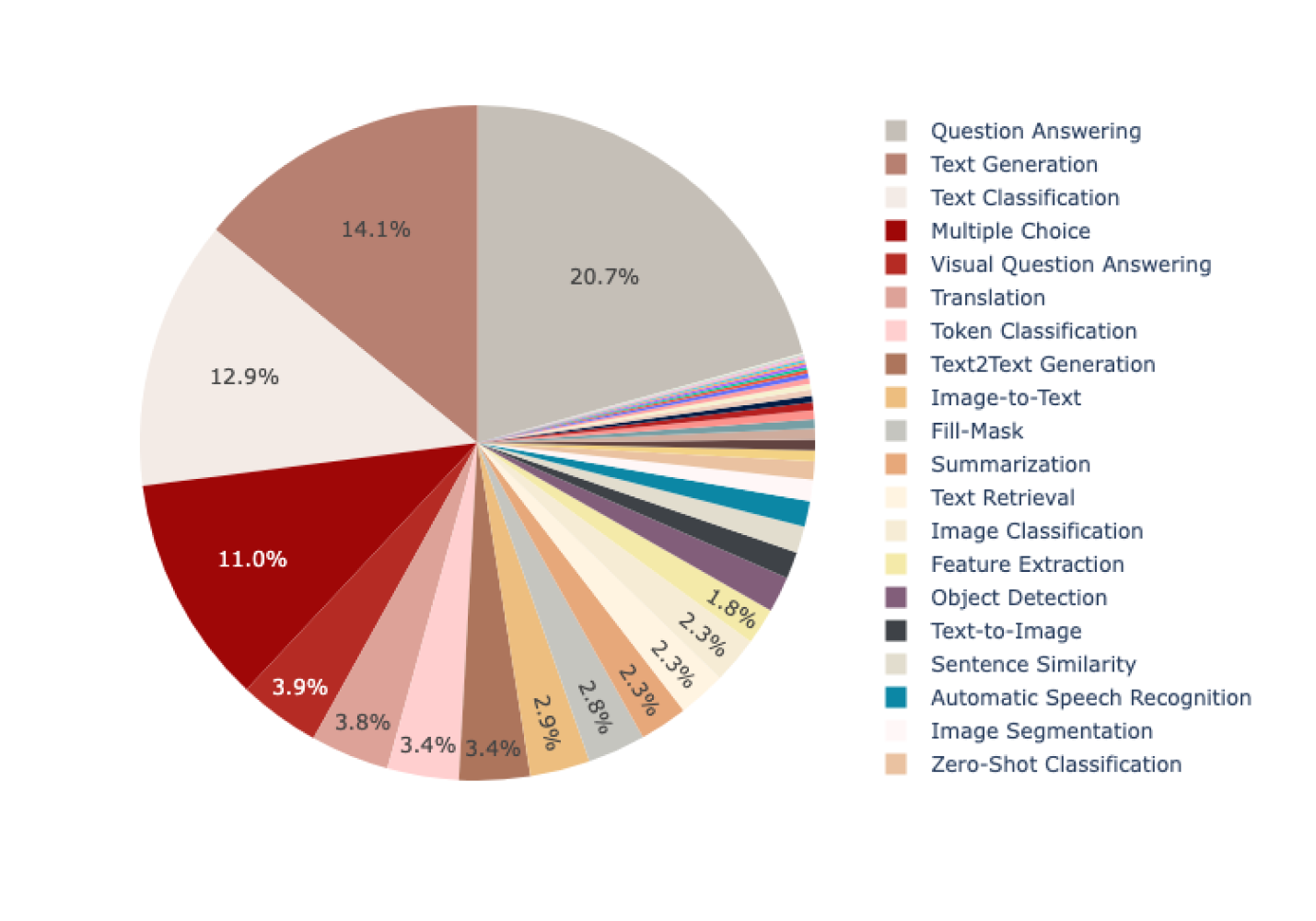}
    \vspace{-10mm}
    \caption{Distribution of task subtypes.}
    \label{fig:dist_task_subtypes}
\end{figure*}

Tasks represent metadata that helps identify the nature of the model, categorized into 47 subtypes tailored to various problem-solving formats. A single dataset can encompass multiple Tasks if it is designed to solve multiple problems simultaneously. Additionally, most recent training datasets are focused on addressing generative AI challenges rather than traditional machine learning or deep learning tasks, resulting in a higher concentration of data related to inference, response generation, and conversational AI.

\paragraph{Languages.}
Abau, Abaza, Abkhaz, Acehnese, Achinese, Acholi, Acoli, Adyghe, Afaan Oromo, Afar, African, African Languages, African-English, Afrikaans, Ahuacatlán, Akan, Akan (Twi), Akkadian, Akuntsu, Albanian, Alemannic, Algerian Arabic, Algerianarabic, Alsatian, Amazigh, Ambon, Ambonese Malay, Amharic, Ancient Greek, Ancient Hebrew, Ancientgreek (To 1453), Ankave, Apurina, Apurinã, Arabian, Arabic, Aragonese, Aramaic, Arapaho, Arifama-Miniafia, Armenian, Aromanian, Arpitan, Asháninka, Asian Languages, Assamese, Assamesee, Assyrian, Assyrian Neo-Aramaic, Asturian, Ateso, Atikamekw, Australian, Avar, Avaric, Aymara, Azerbaijani, Bactrian, Balinese, Balochi, Bambara, Bangla, Bangladeshi Bengali, Banjar, Banjarese, Bashkir, Basque, Batak, Batak Toba, Bavarian, Beja, Belarusian, Bengali, Betawi, Bhojpuri, Bihari, Bima, Bishnupriya, Bishnupriya Manipuri, Bislama, Bodo, Bodo (India), Bokmål, Bornholmsk, Bororo, Borôro, Bosnian, Brazilian, Brazilian Portuguese, Breton, Bribri, Buginese, Bulgaria, Bulgarian, Bulgarian (Latin), Burmese, Buryat, Canadian, Cantonese, Cantonese Chinese, Cappadocian, Catalan, Cebuano, Central Bicolano, Central Bikol, Central Khmer, Central Kurdish, Central Siberian Yupik, Chamorro, Chavacano, Chechen, Cherokee, Cheyenne, Chhattisgarhi, Chichewa, Chinese, Chinese (Chinese), Chinese (Hong Kong), Chinese (Latin), Chinese (Simplified and Traditional), Chinese (Simplified), Chinese (Taiwan), Chinese (Traditional), Chinese(Traditional), Chinese-Simplified, Chishona, Choctaw, Chukchi, Chukot, Churchslavic, Chuvash, Cimbrian, Classical Arabic, Classical Armenian, Classical Chinese, Classical Syriac, Coptic, Cornish, Corsican, Cree, Creek, Crimean Tatar, Croatia, Croatian, Culturax, Cusco Quechua, Czech, Danish, Dari, Dhivehi, Dimli, Dimli (Individual Language), Dinka, Divehi, Dogri, Dotyali, Dungan, Dutch, Dzongkha, Eastern Apurímac Quechua, Eastern Chatino, Eastern Mari, Egyptian, Egyptian Arabic, Emerillon, Emiliano-Romagnolo, Eml, English, Englsih, Erzya, Esperanto, Estonian, Ethiopian Tigrinya, Eu Languages, Ewe, Extremaduran, Faroese, Farsi, Fiji Hindi, Fijian, Filipino, Filipino (Tagalog), Finnish, Flemish, Fon, Fonge, French, Frisian, Frisian Dutch, Friulian, Fula, Fulah, Fulfulde, Gagauz, Gahuza, Galician, Gan Chinese, Ganda, Geez, Georgian, German, Gheg, Ghegalbanian, Ghomálá, Gilaki, Goan Konkani, Gorontalo, Gothic, Greek, Greek (Latin), Greenlandic, Guajajara, Guarani, Gujarati, Gwichin, Haida, Haitian, Haitian Creole, Hakha Chin, Hakka Chinese, Hausa, Hawaiian, Hebraic, Hebrew, Herero, Highland P Nahuatl, Hill Mari, Hindi, Hindi (Latin Script), Hindi English, Hiri Motu, Hittite, Hmong, Hungarian, Hungary, Icelandic, Ido, Igbo, Ilocano, Iloko, India, Indic Languages, Indonesian, Ingush, Interlingua, Interlingue, Inuktitut, Inupiaq, Iranian, Irish, Isixhosa, Italian, Iw, Jamaican Creole English, Japanese, Japanese (Latin), Javanese, Jèrriais, K’iche’, Kaapor, Kabardian, Kabuverdianu, Kabyle, Kalaallisut, Kalmyk, Kamba, Kangri, Kannada, Kanuri, Kapampangan, Kara-Kalpak, Karachay-Balkar, Karelian, Karo, Karo(Brazil), Kashmiri, Kashubian, Kazakh, Khaling, Khmer, Khunsari, Kiche, Kikuyu, Kildin Sami, Kinyarwanda, Kirghiz, Kirundi, Kiswahili, Klingon, Komi, Komi Permyak, Komi Zyrian, Kongo, Konkani, Korean, Kurdish, Kurdish Kurmanji, Kurdish Sorani, Kurmanji, Kwanyama, Kyrgyz, Kölsch, Ladino, Lak, Lao, Laos, Latgalian, Latin, Latvian, Lezghian, Lezgian, Ligurian, Limburgish, Lingala, Lingua Franca Nova, Literary Chinese, Lithuanian, Livvi, Llingala, Lojban, Lombard, Low German, Low Saxon, Lower Sorbian, Luganda, Lugbara, Luo, Luxembourgish, Macedo-Romanian, Macedonian, Madi, Madurese, Magahi, Maghrebi Arabic French, Mai, Maithili, Makassarese, Makurap, Malagasy, Malay, Malayalam, Maltese, Mandarin (Simplified and Traditional), Mandarin Chinese, Manipuri, Manx, Maori, Mapudungun, Marathi, Marshallese, Mazanderani, Mbya Guarani, Meadow Mari, Meitei, Mesopotamian Arabic, Mexicanero, Middle French, Min Dong Chinese, Min Nan Chinese, Minangkabau, Mingrelian, Mirandese, Mizo, Modern Greek, Modern Standard Arabic, Moksha, Mongolian, Moroccan Arabic, Mossi, Mozambican Portuguese, Munduruku, Musi, Myanmar, Nah, Nahuatl, Naija, Nauru, Navajo, Nayini, Ndonga, Neapolitan, Nepali, Newar, Newari, Ngaju, Nheengatu, Nigerian Fulfulde, Nigerian Pidgin, Norse, North Azerbaijani, North Frisian, North Sami, Northern Frisian, Northern Luri, Northern Ndebele, Northern Sami, Northern Sotho, Northwest Gbaya, Norwegian, Norwegian Bokmål, Norwegian Nynorsk, Novial, Nuer, Nuosu, Nyanja, Nyankole, Occitan, Odia, Official Aramaic, Old Church Slavonic, Old East Slavic, Old English, Old French, Old Irish, Old Turkish, Oriya, Oromo, Ossetian, Ottoman Turkish, Pampanga, Pangasinan, Panjabi, Papiamento, Pashto, Pedi, Pennsylvania German, Persian, Persian (Farsi), Philippine, Phrygian, Picard, Piedmontese, Polish, Pomak, Pontic, Portuguese, Punjabi, Pushto, Quechua, Rarómuri, Rejang, Romania, Romanian, Romansh, Rundi, Runyankole, Russia Buriat, Russian, Rusyn, Sakha, Samoan, Samogitian, Sango, Sanskrit, Santali, Sardinian, Scots, Scottish Gaelic, Serbian, Serbo-Croatian, Sesotho, Setswana, Shan, Shona, Sicilian, Silesian, Sindhi, Sinhala, Sinhalese, Skolt Sami, Slovak, Slovene, Slovenian, Somali, Sorani Kurdish, Sorbian, South Azerbaijani, South Levantine Arabic, Southern Altai, Southern Sotho, Spanish, Spanish Sign Language, Sranan Tongo, Standard Arabic, Sundanese, Swahili, Swati, Swedish, Tagalog, Tahitian, Tajik, Tamasheq, Tamil, Tarantino, Tatar, Telugu, Tetum, Thai, Tibetan, Tigrinya, Tok Pisin, Tonga, Tosk Albanian, Traditional Chinese, Tsonga, Tswana, Tulu, Tumbuka, Tunisian Arabic, Turkish, Turkmen, Tuvan, Twi, Udmurt, Uighur, Ukrainian, Umbrian, Umbundu, Upper Sorbian, Urdu, Uyghur, Uzbek, Vedic Sanskrit, Venda, Venetian, Veps, Vietnamese, Vlaams, Vlax Romani, Volapük, Votic, Võro, Walloon, Waray, Welsh, Western Armenian, Western Frisian, Western Mari, Western Panjabi, Western Punjabi, Western Sp Nahuatl, Wixarika, Wolaytta, Wolof, Wu Chinese, Xavante, Xhosa, Xibe, Xitsonga, Yakut, Yiddish, Yorem Nokki, Yoruba, Yue Chinese, Yupik, Zaar, Zaza, Zazaki, Zealandic, Zeeuws, Zhuang, Zulu.

\begin{figure*}[h]
    \centering
    \vspace{-4mm}
    \includegraphics[width=1\textwidth]{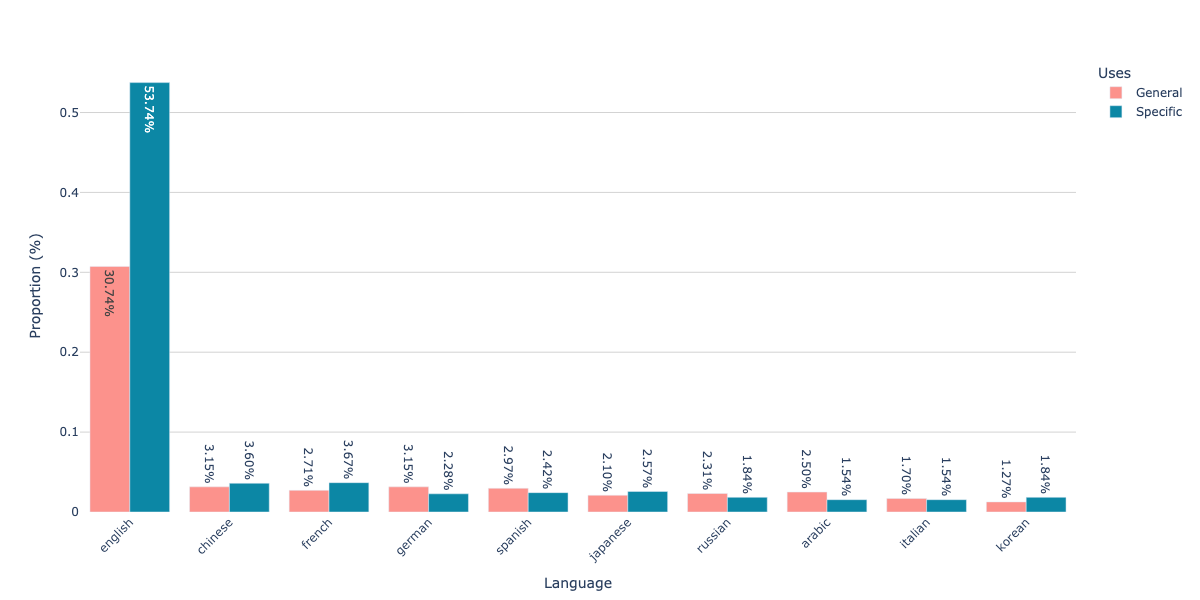}
    \vspace{-5mm}
    \caption{Distribution of languages by use.}
    \label{fig:dist_languages}
\end{figure*}

This chart displays the distribution of the top 10 most frequently used languages for both General-purpose and Specific-purpose datasets. Language data was collected exclusively from the Text modality. English accounted for the highest proportion in both purposes. However, it is notable that Specific-purpose datasets, which focus on specialized knowledge, tend to use more commonly understood languages like English. This likely contributes to the relatively higher proportion of English in such datasets.

\paragraph{Applications (specific-purpose)}
Applied Science(Engineering, Health Sciences, Technology), Formal Science(Computer Science, Mathematics and Logics), Humanity(Arts and Culture, Language and Literature, Philosophy and History), Natural Science(Biological Sciences, Chemical Sciences, Earth and Environmental Sciences, Physical Sciences), Social Science(Business and Economics, Law and Political Science, Sociology and Psychology).

\begin{figure*}[h]
    \centering
    \includegraphics[width=1\textwidth]{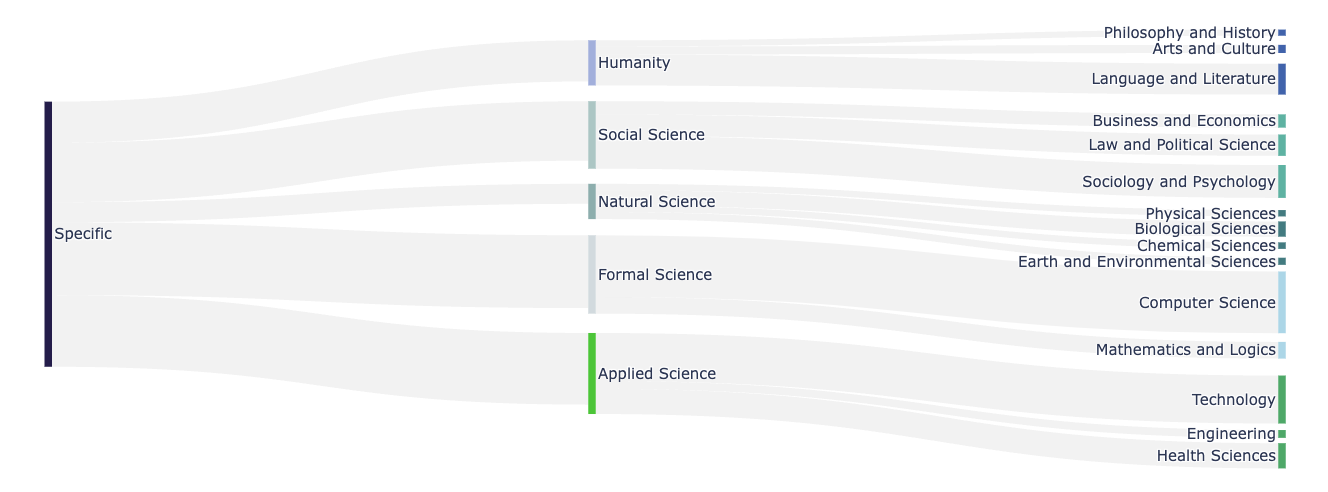} 
    \vspace{-5mm}
    \caption{Domains of specific-purpose datasets.}
    \label{fig:dist_specific_purpose_datasets}
\end{figure*}

Specific-purpose Dataset applications allow multiple selections, resulting in counts exceeding total dataset numbers. Technology, Sociology, and Language fields show relatively high representation. As shown in Figure 5-5, Formal Science datasets dominate, indicating active AI research in mathematics, logic, and computer science. Notable trends include: i) Computer Science: Concentration on generative AI coding solutions; ii) Sociology: Increased AI applications in education; iii) Legal: Development of AI-based legal services; and iv) Bio: Acceleration of AI-driven life science research.

The domain distribution of Specific-purpose Datasets reflects current AI research trends while necessitating domain-specific legal review procedures due to their bounded nature.

\section{Target Datasets for Massive-Scale Analysis}
\label{sec:datalist}
A list of 3,612 datasets was selected as the target datasets.
Only the names of the datasets are provided, so there are instances where different datasets share the same name, resulting in duplicates. You can check it all at the NEXUS web repository.

\paragraph{List of datasets.} A-Bench-HF, A-OKVQA, AAAI-21-SDU-shared-task-1-AI, ACL-ARC, ACL2023-papers, ACL2024-papers, ACVA, ACVA-10percent, ACVA-Arabic-Cultural-Value-Alignment, ADE-Corpus-V2, AESLC, AFHQ, AFQMC, AG's corpus of news articles, AGIEval, AI2 Science Questions, AILA\_casedocs, AILA\_statutes, AIVision360-8k, ALGES, ALLaVA-4V, ALMA-Human-Parallel, ALMA-R-Preference, AMF-PDF, APEACH, AQuA, AR-LSAT, ARC-Challenge, ARC-V1-Feb2018, ARC\_Easy\_SWH, ARPA-Armenian-Paraphrase-Corpus, ATC\_combined, ATEC, AToMiC-Images-v0.2, AURORA, AbductionRules, ActivityNetQA, Add-One-RTE, AdvBench, Advice-EMNLP2020, Aegis-AI-Content-Safety-Dataset-1.0, AfriSenti-Twitter, AfriSenti-twitter-sentiment, Afrikaans\_parakeet\_50hr, AlGhafa-Arabic-LLM-Benchmark-Native, AlGhafa-Arabic-LLM-Benchmark-Native-10percent, AlGhafa-Arabic-LLM-Benchmark-Translated, AlGhafa-Arabic-LLM-Benchmark-Translated-10percent, Aligned Code and Natural Language Pairs, All\_3\_Bug\_report\_elements, AlpaCare-MedInstruct-52k, Alpaca-Star, AlphaMath-Trainset, Alzheimer\_MRI, Amazon Fine Food Reviews, Amazon-C4, Amazon-Reviews-2023, AmericanStories, Amharic\_Instruction\_dataset, AnEM, Anime-dataset, Anthropic\_HH\_Golden, ArabicMMLU, Arabic\_Aya, Arabic\_EXAMS, Arabic\_MMLU, Arabic\_MMLU-10percent, ArcChallenge\_de, Arguana, Arguments Relating to Consequences, Arithmo-Data, Articles\_Constitution\_3300\_Instruction\_Set, ArxivP2P, Asian Language Treebank, AttaQ, AttributionBench, Audience-v1, AudioSet, AutoMathText, AutoMultiTurnByCalm3-22B, AutoTNLI-code, BABE, BBQ, BCCard-Finance-Kor-QnA, BLESS, BLINK, BLUEX, BQ, BRIGHT, BUSTER, BYU-Analogical-Reasoning-Dataset, Bactrian-X, Bahuvachak, Balanced-COPA, BeaverTails, BeaverTails-Evaluation, Bengali-Restaurant-Reviews, BertaQA, BiGGen-Bench, BiRdQA, BiSECT, BiasShadesRaw, BigDoc-MultiTurn-v0.3, BioCreative II Gene Mention Recognition, BioCreative V Track 3: CDR, BirdSet, Bitext-customer-support-llm-chatbot-training-dataset, Bitext-events-ticketing-llm-chatbot-training-dataset, Blog Authorship Corpus, BookMIA, Brains, Breaking\_NLI, Brown Corpus, BrushData, Buzz-V1.2, CAIMAN-ASR-BackgroundNoise, CBSD68, CC-NEWS-ES, CC100, CDial-Bias, CEBaB, CH-CS-Trends, CICERO, CIFAKE-image-dataset, CLAP\_freesound, CLARIAH-CZ Repository, CLAS\_backdoor\_recovery, CLERC, CLIP-Kinetics700, CLIcK, CLSClusteringP2P, CLSClusteringS2S, CLUTRR, CLadder, CMMMU, CMNLI, CMedQAv1-reranking, CMedQAv2-reranking, COCO, COCO-Caption, COCO-Caption2017, COD3S, CODA-19, COIG-CQIA, COLING-20, COLING2024-papers, CONDAQA, COPAL, CR, CREAK, CSL, CSSRS-Suicide, CT-RATE, CUB\_train, CV-Bench, CVECPEAPIBenchmark, CVRR-ES, Calc-ape210k, Calc-asdiv\_a, Calc-mawps, Calc-svamp, Camelyon17-WILDS, Cantonese\_English\_Translation, Captioned\_ADE20K, Capybara, Capybara-Preferences-Filtered, CaseHOLD, CelebA-HQ, CelebA-attrs, CelebA-faces, ChEBI-20-MM, ChartQA, ChartQA, ChartQA, ChartQADatasetV2, ChatDoctor-HealthCareMagic-100k, ChatGPT-Jailbreak-Prompts, ChatQA-Training-Data, ChatRAG-Bench, ChemProt, ChnSentiCorp, ChnSentiCorp, Chunking, ClariQ, Clarity-v1, ClashEval, ClassEval, ClimateAPIBenchmark, CmedqaRetrieval, CmedqaRetrieval-qrels, CoLA, CoNLL-2003, CoSQA, CoT-Collection, CoT-Collection, CoTaEval, Code-Feedback, Code-Functions-Level-Cyber, Code-Functions-Level-General, CodeAlpaca-20k, CodeAlpaca-20k, CodeAlpaca\_20K, CodeComple, CodeEditSearch, CodeExercise-Python-27k, CodeFeedback-Filtered-Instruction, CodeSearchNet-ccr, CodeSearchNet-ccr-go-qrels, CodeSearchNet-ccr-go-queries-corpus, CodeSearchNet-ccr-java-qrels, CodeSearchNet-ccr-java-queries-corpus, CodeSearchNet-ccr-javascript-qrels, CodeSearchNet-ccr-javascript-queries-corpus, CodeSearchNet-ccr-php-qrels, CodeSearchNet-ccr-php-queries-corpus, CodeSearchNet-ccr-python-qrels, CodeSearchNet-ccr-python-queries-corpus, CodeSearchNet-ccr-ruby-qrels, CodeSearchNet-ccr-ruby-queries-corpus, CodeSearchNet-go-qrels, CodeSearchNet-go-queries-corpus, CodeSearchNet-java-qrels, CodeSearchNet-java-queries-corpus, CodeSearchNet-javascript-qrels, CodeSearchNet-javascript-queries-corpus, CodeSearchNet-php-qrels, CodeSearchNet-php-queries-corpus, CodeSearchNet-python-qrels, CodeSearchNet-python-queries-corpus, CodeSearchNet-ruby-qrels, CodeSearchNet-ruby-queries-corpus, CodeUltraFeedback\_binarized, CodeXGLUE, Code\_Vulnerability\_Security\_DPO, Collection of 19,320 blogs, Com2Sense, ComQA, CommitmentBank, ConTRoL-dataset, ConjNLI, ConvFinQA, Cornell Newsroom Summarization Dataset, Counterfactual-StoryRW, Countries by Barcode Prefix, Country Calling Code, Country by Abbreviation, Country by Capital City, Country by Currency-name, Country by Domain TLD, CovidRetrieval, CovidRetrieval-qrels, Crepe, Cross-Language Text Classification using Structural Correspondence Learning, CrossSum, CulturaX, Customer\_Reviews-Second\_Hand\_Apparels, Cyber-Logic, CyberDataset1, CyberNative\_Code\_Vulnerability\_Security\_DPO-PreferenceShareGPT, CyberSecurityEval, Cyberdatasetjson, DDO dataset, DGen, DNC, DNC, DPO-En-Zh-20k, DRCD, DROP, DS-1000, DSTC3-Corpus, DSV, DTS\_session\_datasets, DUDE\_loader, Darija\_Dataset, Daring-Anteater, Datasetclinicalv2, Datasets for Noun Compound Interpretation, Deceptive Opinion Spam Corpus, DetailCaps-4870, Detoxifying Language Models Risks Marginalizing Minority Voices, Dialogue NLI, DiscoEval, Diseases\_Symptoms, Disfl-QA, Div2k, DocLayNet, DocLayNet-base, DocVQA, Docmatix, DocumentVQA, DomainNet, DuRetrieval, DuRetrieval-qrels, DualRL, Dynabench Task 3, Dynamically-Generated-Hate-Speech-Dataset, Dynasent\_Disagreement, ECQA-Dataset, ELYZA-tasks-100, EMNLP2023-papers, EQ-Bench, ES-VA\_translation\_test, ESC50, EVALution 1.0, EarthView, EasyJailbreak\_Datasets, EcomRetrieval, EcomRetrieval-qrels, EffectiveFeedbackStudentWriting, EgoExoLearn, Emilia-Dataset, English-Chinese, EnronSpam, EsportsBench, Europarl Parallel Corpus, Eurovoc, EusExams, EusProficiency, EusReading, EusTrivia, Eval\_Pref\_Dataset\_with\_stella\_400M\_v5\_embeddings, Evol-Instruct-Code-80k-v1, Experimental Data for Question Classification, Explain the Joke, FGVC-Aircraft, FGVC\_Aircraft\_test, FGVC\_Aircraft\_train, FLAN, FLD.v2, FOLIO, FOLIO, FOLIO, FRENK-hate-en, FRENK-hate-hr, FUNSD, FaceCaption-15M, Facts2Story-data, FairFace, FairytaleQA, FarsTail, FarsTail, FeTaQA, Fever, FewSum, FiQA, FiQA, FiQA, FigLang2022, FineTome-100k, FirstnameAndLastnameOnlyAug06, FlashRAG\_datasets, Flickr, FoodSeg103, Format-v1, GAIA, GLOBE, GLUE Diagnostic Dataset, GPT-4-LLM, GPT4-LLM-Cleaned, GPTeacher, GQA, GQA-ru, GRAB, GSM-Plus, GSM8k-bgeval, GTSRB, GUE, GWStance, Gemma\_Cyber, GenAI-Bench, GenAI-Bench-1600, GenQA, Genia Project, Glot500, GocReport-QS, GrailQA, GrandMaster-PRO-MAX, Gryphe-3.5-16k-Subset, Guacamol, HC3, HC3, HCRC Map Task Corpus, HELP, HH-RLHF-Harmless-and-RedTeam-standard, HH-RLHF-Helpful-standard, HPDv2, HQ-Edit, HallusionBench, HaluBench, HaluEval, Harmless, HateEvalTeam, HebDB, Heliconius-Collection\_Cambridge-Butterfly, HellaSwag\_de, HelpSteer, HelpSteer2, HelpSteer2-DPO, HelpSteer2-binarized, Helpful, Helpsteer-preference-standard, Helpsteer2-standard, HindiEnglish Corpora, Hippocorpus, Hopper-v3, Hopper-v3-position, HotpotQA, Huatuo26M-Lite, Hugging Face Datasets, Hugging Face Datasets Contribution Guide, Human Ratings of Natural Language Generation Outputs, HungarianDocQA\_IT\_SynQA, I2D2, IAC 2.0, IAM-line, ICON-QA, IDMGSP, IEMOCAP, IFEval, IFlyTek-classification, IIRC, IIT-CDIP, IL-TUR, ILUR-news-text-classification-corpus-formatted, IMDB, IN22-Gen, IRFL, ImageNet1K-val, ImageRewardDB, InFoBench, Indian Food 101, IndicCOPA, IndicGenBench\_flores\_in, IndicGenBench\_xorqa\_in, IndicNLPSuite, IndicQA, IndicSentiment, IndicSentiment, IndicXParaphrase, IndoMMLU, Infinity-Instruct, Infinity-Instruct, Inkuba-instruct, JA-VG-VQA-500, JA-VLM-Bench-In-the-Wild, JAQKET, JBB-Behaviors, JDReview-classification, JDocQA-binary, JGEW1z1zKopSk7m, JGLUE, JGLUE, JMMLU, JMTEB, JMedBench, JQaRA, JaCWIR, JailBreakV-28k, JapaneseDocQA\_IT, Jigsaw Unintended Bias in Toxicity Classification, John Wieting, KAGL, KMMLU, KMMLU-HARD, KOREAN-WEBTEXT, KOpen-HQ-Hermes-2.5-60K, KPA\_2021\_shared\_task, Kallaama-Wolof-large-v2-prepared, KnowUnDo, Ko-StrategyQA, KoAlpaca-v1.1a, KorMedMCQA, Koumankan\_mt\_dyu\_fr, LAMBDA, LCQMC, LEDGAR, LEval, LIBRA, LLM-AggreFact, LLM\_compression\_calibration, LLaVA-Bench-Wilder, LLaVA-Human-Preference-10K, LLaVA-Instruct-150K, LLaVA-Instruct-150K-JA, LLaVA-NeXT-Data, LLaVA-NeXT-Interleave-Bench, LLaVA-OneVision-Data, LLaVA-ReCap-118K, LLaVA-ReCap-558K, LLaVA-ReCap-CC3M, LMMs-Eval-Lite, LP-MusicCaps-MC, LVEval, LVIS-Instruct4V, LaMini-instruction, Laion\_aesthetics\_5plus\_1024\_33M, Language-v1, Large Movie Review Dataset, LegalLensNER, LegalLensNER-SharedTask, LegalLensNLI, LegalQuAD, Length-v1, ListUltraFeedback, LitSearch, LiveBenchResults, Llama-3-Magpie-Pro-1M-v0.1, Llama-3.1-405B-Instruct-evals, Llama-3.1-70B-Instruct-evals, Llama-3.1-8B-Instruct-evals, Llama-3.1-8B-evals, LoNLI, LogiQA-dataset, LogiQA2.0, Logic2Text, LogicNLI, LongAlign-10k, LongAlpaca-12k, LongBench, LongEmbed, LongForm, LongRAG, LongVideoBench, LongWriter-6k, LrvInstruction, M3IT, M3IT\_ML, MAGE, MASSIVE, MATH, MATH, MATH-Hard, MATRES, MELD, MERA, MFRC, MG-Verilog, MIMICIT, MINT-1T-HTML, MIT\_environmental\_impulse\_responses, ML-ArXiv-Papers, MLDR, MLMMLU, MLRSNet, MLSP2024, MM-UPD, MMBench, MMBench-Video, MMBench-ru, MMBench\_EN, MMBench\_dev, MME, MMHal-Bench, MMLU-Pro, MMLU-SR, MMLU-STEM, MMLU-bgeval, MMLU\_de, MMMU, MMMU, MMMU, MMMU, MMMU\_Pro, MMStar, MMVet, MMarcoRetrieval, MMarcoRetrieval-qrels, MNBVC, MNIST-M, MOH, MP-DocVQA, MSCOCO, MSMarco, MS\_COCO\_2017\_URL\_TEXT, MT-Bench\_Evaluated, MTbenchJapanese, MUIRBENCH, MUSE-News, MVBench, MagicBrush, Magicoder-Evol-Instruct-110K, Magicoder-OSS-Instruct-75K, Magpie-Air-300K-Filtered, Magpie-Llama-3.1-Pro-DPO-100K-v0.1, Magpie-Pro-300K-Filtered, Magpie-Pro-300K-Filtered-H4, Magpie-Pro-MT-300K-v0.1, Magpie-Reasoning-150K, Mantis-Eval, Mantis-Instruct, MatText, Math-Shepherd, Math-Step-DPO-10K, Math-llama2-200k, Math23K, MathInstruct, MathVerse, MathVerse-lmmseval, MathVision, MathVista, MedConceptsQA, MedMCQA, MedQA-Reason, MedQA-USMLE-4-options, MedQA-USMLE-4-options-hf, MedQuad-MedicalQnADataset, MedTrinity-25M, MedicalRetrieval, MedicalRetrieval-qrels, Mercury, MetaMathQA, MetaMathQA, MetaMathQA-40K, Microsoft Research Paraphrase Corpus, Mind2Web, MixEval, Mixed-Arabic-Datasets-Repo, Mmarco-reranking, Mol-Instructions, Mostly Basic Python Problems, Movie Review Data, MuDoConv, MuSR, MuTual, Multi-Genre Natural Language Inference, Multi-Step-Deductive-Reasoning-Over-Natural-Language, MultiArith, MultiHopRAG, MultiJail, MultiPL-E, MultiPremiseEntailment, Multi\_Legal\_Pile, Multilingual Reward Bench, MultilingualSentiment-classification, Multimodal-Mind2Web, MusicCaps, NAACL2024-papers, NCBI Disease Corpus, NExTQA, NFCorpus, NIH-Chest-X-ray-dataset, NLI Style FEVER, NLI\_datasets, NQ, NQ-Swap, NTREX, NVDLibraryBenchmark, NYC-Airbnb-Open-Data, NeQA, Nectar, Neural-Code-Search-Evaluation-Dataset, Neural-Natural-Logic, News-Headlines-Dataset-For-Sarcasm-Detection, NinjaMasker-PII-Redaction, NoCaps, NoCaps, NopmWritingStruct, NuminaMath-CoT, NuminaMath-TIR, NusaX-senti, OCNLI, OCR-VQA, OCRBench, OIG, OIG(unified\_chip2), OK-VQA, OK-VQA\_train, OLID, OLID, OMG, OPP-115 Corpus, OPUS Books, OSCAR-2201, OSCAR-2301, OlympiadBench, OnlineShopping-classification, Open-Assistant, Open-Critic-GPT, Open-Platypus, Open-Sora-Plan-v1.2.0, OpenHermes-2.5, OpenHermes-2.5-H4, OpenHermesPreferences, OpenOrca, OpenOrca, OpenOrca-gugugo-ko, OpenOrca-tr, OpenSubtitles, OpenX-Embodiment, Opus\_Instruct\_25k, P3, PARADE\_dataset, PAWS-gl, PAWSX, PES-2018-2022, PEYMA, PILE\_Wikipedia\_Pretraining\_subset\_valid\_ret\_tokens\_syn\_knowledge\_filtered, PILE\_Wikipedia\_validation\_set\_insert\_ret\_tokens-wikipedia-dpr-k-1-OP-True, PILE\_Wikipedia\_validation\_set\-\_synthetic\_knowledge, PILE\_wikipedia\_synthetic\_knowledge\_filtered, PKU-SafeRLHF, PKU-SafeRLHF-10K, PKU-SafeRLHF-30K, PKU-SafeRLHF-30K-standard, PKU-SafeRLHF-Processed, POPE, POVID\_preference\_data\_for\_VLLMs, PQA, PUGG\_KG, Paper Reviews, PasswordMatch, PatchCamelyon, Pendulum-v1, PerSenT, PerceptionTest\_Val, PersianQA, PersonaHub, PersonalLLMTestSet, PixelBytes-Pokemon, Places\_in\_Japan, PoKi-Poems-by-Kids, PokemonCards, PolishCyberbullyingDataset, PopQA, PostgreSQL, Preference-Collection, Pretraining\_Dataset, Probing Tasks, Products-10k-BLIP-captions, Prometheus2-preference-standard, ProstT5Dataset, ProteinLMBench, ProverbEval, PtBrVId, PubMedQA, PubMedQA, PubMedQA, PubMedQA\_instruction, Puffin, PwC, PythonTutor-Evol-1k-DPO-GPT4\_vs\_35, Q-Bench-HF, Q-Bench2-HF, QANTA, QBQTC, QMDSCNN, QuRating-GPT3.5-Judgments-Test, Quora Insincere Questions Classification, Quora Question Pairs, Qwen\_\_Qwen2-72B-Instruct-details, RACE, RAG\_Multilingual, REDFM, REFreSD, RES-Q, RICO-ScreenQA-Short, RLAIF-V-Dataset, RLHF-V-Dataset, ROCStories, ROCStories and the Story Cloze Test, ROCStories and the Story Cloze Test, ROOT9, RPGPT\_PublicDomain-alpaca, RareBench, ReCoRD, ReDis-QA, RealSynth, RealWorldQA, RealworldQA, Recap-COCO-30K, Recap-DataComp-1B, Recognizing Textual Entailment, RedPajama-Data-1T, RedPajama-Data-1T-Sample, RedPajama-Data-V2, RefCOCO, RefCOCOg, RefCOCOplus, Reflection-Dataset-v2, RiddleSense, RoG-cwq, RoG-webqsp, RobustLR, SAM-LLaVA-Captions10M, SAM\_finetune\_dataset, SATIN, SCAN, SCIERC, SCP-096, SDCNL, SEED, SEED-Bench, SEED-Bench-2, SEED-Data-Edit-Part1-Openimages, SELM-Llama-3-8B-Instruct-dataset, SFT-OpenHermes-2.5-Standard, SHP, SLF5K, SMCalFlow, SMS Spam Collection, SMS Spam Collection, SMolInstruct, SOCKET, SPA-VL, SPHERE\_100K, SPIDER, SPIN\_iter0, SPL, SPML\_Chatbot\_Prompt\_Injection, SQAC, SQuAD, SQuAD 2.0, SQuADDS\_DB, SREDFM, STS-B, STSB, STSbenchmark, SVAMP, SVAMP, SWE-bench, SWE-bench\_Lite, SWE-bench\_Lite\_filtered, SWE-bench\_Lite\_oracle, SWE-bench\_Verified, SWE-bench\_Verified\_50, SWE-bench\_bm25\_13K, SWE-bench\_oracle, Salad-Data, Sarcasm in Twitter, SciFIBench, SciFact, SciNLI, ScienceQA, ScienceQA, ScienceQA\_text\_only, ScreenSpot, SeaExam, Security-TTP-Mapping, SeeTRUE-Feedback, Self-instruct, SemEval 2019 Task 10: Math Question Answering, SemEval-2018 Task 1: Affect in Tweets, SemEval-2018 Task 3: Irony Detection in English Tweets, SemEval2020-Task4-Commonsense-Validation-and-Explanation, SemRel2024, Sen-Making-and-Explanation, SentEval-CR, Sentiment Treebank, Shakespeare\_sonnet154, Shanghai\_Dialect\_TTS\_openai, ShareGPT, ShareGPT4Video, ShareGPT90K, ShareGPT\_Vicuna\_unfiltered, ShareGPT\_Vicuna\_unfiltered, SherLIiC, SherLIiC, SimpleSafetyTests, Skylion007-openwebtext-tokenizer-gpt2, SlimOrca, SlimOrcaDedupCleaned, SlimPajama-627B, SlimPajama-6B, SocialStigmaQA, Sonnet3.5-Charcard-Roleplay, Sonnet3.5-SlimOrcaDedupCleaned, Source Blending in NLG, Source-v1, Spanish Question Answering Corpus, Splash, StArCon, Stable-Diffusion-Prompts, Stack Exchange Data Explorer, Stack-Exchange-April, StanfordCars\_test, StanfordCars\_train, Stheno-Data-Filtered, StockImages-CC0, StrategyQA, StrategyQA, StreamBench, Strong Password Checker, StrongREJECT, StudentEval, Sub\_test\_diff\_length, Summarize-from-Feedback Dataset, SupremeCourtOfIsrael, SynthRP-Gens-v1.1-Filtered-n-Cleaned, Synthetic-Persona-Chat, Synthstruct-Gens-v1.1-Filtered-n-Cleaned, SystemChat, T2Reranking, T2Retrieval, T2Retrieval-qrels, TACO, TACT, TAIDE-14-tasks, TCEval-v2, TCGA, TGQA, THUCNewsText, TNews-classification, TOFU, TORQUE, TR-News, TREC-COVID, TREC-QC, Table-GPT, TapeCam-Mirflickr-Ambient, Tatoeba-Challenge, Taur CoT Analysis Project meta-llama Llama-2-7b-chat-hf, Taur\_CoT\_Analysis\_Project, Taur\_CoT\_Analysis\_Project\_\_\_google\_\_gemini-1.5-pro-001, TechnicalSupportCalls, Telugu\_sentences, TempCompass, Text-to-sql-v1, TextCaps, Thaqalayn-Classical-Arabic-English-Parallel-texts, The SEMAINE Database, The dataset does not have a specific name mentioned in the provided text, but it is referred to as ``a dataset of 40,000 entries'' created for improving online hate detection., The dataset does not have a specific name mentioned, but it is referred to as a ``new dataset with 4,959 questions'' labeled based on the new question type ontology., ThuNewsClusteringP2P, ThuNewsClusteringS2S, TinyModelTokIds, TinyStories, TinyStoriesV2, TinyStoriesV2\_cleaned, TinyStories\_stlm\_training\_progress, Tome-1.5-1m, ToolBench, TopiOCQA, Touche2020, Touche23-ValueEval, Touché23-ValueEval, Toxicity Classification Jigsaw, Tracie, TravelPlanner, Treebank-3, TriangleCOPA, TriviaQA-in-SQuAD-format, TroFi, TruthfulQA\_de, Tuxemon, TvTroper, Twitter-Sentiment-Analysis, UD\_English-EWT, UFSCdatabase, US\_Airline\_Sentiment, UTSD, UltraChat, UltraFeedback, UltraFeedback-preference-standard, UltraInteract\_pair, UltraInteract\_sft, UnifiedSKG, UnsafeBench, VIDIT-Depth-ControlNet, VLFeedback, VQAv2, VQAv2, VQAv2, VUAC, ViMedNLI, ViNLI, VibeEval, Video-MME, VideoChatGPT, VideoRetrieval, VideoRetrieval-qrels, VirusTotalBenchmark, Vista, VisualWebBench, Visual\_Emotional\_Analysis, VitaminC, VizWiz-VQA, Voice-Data-New-Schema-Processed-whisper-medium, WANLI, WIKI\_QA\_Near\_dedup, WMT14 Translation Task Dataset, WMT22-Test, WMT23-Test, WRN-Chapter-1, WRN-Chapter-2, WTO-PDF, WangchanThaiInstruct, WebInstructSub, WebLINX, WebSight, WiC, WikiEval, WikiMIA, WikiPlots, WikiQA-Free\_Form\_QA, WikiSQE, WikiText, Wikipedia\_Person\_Unlearn, Wikitext-TL39, WildBench, WildBench-V2-Model-Outputs, WildChat, WildChat-1M, WildChat-1M-Full, WinoWhy, Winograd Schema Challenge, WizardLM\_evol\_instruct\_V2\_196k, WordLength, WordNet, XCSR, XCSR, XFUN, XL-WiC, XLCoST, XSTest, Yelp Dataset, ZINC20, ZebraLogicBench, ZebraLogicBench-private, Zero-Shot Relation Extraction, ZeroEval, abalone, about, absa, aclue, ade20k, ade20k-mini, ade20k-nano, ade20k-panoptic-demo, ade\_corpus\_v2, adult-census-income, adulteration\_dataset\_26\_08\_2021.csv, advanced\_ai\_risk, adversarial\_fever\_nli, adversarial\_qa, adversarial\_qa, aes\_enem\_dataset, aexams, afrihate, afrimgsm, afrimmlu, afrimmlu-translate-test, afrixnli, afrixnli-translate-test, ag\_news, ag\_news, ag\_news, ag\_news\_pt, agieval-aqua-rat, agieval-aqua-rat, agieval-gaokao-biology, agieval-gaokao-chemistry, agieval-gaokao-chinese, agieval-gaokao-english, agieval-gaokao-geography, agieval-gaokao-history, agieval-gaokao-mathcloze, agieval-gaokao-mathqa, agieval-gaokao-physics, agieval-jec-qa-ca, agieval-jec-qa-kd, agieval-logiqa-en, agieval-logiqa-en, agieval-logiqa-zh, agieval-lsat-ar, agieval-lsat-ar, agieval-lsat-lr, agieval-lsat-lr, agieval-lsat-rc, agieval-lsat-rc, agieval-math, agieval-sat-en, agieval-sat-en, agieval-sat-en-without-passage, agieval-sat-en-without-passage, agieval-sat-math, agieval-sat-math, ai-arxiv-chunked, ai-arxiv2, ai-arxiv2-semantic-chunks, ai-medical-chatbot, ai-text-detection-pile, ai2\_arc, ai2\_arc, ai2\_arc-hi, ai2d, aihub\_retriever\_commonsense, aihub\_retriever\_news, aimo-validation-math-level-5, aio, air-bench-2024, airbnb\_embeddings, airdialogue, airoboros, airoboros-3.2, ajgt\_twitter\_ar, al\_tadmoreyyah, all-nli, all\_nli\_angle\_format\_b, allegro\_reviews, allocine, allocine, alloprof, aloha\_sim\_insertion\_human, aloha\_sim\_transfer\_cube\_human, alpaca, alpaca, alpaca-cleaned, alpaca-data-gpt4-chinese, alpaca-data-pt-br, alpaca-gpt4, alpaca-gpt4-data-zh, alpaca-zh, alpaca2, alpaca\_2k\_test, alpaca\_en, alpaca\_eval, alpaca\_farm, alpaca\_gpt4\_en, alpaca\_gpt4\_zh, alpaca\_messages\_2k\_dpo\_test, alpaca\_messages\_2k\_test, alpaca\_subset\_1, alpaca\_zh, alphanli, amazon, amazon\_counterfactual, amazon\_massive\_intent, amazon\_massive\_scenario, amazon\_polarity, amazon\_polarity, amazon\_polarity, amazon\_reviews\_multi, amazon\_reviews\_multi, amazon\_us\_reviews, ambient, ambig\_qa, ambig\_qa, americas\_nli, amharic-qa, ami, ami-ihm, ammlu, amnesty\_qa, analogy\_questions, ancora-ca-ner, animals-ij5d2, anli, anli, anli, anonymous-working-histories, aozorabunko-clean, apex-instruct-for-annealing-sup, app\_reviews, app\_reviews, apps, apps, apps-qrels, apps-queries-corpus, aqua\_rat, aquamuse, ar, arabic\_xvector\_embeddings, arc-challenge-bgeval, arc-easy-bgeval, arc-tr-v0.2, arc\_ca, arc\_challenge\_mt, arc\_italian, arcd, arena-arxiv-7-2-24, arena-human-preference-55k, arena\_battles\_embeddings, argilla-dpo-mix-7k-refined-critic-reformat, argilla-ultrafeedback-binarized-preferences-cleaned, args.me corpus, arguana, arguana, argument-reasoning-comprehension-task, argument\_quality\_ranking\_30k, arithmetic, ark\_example, arxiv-abstracts-2021, arxiv-classification, arxiv-clustering-p2p, arxiv-clustering-s2s, arxiv-march-2023, arxiv-summarization, arxiv\_alltime, arxiv\_dataset, arxivqa\_test\_subsampled, asdiv, ashraq-esc50-1-dog-example, askubuntudupquestions-reranking, asleep\_keyboard, aspectemo, asset, assin2, astro-bench-test, atco2-asr, atco2-asr-atcosim, atis, atlas, attempto-nli, audio-diffusion-256, audio-labeled, audio-set-16khz, audio-vad, audio\_test\_dataset, audioset, auto-mpg, avicenna, awesome-chatgpt-prompts, aya\_collection, aya\_collection\_language\_split, aya\_dataset, aya\_evaluation\_suite, aya\_redteaming, azaria-mitchell-diff-filtered-2, bAbI-tasks, babi, babi\_nli, babi\_qa, babilong, babilong-1k-samples, babylm, baidu-ultr\_baidu-mlm-ctr, baihe-private, baize-chatbot, balanced-copa, banking77, banking77, banking77, base-security-qa, based-fda, based-squad, basic-knowledge-test, basic\_arithmetic, basqueGLUE, bbc-news, bbc\_news\_alltime, bbh, bbh, bbh, bbh-cot, bbq, bbq, bbq, bc5cdr, beans, beethoven, beir, beir-corpus, belebele, bert-cloth, bert\_pretrain\_phase2, bgglue, bianet, bias\_in\_bios, biblenlp-corpus-mmteb, big-patent-clustering, big\_bench\_hard, bigbench, bigbench, bigbench, bigbenchhard, bigcode-pii-dataset, bigcodebench, bigcodebench-hard, bigvul, billsum, billsum, binding\_affinity, biorxiv-clustering-p2p, biorxiv-clustering-s2s, biorxivP2P, biosses-sts, bird\_text\_to\_sql, bitext\_nusatranslation\_miners, bitext\_nusax\_miners, bizbench, blimp, blimp, blimp, blurb, blurbs-clustering-s2s, bn\_hate\_speech, bnb, bold, bonito-experiment, bonito-experiment-eval, bookcorpus, booksum, booksum, booksum-complete-cleaned, boolq, boolq, boolq-audio, boolq-natural-perturbations, boolq\_helm, boolq\_italian, boolq\_pt, bornholmsk\_parallel, br-quad-2.0, br\_quad\_20, brain\_data\_corr\_balanced, brain\_mri\_train\_test\_split, break\_data, breast-cancer, breast-cancer-wisconsin, breast-histopathology-images, breastcancer-ultrasound-images, bright-expanded, broad\_twitter\_corpus, bsard, bt\_modified, bucc-bitext-mining, c3, c4, c4-10k, c4-code-20k, c4-code-tokenized-2b, c4-en-10k, c4-en-validation, c4-filter-small, c4-subsets, c4-tokenized-2b, c4\_calibrate\_mini, cad\_naacl2021, cail2018, callfriend, camel, camel-bio, captcha-images, cartoon-blip-captions, casehold, casino, catalanqa, catalonia-independence-corpus, cats-image, cats\_vs\_dogs, cats\_vs\_dogs\_sample, cc100-yue-tagged, cc12m-wds, cc3m-wds, cc\_news, ccaligned\_multilingual, ccmatrix, ccnews, cdt, cedr-classification, cedr\_v1, celeb-a-hq, celeba-hq, celeba-hq-256x256, certifications, ceval-exam, ceval-exam, cfq, charades, chat-formatted-magicoder, chat-formatted-metamath, chat\_formatted\_examples, chat\_malformatted\_examples, chatbot\_arena\_completions, chatbot\_arena\_conversations, cherry\_picked\_prompts, chess-evaluations, chest-xray, chest-xray-classification, chest-xray-pneumonia, chest\_xray, chestx, chinese\_conversation\_and\_spam, chinese\_ner\_sft, chirp-v2-dataset, chr\_en, chronos\_datasets, chronos\_datasets\_extra, chunked-shuffled-wikipedia20220301en-bookcorpusopen, churn-prediction, cifar10, cifar100, cifar10\_10pct\_plus\_cifar100\_humans, cinepile, circa, cirr\_imgs, cirr\_val, citation\_intent, cities\_wiki\_clustering, civil\_comments, civil\_comments, claim\_stance, clapnq, clcd-english, cleaned-mining-deepseek-llm-python-binarized-gt-replace, cleaned-mining-deepseek-llm-python-binarized-gt-replace, cleaned-mining-deepseekllm-base-all-binarized, clickbait\_news\_bg, climate, climate-1day, climate-2day, climate-2day, climate-3day, climate-fever, climate-fever, climate\_detection, climate\_sentiment, cloudops\_tsf, clue, cmath, cml-tts, cmmlu, cmrc2018, cmu-arctic-xvectors, cnli, cnn-dailymail, cnn\_dailymail, cnn\_dailymail, coached\_conv\_pref, coco, coco-30-val-2014, coco-test, coco2017, coco2017, coco\_captions\_1107, coco\_dataset\_script, coco\_test, coco\_test\_function, coco\_train, coco\_val, coconot-sft, codah, code-chat-assistant-v1, code-comprehension, code-docstring-corpus, code-feedback-10k-deepseekv2-critic, code\_contests, code\_contests, code\_exercises, code\_generation, code\_generation\_lite, code\_instructions\_122k\_alpaca\_style, code\_search\_net, code\_x\_glue\_ct\_code\_to\_text, codefeedback-mt-qrels, codefeedback-mt-queries-corpus, codefeedback-st-qrels, codefeedback-st-queries-corpus, codeparrot-ds-train, codeparrot-ds-valid, codeparrot\_clean\_subset\_train, codetrans-contest-qrels, codetrans-contest-queries-corpus, codetrans-dl-qrels, codetrans-dl-queries-corpus, codexglue\_code2text\_go, codexglue\_code2text\_java, codexglue\_code2text\_javascript, codexglue\_code2text\_php, codexglue\_code2text\_python, coding, coedit, colorization\_dataset, colpali\_train\_set, combined, comet-atomic-2020, commavq, commitpackft, common\_gen, common\_gen, common\_gen, common\_voice\_11\_0, common\_voice\_12\_0, common\_voice\_13\_0, common\_voice\_14\_0, common\_voice\_15\_0, common\_voice\_16\_0, common\_voice\_16\_1, common\_voice\_17\_0, common\_voice\_17\_0, common\_voice\_17\_0, common\_voice\_18\_0, common\_voice\_1\_0, common\_voice\_2\_0, common\_voice\_4\_0, common\_voice\_6\_1, common\_voice\_7\_0, common\_voice\_8\_0, common\_voice\_9\_0, commongen\_lite, commonsense\_qa, commonsense\_qa, commonsense\_qa\_2.0, community-science-paper-v2, comparisons, competition\_math, complex\_web\_questions, conala, conceptual-captions-12, conceptual\_captions, conll04, conll2002, conll2003, conll2003, conll2012\_ontonotesv5, conllpp, contents, contents, conv\_ai\_2, copa, copa\_hr, copycolors\_mcqa, copycolors\_mcqa\_mib, coqa, coqa, coqa-stories, coral, cord-layoutlmv3, cord-v1, cord-v2, core17-instructions, corrected\_ifeval, cos\_e, cosmopedia, cosmopedia-100k, cosmos\_qa, cosmos\_qa, cosmos\_qa, cosmos\_qa, cosqa-qrels, cosqa-queries-corpus, cot-eval-traces-2.0, counter-strike-001, counterfactually-augmented-imdb, counterfactually-augmented-snli, country-by-national-dish, country-continent.json, country-government-type.json, country-independence-date.json, country-iso-numeric.json, country-region-in-world.json, coverbench, covost2, covost2, cowrie\_dataset, coyo-700m, coyo-hd-11m-llavanext, cppe-5-sample, cqadupstack-android, cqadupstack-english, cqadupstack-gaming, cqadupstack-generated-queries, cqadupstack-gis, cqadupstack-mathematica, cqadupstack-physics, cqadupstack-programmers, cqadupstack-stats, cqadupstack-tex, cqadupstack-unix, cqadupstack-webmasters, cqadupstack-wordpress, craigslist\_bargains, crates-20240903, creak, credit-card-clients, crello, cresa-identity-train-1, cropped-vggface2-224, cross\_code\_eval\_python, crowdflower, crows\_pairs, crows\_pairs\_multilingual, cruxeval, csatqa, csgo-object-detection, csqa\_korean\_val, cti-bench, cuad-qa, cuad-qa, cuad-qa, cultural\_awareness\_mcq, curiosity\_dialogs, cvqa, cvss, cyber-sharters, cyberQA, cyber\_MITRE\_CTI\_dataset, cyber\_MITRE\_attack\_tactics-and-techniques, cyberattack, cyberattack2, cyberllamadataset, cybersec, cybersec, cybersec\_mitre\_attack\_tactics\_techniques\_instruction\_data, cybersecurity-rules, cybersharter-v3, cybersharterv2, cycic\_classification, cycic\_multiplechoice, da-hashtag-twitterhjerne, dadc-limit, daily-papers, daily-papers-stats, daily\_dialog, daily\_dialog, daily\_dialog, dailydialog, dalle-3-dataset, dapr, daring-anteater-specialized, dart, darumeru, data, data-diversity-nli-id, data-mistral-7b-instruct-sppo-iter1, data-mistral-7b-instruct-sppo-iter2, data-mistral-7b-instruct-sppo-iter3, data\_analysis, data\_jobs, databricks-dolly-15k, databricks-dolly-15k-curated-en, databricks-dolly-15k-ja, databricks-dolly-15k-ja-gozarinnemon, databricks-dolly-15k-ja-gozaru, databricks\_dolly\_15k, datacomp\_1b, dataset, dataset, dataset-tldr-preference-dpo, datatrove-tests, datetime-dataset, dbpedia, dbpedia-entities-openai-1M, dbpedia-entity, dbpedia\_14, dbpedia\_14, dclm-baseline-1.0, dclm-baseline-1.0-parquet, dclm-micro, de\_ifeval, deal\_or\_no\_dialog, decision\_transformer\_gym\_replay, decomp, deepfashion-inshop, deepfashion-multimodal, defeasible-nli, defeasible-nli, definite\_pronoun\_resolution, deita-10k-v0-sft, demmlu\_no\_train, demo1, demo\_data, descriptiveness-sentiment-trl-style, descriptors-text-davinci-003, details\_silma-ai\_\_SILMA-9B-Instruct-v1.0, dialog-inpainting, dialog-studio-sup, dialog\_re, dialogstudio, dialogsum, dialogsum-test, diamonds, diffusiondb, diffusiondb-pixelart, diplomacy\_detection, dippy\_synthetic\_dataset, discofuse, discosense, discovery, discrim-eval, disrpt, distilabel-capybara-dpo-7k-binarized, distilabel-intel-orca-dpo-pairs, distilabel-intel-orca-dpo-pairs-binarized, distilabel-math-preference-dpo-de, diversevul, dkhate, dl-doc-search, dl-mappings, dm-mathematics-sup, do-not-answer, doc-vqa, docci, doclaynet\_bench, docprompting-conala, docred, docvqa-single-page-questions, docvqa\_1200\_examples, docvqa\_1200\_examples\_donut, docvqa\_test\_subsampled, dog-food, dolly-15k-instruction-alpaca-format, dolly-15k-oai-style, dolly-chatml-sft, dolly\_hhrlhf, dolma, doqa, dpo-mix-7k, dreaddit, dream, drop, drop, ds\_robot\_33\_large, dsa\_hk231\_records, dsdm-candidate-c4, dsir-pile-100k, dsir-pile-1m, dsprite-counterfactual, dt-mappings, dtd, dummy\_image\_class\_data, dummy\_image\_text\_data, duorc, dynamic\_sonnet\_llama3, dynasent, e-care, e2e\_nlg, e2e\_nlg, e2e\_nlg, e3c-sentences, eQASC, eQASC-perturbed, and eOBQA, easynum, ecthr\_cases, edacc, edgar-corpus, egoschema, ekar\_english, eli5\_category, elix\_latent\_cleaned, elix\_latent\_preferences\_gpt4, emb, embedded\_movies, embeddings-dataset-paraphrase, emea, emo, emodb, emotion, emotion, emotion, emotion, emotion-balanced, emotion\_max\_500, empathetic\_dialogues\_llm, en\_MIRACL, en\_MrTidy, enem, eng\_guj\_parallel\_corpus, english\_dialects, english\_quotes, english\_quotes\_copy, enron\_aeslc\_emails, enron\_qa\_0805, enron\_qa\_0822, enron\_spam, eqanun, equate, equity\_evaluation\_corpus, es\_ifeval, esb-datasets-test-only-sorted, esc50, esmmlu\_no\_train, esnli, esnli, essays-with-instructions, essential-terms, ethics, ethics, ethos, ethos, etth1, eurlex, eurlex, eurlex-multilingual, europa\_ecdc\_tm, europarl, europarl-mono, europarl\_en-it, eurosat, eurosat-demo, eurosat-rgb, eval\_gerv, event2Mind, events\_classification\_biotech, everyday-conversations-llama3.1-2k, evo-circuit-breaker, evol-codealpaca-v1, evol-instruct, example-documents, exams, exebench, execution-v2, expresso, eye-disease-dataset, fact-qa-iclr, fair-rationales, fairface, fake\_news, falcon-refinedweb, faroese-sts, fashion-product-images-small, fashion-products-small, fashion-style-instruct, fashion200k, fashion\_mnist, fashioniq\_val, fashioniq\_val\_imgs, fashionpedia, fast-flash-hackernews-posts, fast\_food\_image\_classification, feb, feedbackQA, femnist, fer-2013, fever, fever, fever, fever-evidence-related, few-nerd, few-nerd, ffhq-256, ffhq-256, fig-qa, fill10, fill50k, filtered-wit, finance-alpaca, finance-tasks, finance-unsup, financebench-test, financial-classification, financial-qa-10K, financial-reports-sec, financial\_phrasebank, financial\_phrasebank, financial\_phrasebank\_split, finanical-rag-embedding-dataset, fineweb, fineweb-100k\_en-med, fineweb-100m-sample, fineweb-1B, fineweb-edu, fineweb-edu-fortified, fineweb-edu-score-2, fineweb-sample, fingpt-convfinqa, fingpt-finred, fingpt-fiqa\_qa, fingpt-headline-cls, fingpt-headline-cls, fingpt-ner, fingpt-ner-cls, fingpt-ner-cls, fingpt-sentiment-cls, fingpt-sentiment-train, fingpt-sentiment-train, finqa, finqa, finred, fiqa, fiqa, fiqa, fiqa-sentiment-classification, firefly-train-1.1M, fixtures\_ade20k, fixtures\_docvqa, fixtures\_image\_utils, flan, flan-10k-flat, flan-v2, flan\_labeled, flare-cfa, flare-finqa, fleurs, fleurs, flickr-megalith-10m-internvl2-multi-caption, flickr30k, flickr30k, flickr30k\_test, flickr8k, flickr8k-dataset, flickr\_1k\_test\_image\_text\_retrieval, flores, flores, flores200, flores\_101, flores\_plus, flowers-102-categories, fma, folio, folio, food101, food\_img\_caption\_small, football-dataset, forai\_ml\_masakhane\_mafand, fqa, fquad2\_test, fr\_ifeval, freebase\_qa, freesolv, french-bench-grammar-vocab-reading, french\_bench\_arc\_challenge, french\_bench\_hellaswag, frmmlu\_no\_train, fsdkaggle2019, ft-instruction-synthesizer-collection, full-hh-rlhf, funsd-layoutlmv3, fusion-image-to-latex-datasets, galcola, gandalf\_ignore\_instructions, gap, gc\_fine\_ultrafeedback\_nosys, gcbinarized\_fine\_ultrafeedback, gcbinarized\_ultrafeedback\_nosys, gcc\_caption\_only, gemini\_result\_kospi\_0517\_jsonl, gemma2-ultrafeedback-armorm, gemma2-ultrafeedback-armorm-add-distance, gemma2-ultrafeedback-armorm\_multi\_pairs, gen-debiased-nli\#training-with-our-datasets, general\_data\_v1, generated\_reviews\_enth, generics\_kb, geneturing, genies\_preferences, genomics-long-range-benchmark, genshin-voice, geo\_dataset, georeview-classification, german-traffic-sign-detection, germanquad, giga\_fren, gigaspeech, github-code, github-code-clean, github-code-duplicate, github-issues, github-jupyter-text-code-pairs, glaive-code-assistant-v3, glaive-function-calling-v2, glaive-function-calling-v2-sharegpt, glaive-function-calling-v2-sharegpt, glaive\_toolcall\_en, glaive\_toolcall\_zh, glosslm-corpus-split, glucose, glue, glue, glue-mnli-train, glue-ptpt, gnad10, go\_emotions, go\_emotions, gooaq, gooaq, goodwiki, google-argentinian-spanish, google-chilean-spanish, google-colombian-spanish, google-gujarati, google-marathi, google-tamil, google\_\_recurrentgemma-9b-it-details, google\_wellformed\_query, gorilla, govreport-summarization, gpqa, gpt4\_evol\_1.3k, gpt4\_judge\_battles, gpt4\_long\_context\_train\_data\_100k, gpt4all-j-prompt-generations, gpt4v-emotion-dataset, gpt\_alpaca, gpt\_roleplay\_realm, gpt\_teacher, gqa, great, gsm-hard, gsm8k, gsm8k, gsm8k, gsm8k-ja-test\_250-1319, gsm8k-ko, gsm8k\_tr-v0.2, gtsrb, gtzan, gtzan, guanaco-llama2, guanaco-llama2-1k, guanaco-llama3-1k, guanaco-sharegpt-style, guidelines, gutenberg-dpo-v0.1, hackaprompt-dataset, hackercup, hagrid, hands-images, hans, harmful\_behaviors, harmful\_harmless\_instructions, harmless\_alpaca, has\_part, hate-speech-dataset, hate\_speech\_offensive, hate\_speech\_offensive, hatexplain, hatexplain, head\_qa, head\_qa, headline-classification, health\_care\_german, health\_fact, hellaswag, hellaswag, hellaswag-bgeval, hellaswag\_tr-v0.2, helsteer\_correctness\_combined\_format, hendrycks\_ethics, hendrycks\_math, hendrycks\_math, herman-json-mode, hermes-function-calling-v1, hest, hf-stack-v1, hh-rlhf, hh-rlhf, hh-rlhf-12k-ja, hh-rlhf-helpful-base-trl-style, hh-rlhf-safety-v3, hh-rlhf-safety-v3-dpo, hh-rlhf-trl-style, hifi-tts, higgs, high\_quality\_public\_evaluations, hindsight-neglect-10shot, history-mcq, hkcancor, hlgd, hope\_edi, hotel\_datasets, hotpot\_qa, hotpot\_qa, hotpotqa, hotpotqa, hotpotqa-qrels, hpo\_finetune\_data\_4way\_mc\_train\_max\_10k\_per\_task, hpprc\_emb\_reranker\_score, hug\_stack, huggingface\_doc, huggingface\_doc\_qa\_eval, human\_assistant\_conversation\_deduped, human\_parsing\_dataset, humaneval-x, humanevalpack, humanevalplus, humicroedit, hupd, hybrid\_qa, hyperpartisan\_news\_detection, i2p, ichikara-instruction, id-csqa, idl-wds, iitb-english-hindi, image-text-demo, imagefolder\_with\_metadata, imageinwords, imagenet-100, imagenet-1k, imagenet-1k-standardized, imagenet-1k-vl-enriched, imagenet-1k\_tiny, imagenet-hard, imagenet-r, imagenet2012-1k-subsampling-50, imagenet\_1k\_resized\_256, imagenette, imagenette, imagenette2-320, imagetrain, imagewoof, imbd-2024-training, imdb, imdb, imdb, imdb, imdb-dpo, imdb-movie-reviews, imdb-truncated, imdb\_pt, imdb\_small, img\_sketch\_100k, imgsys-results, implicatures, implicit-cot-math, implicit-hate, imppres, imppres, imppres, inappropriateness-classification, indian\_legal\_corpus, indicxnli, indonlu, infovqa\_test\_subsampled, inquisitive\_qg, instruct-data-basics-smollm-H4, instruct-v3, instructhumaneval, instruction-dataset, instruction-dataset, instruction-dataset-mini, instruction-following-eval, instruction\_following, instructpix2pix-10-samples, instructpix2pix-1000-samples, instructpix2pix-clip-filtered, intent\_classification, into-the-unknown, invoices-and-receipts\_ocr\_v1, invoices-donut-data-v1, ioi\_dataset, iris, iris, irmas, iterative-prompt-v1-iter1-20K, iterative-prompt-v1-iter2-20K, iterative-prompt-v1-iter3-20K, iterative-prompt-v1-iter4-20K, iterative-prompt-v1-iter5-20K, iterative-prompt-v1-iter6-20K, iterative-prompt-v1-iter7-20K, iterative-prompt-v1-iter8-20K, iterative-prompt-v1-iter9-20K, iwslt2017, jailbreak-classification, japan-law, jat-dataset, javascript, jenny-tts-6h, jfleg, jhumaneval, jnlpba, job-descriptions, job-descriptions-public, joke\_training, json-mode-eval, just-eval-instruct, kalo-opus-instruct-22k-no-refusal, kde4, kenetoszto\_cumulative\_distribution, kesalahan-tatabahasa-instructions, keywords, kilt-128, kilt\_tasks, kinopoisk-sentiment-classification, kk, klej-cdsc-e, klej-dyk, klej-nkjp-ner, klej-polemo2-in, klej-polemo2-out, klej-psc, klue, kmZQBkk558WWAGV2, kmhas\_korean\_hate\_speech, kmmlu\_subset100, know\_sql, knowledge\_application, ko\_text2sql, ko\_truthful\_qa, kobest\_v1, kohatespeech, kor\_unsmile, korean\_safe\_conversation, korean\_textbooks, korquad, kto-mix-14k, kullm-v2, lab-bench, laion-art, laion-coco, laion-coco-aesthetic, laion-coco-nllb, laion\_improved\_aesthetics\_6.5plus\_with\_images, lambada, lambada, lambada\_multilingual, lambada\_openai, lamini\_docs, lamini\_docs\_evaluation, language, language-identification, lasr-cv-eval, latex-formulas, law-stack-exchange, law-tasks, lbox\_open, lbpp, lca-project-level-code-completion, leafy\_spurge, leetcode, legal\_lama, legal\_summarization, legalbench, legalbench-entire, legalbench\_consumer\_contracts\_qa, legalbench\_corporate\_lobbying, legalbench\_instruct, lemexp, lex-glue, lex\_glue, lfw, liar, librispeech10h, librispeech\_asr, librispeech\_asr, librispeech\_asr, librispeech\_asr, librispeech\_asr\_demo, librispeech\_asr\_dummy, librispeech\_asr\_dummy, librispeech\_asr\_dummy, librispeech\_asr\_for\_optimum\_habana\_ci, librispeech\_long, libritts, libritts-r-filtered-speaker-descriptions, libritts\_r, libritts\_r\_filtered, libritts\_r\_tags\_tagged\_10k\_generated, librivox-full-catalog-archive, license-plate-object-detection, lighteval-ceval-exam, lighteval-cmmlu, lighton-ms-marco-mini, lila, lima, lima, limit, lin\_reg, ling\_in\_loop, linnaeus, livedoor-news-corpus, livedoor-news-corpus, liveqa, lj\_speech, lj\_speech, llama-2-arxiv-papers-chunked, llama-2-finance, llama3-jailbreaks, llama3-ultrafeedback, llama3-ultrafeedback-armo-1024-test\_harvard, llama3-ultrafeedback-armo-1024\_harvard, llama3-ultrafeedback-armorm, llama3-ultrafeedback-armorm, llama3\_ultrafeedback\_with\_tie\_armorm, llama3\_value\_dataset, llava-bench-coco, llava-bench-in-the-wild, llava-en-zh-2k, llava-instruct-mix-vsft, llm-jp-eval, llm\_global\_opinions, llm\_merging, llmrepair, lmsys-arena-human-preference-55k-thresholds, lmsys-chat-1m, lmsys-chat-1m, lnqa, logical-fallacy, logikon-bench, logiqa, logiqa2, long-doc\_book\_en, long\_context\_eval, longform\_article\_summarization, lotsa\_data, lotte\_passages, lsat\_qa, lsun-bedrooms, lsun\_church\_train, m2d2, m3exam, mCSQA, mOSCAR, m\_arc, m\_hellaswag, m\_mmlu, m\_truthfulqa, magpie, magpie-ultra-v0.1, malicious-smart-contract-dataset, marvl, masakhaner, masakhaner2, masakhanews, masakhanews, masakhapos, massive, math, math, math\_alltime, math\_dataset, math\_dataset, math\_qa, math\_qa, math\_qa, mathqa-bgeval, max-of-10-proofs, max-of-4-proofs, maxm, mbib-base, mbpp, mbppplus, mc\_taco, mc\_taco, mcscript, mctest, md\_gender\_bias, meanwhile, measuring-hate-speech, med\_qa, medical-dialogue-to-soap-summary, medical-qa-datasets, medical-qa-shared-task-v1-toy, medical-question-answering-datasets, medical-question-pair-dataset, medical\_knowledge\_from\_extracts, medical\_meadow\_medical\_flashcards, medical\_meadow\_medqa, medicationqa, medicine-tasks, medmcqa, medmcqa\_jp, medmentions, medmnist-v2, medqa, medqa, medqa-MedGENIE, medrxiv-clustering-p2p, medrxiv-clustering-s2s, meetingbank, megalith-10m-florence2, memo-trap, mental\_health\_chatbot\_dataset, mental\_health\_counseling\_conversations, menyo20k\_mt, meta-imagine-dataset, meta-llama\_\_Meta-Llama-3.1-70B-Instruct-details, meta-llama\_\_Meta-Llama-3.1-8B-Instruct-details, meta\_woz, mfaq, mgb1, mgsm, mgsm\_gl, miam, microsoft\_\_Phi-3-medium-4k-instruct-details, microsoft\_\_Phi-3-mini-4k-instruct-details, midjourney-prompts-only, mimir, mind\_small, mindgames, minds14, mini-fineweb, mini\_imagenet, mini\_pile\_cc, minif2f-lean4, minipile, minipile, mintakaqa, miracl, miracl-corpus, miracl-en-queries-22-12, miracl\_bm25\_negative, mirage-eval-rag-output, mistral-instruct-ultrafeedback, mistral-sft-iter1-eval, mix-instruct, mix\_copy, mkb, mkqa, ml-qrecc, mlabonne-chatml-dpo-pairs-copy, mlb\_data, mlqa, mlqa, mls10k\_nemo, mls\_eng, mls\_eng\_10k, mls\_eng\_10k, mlsum, mmarco, mmlu, mmlu, mmlu, mmlu, mmlu, mmlu-computer\_security-neg, mmlu-computer\_security-original-neg, mmlu-computer\_security-original-neg-prepend, mmlu-computer\_security-rule-neg, mmlu-computer\_security-rule-neg-prepend, mmlu-cs, mmlu-redux, mmlu-tr, mmlu\_custom, mmlu\_italian, mmlu\_no\_train, mmlu\_ru, mmlu\_tr-v0.2, mmteb-miracl, mmteb-miracl-reranking, mnist, mnist-text-small, mnli, mnli\_dataset\_genetic, mocha, model-repos-stats, model-written-evals, modus-tollens, monkey\_business, monology-pile-uncopyrighted-tokenizer-gpt2, monotonicity-entailment, moral\_stories, movie\_rationales, movies, mozilla\_foundation\_common\_voice\_corpus\_18\_0, mptrj, mr-tydi, mrpc, mrqa, mrqa, ms-marco-en-bge, ms\_marco, ms\_marco, ms\_ninespecies\_benchmark, mscoco-small, mscoco\_2014\_5k\_test\_image\_text\_retrieval, msmarco, msmarco, msmarco-bm25, msmarco-corpus, msmarco-passage, msmarco-passage, msmarco-passage-aug, msmarco-passage-corpus, msmarco-qrels, msmarco-v2, msr\_sqa, msr\_text\_compression, mt-bench, mt\_bench\_en, mt\_bench\_human\_judgments, mt\_bench\_prompts, mt\_gender, mt\_geneval, mteb-fr-reranking-alloprof-s2p, mteb-fr-reranking-syntec-s2p, mteb-fr-retrieval-syntec-s2p, mtop\_domain, mtop\_intent, multi-humaneval, multi-session\_chat, multi-wiki-clustering-p2p, multi30k, multi\_dir\_dataset, multi\_eurlex, multi\_lexsum, multi\_news, multi\_news, multi\_nli, multi\_nli, multi\_task\_multi\_modal\_knowledge\_retrieval\_benchmark\_M2KR, multi\_woz\_v22, multi\_woz\_v22, multi\_x\_science\_sum, multiclass-sentiment-analysis-dataset, multilingual, multilingual-llava-bench, multilingual-sentiments, multilingual-wikihow-qa-16k, multilingual\_advbench, multilingual\_advbench\_llama31\_generated, multilingual\_cc\_news, multilingual\_librispeech, multimodal-m3exam, multinerd, multirc, multiturn-Calm3-manual, multiun, mushroom, must-c-en-es-02, mutual, mutual, mwp\_basic, mwsc, my-NFT-summer-balanced1, naab, naamapadam, naijavoices-dataset, namuwiki, nan-nli, narrativeqa, narrativeqa, naruto-blip-captions, natural-instructions, natural-language-satisfiability, natural-questions, natural\_questions, natural\_questions, naver-news-summarization-ko, ncbi\_disease, ncbi\_disease, ncbi\_disease, needle-in-a-haystack-biographies-v0, negative-Meta-Llama-3.1-8B-Instruct-OSS-Tool-Instruct, ner-wikipedia-dataset, netflix-shows, neuclir-2022, neuclir-2023, neural-conv-qa, neural\_constructions, neuronovo-utc-data-glue-cola, neuronovo-utc-data-glue-mnli, neuronovo-utc-data-goemotions, neuronovo-utc-hate-speech18-sentences, neuronovo-utc-measuring-hate-speech, neuronovo-utc-persent-doc, neuronovo-utc-tweeteval-emotions, neuronovo-utc-tweeteval-sentiment, neuronovo-utc-unhealthy-conversations, new-title-chinese, new\_base\_rs\_mix, new\_league\_data, new\_league\_data\_bestest, new\_league\_data\_max\_plus, news-data, news21-instructions, news\_commentary, news\_commentary, newyorker\_caption\_contest, nfcorpus, nfcorpus, nfcorpus-qrels, nimrod-uk-1km, nlg-bias, nli-debiasing-datasets, nli-veridicality-transitivity, nllb, nlu-asdiv-dataset, no\_robots, noisyner, nomic-bert-2048-pretraining-data, nomiracl, norec\_agg, norec\_sentence, norne, nouns, npm-20240828, nq, nq, nq\_bm25\_top100\_subset, nq\_bm25\_top100\_subset\_oracle, nq\_corpus\_dpr, nq\_open, nq\_open, nrrqa-string, nsfw\_detect, nsmc, nucleotide\_transformer\_downstream\_tasks, nuggets-kmeans-100, num\_fh, number-pairs, numer\_sense, numerai-datasets, numericnlg, nusaparagraph\_topic, nusatranslation\_emot, nusax\_senti, nyu\_depth\_v2, oak, oasst1, oasst1\_pairwise\_rlhf\_reward, oasst2, oasst2\_dpo, oasst\_top1\_2023-08-25, obqa, octopack, odd-man-out, odex, offenseval\_dravidian, offsetbias, ogbg-molhiv, ogiri-debug, ogiri-test, ohsumed, okapi\_arc\_challenge, okapi\_hellaswag, okapi\_mmlu, ollie, olm-wikipedia-20221220, omega-multimodal, omega-multimodal-ids, onestop\_qa, open-images-v7, open-instruct-v1, open-web-math, openai-moderation-api-evaluation, openai\_humaneval, openai\_summarize\_comparisons, openai\_summarize\_tldr, openassistant-guanaco, openassistant-guanaco-reformatted, openbookqa, openbookqa, openbookqa\_gl, opendevin\_DataDevinator, openhermes, openhermes-2.5-llama3, openpi-dataset, openslr, openwebtext, openwebtext-100k, openwebtext-10k, openwebtext-gemma-1024, openwebtext-tokenized-small, opus-100, opus-100, opus-100, opus\_books, opus\_paracrawl, opus\_paracrawl, opus\_samantha, opus\_tedtalks, opus\_xhosanavy, opusparcus, orca-math-word-problems-200k, orca-math-word-problems-200k, orca-mini, orca\_dpo\_pairs, orca\_dpo\_pairs, orpo-dpo-mix-40k, orthogonal-activation-steering-TOXIC, oscar, owt-processed\_256, owt-processed\_512, owt-processed\_8, oxford-flowers, oxford-iiit-pet, p3-supernatural-sup, pair\_data\_v2\_80K\_wsafety, pair\_data\_v2\_80K\_wsafety\_short, paloma, pangbo, para\_pat, paradetox, parafrases\_gl, parallel-sentences-ccmatrix, parallel-sentences-europarl, parallel-sentences-global-voices, parallel-sentences-jw300, parallel-sentences-news-commentary, parallel-sentences-opensubtitles, parallel-sentences-opus-100, parallel-sentences-talks, parallel-sentences-wikimatrix, parsinlu-multiple-choice, parsinlu\_entailment, parsinlu\_reading\_comprehension, parsinlu\_sentiment, parsinlu\_translation\_en\_fa, parti-prompts, pascal-context, pascal-voc-2012, patent-classification, patfig, path-vqa, pattern-matching-suppression, paws, paws, paws, paws-x, paws-x, pcr\_single\_antecedent, pearl\_benchmark, pec, peer\_read, pendakwah\_teknologi\_yt\_stt\_dataset, peoples\_speech, persian\_news\_dataset, persian\_qa, persona, persona-chat, persona\_gpt4\_paired\_fullscale, personachat\_truecased, personahub\_augmented\_v0, perspectrum, perturbed-wsc, pg-wikiSQL-sql-instructions-80k, pg19, pg19, pg19-4k, pg19-test, pg\_books-tokenized-bos-eos-chunked-65536, phishing-dataset, phrase\_sense\_disambiguation, phrase\_similarity, pianos, pib, pica\_ar, pickapic\_v1, pickapic\_v1\_no\_images, pickapic\_v2, pico-annotation, pii-masking-200k, pii-masking-200k, pii-masking-300k, pile, pile-10k, pile-deduped-pythia-random-sampled, pile-of-law, pile-small-tokenized-2b, pile-uncopyrighted, pile-val-backup, pile\_books3, piqa, piqa, piqa, pira, pixelprose, pku-seg, plant-genomic-benchmark, plant\_species, plasticc-gp, pn\_summary, poem\_sentiment, poem\_sentiment, poetry, pointer-problems-v3, points-of-correspondence, pokemon-blip-captions, pokemon-blip-captions, pokemon-blip-captions-en-zh, pokemon-gpt4-captions, poleval2019\_mt, political, polqa, polynomial, polyvore, popular-deepfakes, poquad, poquad, ppc, pragmeval, prefdedup, preference-test-sets, preference\_700K, preference\_data\_v2\_80K\_wsafety, preference\_dataset\_mixture2\_and\_safe\_pku, preference\_dataset\_mixture2\_and\_safe\_pku, pretokenized\_data\_yeast, pretrain, prh, prism-alignment, privy, prm800k\_passk\_qs1000\_discount0.9, probability words, probability\_words\_nli, programming\_books\_rus, prompt-collection-v0.1, prompt-injections, prompt\_injection\_password, proof-pile, proof-pile-2, proofnet, propainter-object-removal, propara, prosocial-dialog, prost, prost, proto\_qa, publaynet\_bench, pubmed, pubmed-summarization, pubmed\_qa, pure\_pixel\_yt\_stt\_dataset, pusht, pusht\_keypoints, puzzte, pypi-20230724, pypi-20230724, python-bugs, python-code-dataset-500k, python-codes-25k, python-github-code, python-github-code, python-lines, python-state-changes, python\_code\_instructions\_18k\_alpaca, qa\_srl, qasc, qasc, qasper, qasper, qed, qed\_amara, qm9, qnli, qqp, quac, quail, quality, quarel, quartz, query2doc\_msmarco, quora, quora, quora-duplicates, quora-qrels, quora-question-answer-dataset, quoref, quote-repetition, race, race, race-c, raft, rag-dataset-12000, rag-mini-bioasq, rag-mini-wikipedia, rag\_instruct\_benchmark\_tester, ragbench, rai\_gender\_bias\_splitted\_v\_4\_2, rai\_hate\_bias\_splitted\_v\_4\_2, rai\_political\_bias\_splitted\_v\_4\_2, ramdom-to-fixed-multiturn-Calm3, random\_dataset\_1, random\_graph, random\_label, rank\_zephyr\_training\_data, rankzephyr\_longcontext\_merged\_80k, ravdess, raw\_instruction\_en\_ko\_translation, react-code-instructions, real-persona-chat, real-toxicity-prompts, realvis-xl, realworldqa, reasoning, reazonspeech, rebel-dataset, recast, recipe\_nlg, recipe\_nlg\_lite, reclor, red-dot-design-award-product-description, reddit-clustering, reddit-clustering-p2p, reddit-comments-uwaterloo, reddit\_tifu, redefine-math, refcoco\_det, refcocog, refcocop, rejection\_sampling\_18769, rejection\_sampling\_22689, rejection\_sampling\_30566, rejection\_sampling\_4458, relbert, repetitive-algebra, replay, replit-comments-categorized, repobench-r, requested-arxiv-ids-3, resisc45, resisc45, results\_public, results\_v2, resume-job-description-fit, resume\_seven\_class, retro-ascii-art-v1, retweet, reuters21578, review\_preference, reward-bench, reward\_anthropic, rheumatology-biologics-dataset, ria-news-retrieval, riddle\_sense, rlaif-v\_formatted, rlhf-reward-datasets, rm-hh-rlhf, rm-static, ro\_fake\_news, ro\_mmlu, ro\_sts\_parallel, roberta\_pretrain, robust-finetuning, robust04-instructions, robustLR, robust\_nli\_is\_sd, roneneldan-TinyStories-tokenizer-gpt2, ropes, rosetta-code, rotten\_tomatoes, rotten\_tomatoes, rpj-v2-sample, rs-base-mix-L3.1-8B-generations, rtGender, rte, rte, rte\_dataset\_genetic, ru-paraphrase-NMT-Leipzig, ru-reviews-classification, ru-scibench-grnti-classification, ru-scibench-oecd-classification, ru-wikipedia, ru\_sci\_bench\_mteb, rublimp, rubq-retrieval, rubygems-20230301, rucola, rule-security-risks, ruletaker, ruletaker, ruletaker, ruozhiba-llama3, russian\_super\_glue, rvl\_cdip, rvl\_cdip\_10\_examples\_per\_class\_donut, rvl\_cdip\_small, sae-monology-pile-uncopyrighted-tokenizer-gpt2, safe-guard-prompt-injection, saferpaca4\_qa, safety-prompts, safim, salt, salt-corrected, salt-corrected-asr-data-transcriptions, samromur\_children, samsum, samsum, samsum, sangraha, scanned-arxiv-papers, scanned-arxiv-papers-id, schema\_guided\_dstc8, sciarg, scicite, scidocs, scidocs, scidocs-qrels, scidocs-reranking, scientific-calculation-test, scientific-exaggeration-detection, scientific\_lay\_summarisation, scientific\_papers, scifact, scifact, scifact, scifact, scifact-qrels, sciq, sciq, sciq\_italian, scirepeval, scirepeval\_test, scitail, scitail, scitail, scitldr, scqa\_german\_combined, scqa\_german\_new, scrolls, scruples, scruples, sec-data-mini, sec-football-qa, seeds, seggpt-example-data, self-instruct, self-oss-instruct-sc2-H4, self-oss-instruct-sc2-exec-filter-50k, self\_instruct, sem eval 2010 task 8, sem\_augmented\_fever\_nli, sem\_eval\_2010\_task\_8, semantic-segmentation-test-sample, semeval-2010-pre, semeval2017, senryu-debug, senryu-marusen, senryu-shashin, senryu-test, sent-summary, sent\_comp, sentiment-mental-health, sentiment-trl-style, sentiment140, sentiment140, seven-wonders, sft-v1-27k, sft\_depressedLLM, sft\_prm800k\_processed, shadow-alignment, shapes3d-dist, sharegpt-cleaned, sharegpt-english, sharegpt\_gpt4, sharegpt\_v3\_unfiltered\_cleaned\_split, shiftproject\_test, shiji-70liezhuan, sib200, sib200, sick, sickr-sts, sidewalk-semantic, sig-figs, silicone, simple-cooccurrence-bias, simple\_k\_wc, simple\_wikipedia, single\_fact\_data\_v1, siqa, siraj\_gwas\_highpip, skin\_cancer, slim\_pajama\_chunk1, slimorca-deduped-cleaned-corrected, slimpajama-per-source-length-upsample, slimpajama\_Qwen2\_tokenized\_upsample\_4096\_chunk\_256K, slimpajama\_llama\_tokenized\_upsample\_4096\_chunk\_256K, slither-audited-smart-contracts, slovenian-llm-eval-v0, slue-phase-2, slurp, sm, small-the\_pile, small\_wiki\_dataset, smithsonian\_butterflies\_subset, smollm-corpus, sms-spam-collection, sms\_spam, snli, snli, snli, social\_bias\_frames, social\_bias\_frames, social\_i\_qa, social\_i\_qa, social\_i\_qa, soda, soda-audio, soda\_synthetic\_dialogue, sova\_rudevices, spamassassin, sparp, spartqa-mchoice, spartqa-yn, spcorrect, speech\_commands, speech\_robust\_bench, speechocean762, spellcheck\_benchmark, spellcheck\_punctuation\_benchmark, spgispeech, spider, spider-context-validation, spider2-lite, spolin, spotify-tracks-dataset, sprintduplicatequestions-pairclassification, spurious\_nli\_correlation\_0.95, spurious\_nli\_distribution\_shift, sql-create-context, sql-create-context-copy, squad, squad, squad\_20\_ptbr, squad\_kor\_v1, squad\_modified, squad\_v1\_pt\_br, squad\_v2, squad\_v2, squad\_v2, squad\_with\_test, sst2, sst2, sst2, sst2, sst2\_dataset\_genetic, sst2\_pt, sst5, st-vqa, stable-diffusion-prompts, stable-diffusion-prompts-2.47M, stable-diffusion-xl, stack-3b-sample, stack-exchange-instruction, stack-exchange-paired, stackexchange-clustering, stackexchange-clustering-p2p, stackexchange\_titlebody\_best\_voted\_answer\_jsonl, stackoverflow, stackoverflow-qa-qrels, stackoverflow-qa-queries-corpus, stackoverflowdupquestions-reranking, stanford-dogs, stanford\_alpaca, stanford\_cars, starcoderdata, state-of-the-union-addresses, static-analysis-eval, stereoset, stheno-filtered-v1.1, stk-technical-indicators-15min, stk-technical-indicators-1hour, stk-technical-indicators-30min, stk-technical-indicators-4hour, stk-technical-indicators-5min, stock11, stories, story\_cloze, storycommonsense, strain\_selection, strategic\_game\_chess, strategyqa, sts12-sts, sts13-sts, sts14-sts, sts15-sts, sts16-sts, sts17-crosslingual-sts, sts22-crosslingual-sts, stsb, stsb\_multi\_mt, stsbenchmark-sts, student-alcohol-consumption, student\_performance, style, su-csqa, subj, subjectivity, subjqa, subset\_arxiv\_papers\_with\_embeddings, summarization\_eval, summarization\_gl, summarize-from-feedback, summarize\_from\_feedback, summarize\_from\_feedback, summarize\_from\_feedback\_oai\_preprocessing\_1706381144, summarize\_from\_feedback\_oai\_preprocessing\_pythia-6.9b-gold, summarize\_from\_feedback\_tldr\_3\_filtered\_oai\_preprocessing\_1704563162, summarize\_from\_feedback\_tldr\_3\_filtered\_oai\_preprocessing\_1706381144, summedits, summeval, sun397, sun397, super\_glue, super\_glue, superb\_demo, superb\_dummy, svamp, svhn\_cropped, svhn\_cropped\_balanced, swag, swiss\_citation\_extraction, swiss\_criticality\_prediction, syntactic-augmentation-nli, syntaxgym, synth-vuln-fixes, synth\_pass\_open, synthdog-en, synthdog-ja, synthdog-ko, synthdog-zh, synthetic-instruct-gptj-pairwise, synthetic-mapping-simple, synthetic-text2sql-qrels, synthetic-text2sql-queries-corpus, syntheticDocQA\_artificial\_intelligence\_test, syntheticDocQA\_energy\_test, syntheticDocQA\_government\_reports\_test, syntheticDocQA\_healthcare\_industry\_test, synthetic\_pii\_finance\_multilingual, synthetic\_text\_to\_sql, system\_prompt, tab\_fact, tabfquad\_test\_subsampled, tabular-benchmark, taiwan\_company\_revenue, taobao, task-mixture-no-pythia, task-specific-datasets, task039\_qasc\_find\_overlapping\_words, task084\_babi\_t1\_single\_supporting\_fact\_identify\_relevant\_fact, task1198\_atomic\_classification\_owant, task1391\_winogrande\_easy\_answer\_generation, task140\_detoxifying-lms\_classification\_style, task1448\_disease\_entity\_extraction\_ncbi\_dataset, task1605\_ethos\_text\_classification, task1711\_poki\_text\_generation, task247\_dream\_answer\_generation, task275\_enhanced\_wsc\_paraphrase\_generation, task280\_stereoset\_classification\_stereotype\_type, task290\_tellmewhy\_question\_answerability, task391\_causal\_relationship, task620\_ohsumed\_medical\_subject\_headings\_answer\_generation, task636\_extract\_and\_sort\_unique\_alphabets\_in\_a\_list, task705\_mmmlu\_answer\_generation\_high\_school\_macroeconomics, task717\_mmmlu\_answer\_generation\_logical\_fallacies, task742\_lhoestq\_answer\_generation\_frequency, tasksource, tasksource\_dpo\_pairs, tatdqa\_test, tatoeba, tatoeba-bitext-mining, tatoeba\_mt, tau\_srir\_db, taxi, taylor\_swift, td\_pd16, teca, tecla, ted\_talks, ted\_talks\_iwslt, tedlium, telecom-conversation-corpus, tellmewhy, temp, template-generation, temporal-nli, tenkgnad-clustering-s2s, terra-pairclassification, test, test2, testCityDatasetJuly06\_split\_words\_all, testStreetDatasetJuly06\_split\_words\_all, test\_dataset, test\_dataset, test\_generation, test\_imdb\_embedd2, test\_import\_dataset\_from\_hub\_using\_settings\_with\_recordsFalse, test\_import\_dataset\_from\_hub\_using\_settings\_with\_recordsTrue, test\_import\_dataset\_from\_hub\_using\_wrong\_settings\-\_with\_records\_False, test\_import\_dataset\_from\_hub\_using\_wrong\_settings\_with\_records\_True, test\_import\_dataset\_from\_hub\_with\_records\_False, test\_import\_dataset\_from\_hub\_with\_records\_True, test\_librispeech\_parquet, testdataset, testing\_alpaca\_small, testing\_codealpaca\_small, testing\_self\_instruct\_small, text2image-multi-prompt, text8-chunked1024, textbooks, textvqa, thai\_exam, the-stack, the-stack-dedup, the-stack-github-issues, the-stack-smol, the-stack-v2, the-stack-v2-dedup, the-stack-v2-train-full-ids, the-stack-v2-train-smol-ids, the\_cauldron, the\_pile\_00\_arxiv, the\_pile\_arxiv\_50k\_sample, timit, tiny-codes, tiny-imagenet, tiny-imdb, tiny-lessons, tiny-shakespeare, tiny-supervised-dataset, tiny-textbooks, tinyAI2\_arc, tinyGSM8k, tinyHellaswag, tinyMMLU, tinyTruthfulQA, tinyWinogrande, tiny\_shakespeare, tinystories-1k, titanic, tldr, tldr-17, tldr-preference-sft-trl-style, tldr-preference-trl-style, tldr-preference-trl-style, tldr-with-sft-reference, tmlu, tmmluplus, tmmluplus, tokenized\_audio\_dataset\_new\_format, tokenized\_slim6B\_train\_neox\_4096, told-br, tom-qa-dataset, tomi-nli, top\_5\_insurance\_brands\_june\_news\_and\_twitter\_only, touche2020, toutiao-text-classfication-dataset, toxic-chat, toxic-dpo-v0.2, toxic\_conversations\_50k, toxigen-data, toy\_downstream\_tasks\_multilabel, trafilatura-extracted-subset, train\_0.5M\_CN, train\_2M\_CN, train\_3.5M\_CN, train\_dataset, trc\_uniform\_313k\_eval\_45\_filtered\_chat, trec, trec-covid, trec-covid, treino, trivia\_qa, trivia\_qa, trivia\_qa, trivia\_qa\_tiny, triviaqa-bgeval, trl-test-instruction, true-cds-protein-tasks, truthful-dpo, truthful\_qa, truthful\_qa, truthful\_qa-tr-v0.2, truthfulqa\_gl, truthfulqa\_italian, truthy-dpo-v0.1, tsla-historic-prices, tulu-2.5-preference-data, tulu-v2-sft-mixture, tulu-v2-sft-mixture-cot, tulu-v2-sft-mixture-flan, tulu-v2-sft-mixture-lima, tulu-v2-sft-mixture-science, tulu-v2-sft-seed-short-instruct-claude-distill-full-finetuning-v1, tulu-v2-sft-seed-short-instruct-claude-distill-syn-knowledge-finetuning-v1, turk, turkish-instruction-dataset-prepared, turkish-law-chatbot, turkish-sentiment-analysis-dataset, tw-legal-benchmark-v1, tweet-eval-emotion, tweet\_eval, tweet\_eval, tweet\_qa, tweet\_sentiment\_extraction, tweet\_sentiment\_multilingual, tweet\_topic\_multi, tweet\_topic\_single, twentynewsgroups-clustering, twitter-airline-sentiment, twitter-financial-news-sentiment, twitter-financial-news-topic, twitter-sentiment, twitter-sentiment-analysis, twitter\_disaster, twittersemeval2015-pairclassification, twitterurlcorpus-pairclassification, tydi\_xor\_rc, tydiqa, tydiqa, tydiqa-goldp, ucf101, uci-mushrooms, ucmerced, uhura-arc-easy, uhura-truthfulqa, ultra-orca-boros-en-ja-v1, ultrabin\_clean\_max\_chosen\_min\_rejected, ultrachat, ultrachat-10k-chatml, ultrachat\_200k, ultrachat\_2k, ultradistil-intel-orca-dpo-de, ultrafeedback-binarized-preferences, ultrafeedback-binarized-preferences-cleaned, ultrafeedback-binary-classification, ultrafeedback-multi-binarized-preferences-cleaned, ultrafeedback\_binarized, ultrafeedback\_binarized, ultrafeedback\_binarized\_1percent, ultrafeedback\_binarized\_cleaned, ultrafeedback\_binarized\_cleaned\_train, ultrafeedback\_clair\_32k, ultrafeedback\_preference, ultrafeedback\_prompts, ultrafeedback\_subset, umimeto-qa, un\_pc, unarXive\_imrad\_clf, unbiased\_training\_pairs, unified\_sqlv2, uniref50, universal\_dependencies, universal\_dependencies, universal\_ner, unnatural-instructions, unscramble, urbansound8K, us-colleges-universities, usps, v\_niah\_needles, various-supervised-sup, vctk, verbosity-control-training, verdicts, viethw-syn-data, vietnam-normalize-24k, vietnamese\_students\_feedback, viggo, viquiquad, visdial, visual-riddles, visual\_riddles, vitaminc, vivos, vlmu\_eval\_935rows, vocab\_tags, voice-dataset, voxpopuli, vqa, vqa-rad, vqasynth\_spacellava, vqav2-small, vsr\_random, vstar\_bench, waimai-classification, wds\_vtab-cifar100, wds\_vtab-clevr\_count\_all, wds\_vtab-resisc45, weather\_forecast\_japan, web\_nlg, web\_nlg, web\_nlg, web\_nlg, web\_questions, web\_questions, webgpt\_comparisons, webis-touche2020, websrc, websrc-test, wenigpt-agent-sft-1.0.1, wenigpt-agent-validation-1.0.5, whatsup\_all, whisper-training, whisper\_transcriptions.reazonspeech.all.wer\_10.0.vectorized, wider\_face, wiki-2024-new-knowledge-qa, wiki-2024-olympics-knowledge-qa, wiki-auto, wiki-ss-corpus, wiki\_auto\_all\_data, wiki\_bio, wiki\_bio, wiki\_bio\_gpt3\_hallucination, wiki\_dpr, wiki\_hop, wiki\_lingua, wiki\_lingua, wiki\_lingua, wiki\_medical\_terms, wiki\_medical\_terms\_llam2\_format, wiki\_movies, wiki\_qa, wiki\_qa, wiki\_toxic, wikiann, wikiart, wikiart\_recaption, wikidata-en-descriptions-small, wikimedqa, wikimusictext, wikiner\_fr, wikipedia, wikipedia, wikipedia, wikipedia, wikipedia-2023-11-embed-multilingual-v3, wikipedia-2023-11-reranking-multilingual, wikipedia-2023-11-retrieval-multilingual-corpus, wikipedia-2023-11-retrieval-multilingual-qrels, wikipedia-2023-11-retrieval-multilingual-queries, wikipedia-22-12, wikipedia-22-12-en-embeddings, wikipedia-22-12-simple-embeddings, wikipedia-cn-20230720-filtered, wikipedia-en-nov22-1-sentence-level, wikipedia-en-sentences, wikipedia-ja-20230720, wikipedia-korean-20240501-1million-qna, wikipedia-nq, wikipedia-trivia, wikipedia.fr, wikisql, wikitablequestions, wikitext, wikitext-103-filtered, wikitext-103-raw-v1, wikitext-103-raw-v1-rwkv-v5-tokenized, wikitext\_103\_detokenized, wikitext\_\_wikitext-2-raw-v1, wikitext\_alltime, wikitext\_document\_level, wikitext\_fr, wikitext\_latest, wikitext\_linked, wildguardmix, wildjailbreak, wildvision-arena-data, wildvision-chat, wine-quality, winobias\_antistereotype\_test\_v5, winogavil, winograd\_wsc, winogrande, winogrande, winogrande-bgeval, winogrande-tr, winoground, wiqa, wisconsin-breast-cancer, wit, wmdp, wmdp-corpora, wmdp\_fewshot, wmdp\_fewshot\_balanced, wmt14, wmt16, wmt17, wmt19, wnli-es, wnut\_17, word-embeddings-for-nmt, wouldyourather, wrime, wrime-sentiment, wuerstchen-dataset, wura, xCodeEval, xcopa, xcsr, xenium\_25\_lung\_dataset\_update1, xenium\_25\_lung\_dataset\_update2, xglue, xgqa, xlam-function-calling-60k, xlcost-text-to-code, xlsum, xlsum, xlsum, xlsum, xlsum\_ja, xlwic, xm3600, xmarket\_ml, xnli, xnli, xnli-eu, xnli2.0\_arabic, xnli2.0\_swahili, xnli2.0\_thai, xquad, xquad, xquad-ca, xstest-response, xstest-v2-copy, xstory\_cloze, xsum, xsum, xsum, xtreme, xtreme\_s, xvnli, xwinograd, xwinograd-ja, xwjzds-extractive-qa, yahoo\_answers\_topics, yahoo\_answers\_topics, yelp\_review\_full, yfcc100m\_openai\_subset, yodas, yodas2, yolksac\_human, yolochess\_lichess-elite\_2211, yoruba\_bbc\_topics, youtube\_caption\_corrections, zanya-custom-dataset-test, zara\_embeddings\_newcategories, zero\_scrolls, zeroth-korean, zest, zotero-articles, ztf-dr3-m31-features, 01-ai\_\_Yi-1.5-34B-Chat-details, 01-ai\_\_Yi-1.5-9B-Chat-details, 1-sentence-level-gutenberg-en\_arxiv\_pubmed\_soda, 10Kprompts-mini, 10k\_prompts\_ranked, 1dCA\_r2s20T20, 1million-gpt-4, 200,000 Jeopardy Questions in a JSON File, 2016-12-built-in-intents, 20\_newsgroups, 220k-GPT4Vision-captions-from-LIVIS, 2WikiMultihopQA, 3MAD-Tiny-1K, 8k\_tokenized\_audio\_dataset\_20k\_samples, 8k\_tokenized\_audio\_dataset\_2k\_samples, 8tags

\end{document}